\documentclass[twocolumn,aps,prd,groupedaddress,showpacs,nofootinbib]{revtex4}
\usepackage{graphicx}
\usepackage{amsmath}
\usepackage{amssymb}
\usepackage{multirow}
\usepackage{bm}
\usepackage{color}

\usepackage{relsize}
\RequirePackage{xspace}
\usepackage{dcolumn}
\usepackage{bm}

\def\jpsi{{J/\psi}}
\def\psip{{\psi^\prime}}
\def\chicj{{\chi_{cJ}}}
\def\chicz{{\chi_{c0}}}
\def\chico{{\chi_{c1}}}
\def\chict{{\chi_{c2}}}

\def\ss{{\bigl.^3\hspace{-1mm}S^{[1]}_1}}
\def\sps{{\bigl.^1\hspace{-1mm}S^{[8]}_0}}
\def\so{{\bigl.^3\hspace{-1mm}S^{[8]}_1}}
\def\pjs{{\bigl.^3\hspace{-1mm}P^{[1]}_J}}
\def\tpos{{\bigl.^3\hspace{-1mm}P^{[1]}_1}}
\def\tpts{{\bigl.^3\hspace{-1mm}P^{[1]}_2}}
\def\pj{{\bigl.^3\hspace{-1mm}P^{[8]}_J}}
\def\p0{{\bigl.^3\hspace{-1mm}P^{[8]}_0}}

\def\to{\rightarrow}

\def\be{\begin{equation}}
\def\ee{\end{equation}}
\def\bea{\begin{eqnarray}}
\def\eea{\end{eqnarray}}

\def\a{\alpha}
\def\th{\theta}

\begin{document}


\title{\mbox{}\\[10pt]
Spin correlations in polarizations of P-wave charmonia $\chi_{cJ}$
and impact on $J/\psi$ polarization}

\author{Hua-Sheng Shao$^{(a)}$,  Kuang-Ta Chao$^{(a,b,c)}$}
\affiliation{ {\footnotesize (a)~Department of Physics and State Key
Laboratory of Nuclear Physics and Technology, Peking University,
 Beijing 100871, China}\\
{\footnotesize (b)~Collaborative Innovation Center of Quantum Matter, Beijing 100871, China}\\
{\footnotesize (c)~Center for High Energy Physics, Peking
University, Beijing 100871, China}}




\begin{abstract}
Based on a general form of the effective vertex functions for the
decays of P-wave charmonia $\chicj$, angular distribution formulas
for the subsequent decays $\chicj\rightarrow \jpsi \gamma$ decay and
$\jpsi \to \mu^+\mu^-$ are derived. The formulas are the same as
those obtained in a different approach in the literature. Our
formulas are expressed in a more general form, including
parity violation effects and the full angular dependence of $\jpsi$
and muon in the cascade decay $\chicj\to\jpsi\gamma\to\mu^+\mu^-\gamma$. The $\chicj$ polarization observables are expressed in terms of
rational functions of the spin density matrix elements of $\chicj$
production. Generalized rotation-invariant relations for arbitrary
integer-spin particles are also derived and their expressions in
terms of observable angular distribution parameters are given in the
$\chi_{c1}$ and $\chi_{c2}$. To complement
our previous direct-$\jpsi$ polarization result, we also discuss the impact
on the observable prompt-$\jpsi$ polarization. As an illustrative
application of our angular distribution formulas, we present
the angular distributions in terms of the tree-level spin density
matrix elements of $\chi_{c1}$ and $\chi_{c2}$ production in several
different frames at the Large Hadron Collider. Moreover, a
reweighting method is also proposed to determine the entire set of
the production spin density matrix elements of the $\chi_{c2}$, some
of which disappear or are suppressed for vanishing higher-order multipole
effects making the complete extraction difficult experimentally.
\end{abstract}
\maketitle
\section{Introduction}
The polarization of heavy quarkonium in hadroproduction e.g. at the
Tevatron and the LHC, is a long-standing issue in heavy quarkonium
physics~\cite{Brambilla:2010cs}. Non-relativistic QCD
(NRQCD)~\cite{Bodwin:1994jh}, a rigorous effective field theory
founded on the nonrelativistic nature of heavy quarkonium, foresees
that a $Q\bar{Q}$ pair may be formed in a color-octet (CO) state
during the hard reaction at short-distances before it hadronizes
into a color-singlet (CS) physical quarkonium by radiating soft
gluons.
In particular, the $\jpsi$ ($c\bar{c}$ bound state with quantum
number $J^{PC}=1^{--}$), when produced at high transverse momentum
($p_T$), should predominantly originate from gluon fragmentation
into $c\bar{c}[\so]$, then evolve into the observed
$c\bar{c}[\ss]$~\cite{Braaten:1999qk}. The gluon fragmentation
mechanism guarantees that the $\jpsi$ is produced transversely
polarized in the helicity (HX) frame when its $p_T$ is sufficiently
large.
However, the data measured by the
CDF~\cite{Affolder:2000nn,Abulencia:2007us} Collaboration at the
Tevatron indicate that the $\jpsi$ is mainly unpolarized and even
slightly longitudinally polarized at large $p_T$, up to $20$ GeV.
This is the ``polarization puzzle" of heavy quarkonium production.

The understanding of charmonium polarization is also important for
the simulations in the experimental analyses: the detector
acceptance for lepton pairs from the decay of $\jpsi$ (or other
heavy quarkonia) strongly depends on the $\jpsi$
polarization~\cite{Faccioli:2010kd}. The lack of a consistent
description of the polarization in the simulation of quarkonium
production results in one of the largest systematic uncertainties
affecting the precision of cross section measurements.

Experimentally, measurements of direct-$\jpsi$ production at hadron
colliders are incomplete. The measured prompt $\jpsi$ data include
both direct production and feed-down contributions from $\chi_c$ and
$\psip$, through the decays $\chi_c\to \jpsi \gamma$ and $\psip \to
\jpsi \pi \pi$(plus a small contribution of $\psip \to \chi_c \gamma
\to \jpsi \gamma \gamma$). Therefore, in order to compare the
theoretical results with experimental data, the $\chi_c$ and $\psip$
yield and polarizations must also be calculated. Moreover, the
$\chi_c$ meson has its own phenomenological interest. The ratio of
the differential cross sections for the $\chico$ and $\chict$
inclusive productions at the Tevatron has been measured by the CDF
Collaboration~\cite{Abulencia:2007bra}. Their results show that the
ratio disagrees with the spin symmetry expectation from the
leading-order (LO) computation. After including the next-to-leading
(NLO) QCD radiative correction~\cite{Ma:2010vd}, the asymptotic
behavior of $\frac{\rm{d}\hat{\sigma}}{\rm{d}p_T^2}$ changes from
$p_T^{-6}$ at LO to $p_T^{-4}$ at NLO for the $\pjs$ channel and,
hence, becomes comparable to the contribution of $\so$ at large
$p_T$. The result provides an opportunity to solve the contradiction
between the experiment and the theoretical prediction. The recent
LHCb result~\cite{LHCb:2012ac} for
$\frac{\rm{d}\sigma_{\chict}}{\rm{d}\sigma_{\chico}}$ stays within
the error bars of the NLO NRQCD prediction. Surely, as in the
$\jpsi$ case, the investigation of polarization of $\chi_c$ will
also be very helpful in understanding the charmonium production
mechanisms in QCD.

We now briefly review the recent progress in the theory of heavy
quarkonium hadroproduction. In Ref.~\cite{Campbell:2007ws}, it was
found that the NLO prediction for the direct-$\jpsi$ yield in the
$\ss$ channel is 2 orders of magnitude larger than the LO one at
large $p_T$, while NLO corrections for the CO S wave are
small~\cite{Gong:2008ft}. For the P wave, the NLO corrections for
the $\pjs$ channel~\cite{Ma:2010vd} and $\pj$
channel~\cite{Butenschoen:2010rq,Ma:2010yw,Ma:2010jj} are found to
be very large but negative. As for the polarization, the NLO QCD
correction~\cite{Gong:2008zz} for the direct-$\jpsi$ in the CS changes it from
being transverse (LO) to longitudinal (NLO) in the HX frame.This can be
understood in collinear factorization up to the NLO power in
$\frac{m_c^2}{p_T^2}$~\cite{Kang:2011mg}. However, even after
including $\rm{NNLO^{\star}}$ corrections~\cite{Artoisenet:2008fc},
in which only tree-level diagrams at $\a_S^5$ are considered and
infrared cutoffs are imposed to avoid soft and collinear
divergences, theoretical predictions of CS contributions to the
yields and polarizations are still in disagreement with the CDF
data~\cite{Affolder:2000nn,Abulencia:2007us}. Recently, two
groups~\cite{Butenschoen:2012px,Chao:2012iv} have presented their
NLO results for the direct-$\jpsi$ production at hadron colliders,
but drawn very different conclusions due to different treatments for
the fit procedure of the available data and extracted different CO
long-distance matrix elements (LDMEs). The former
group~\cite{Butenschoen:2012px} uses a global fit, while the
latter~\cite{Chao:2012iv} just concentrates on hadroproduction,
including not only the $\jpsi$ yields but also the polarization
data. 
In the latter approach, the predictions of yields at the LHC are in
good agreement with data~\cite{Aad:2011sp,Chatrchyan:2011kc} up to
$70$ GeV, and the $\jpsi$ produced at the Tevatron and the LHC is
found to be almost unpolarized~\cite{Chao:2012iv}.

The angular distributions of the $\chi_c$ decay into $\jpsi+\gamma$
have been studied in Ref.~\cite{Kniehl:2003pc} and the authors have
also calculated the spin density matrix elements (SDMEs) of $\chico$
and $\chict$ at the Tevatron Run I. The formulas of the angular
distributions have also been derived in Ref.~\cite{Faccioli:2011be},
additionally considering the subsequent decay of the $\jpsi$ into a
lepton pair. In the present paper we rederive the same expressions
using a different formalism. Our formulas, like
Eqs.(\ref{eq:WFF}) and (\ref{eq:sdchij}) below, can be easily extended to derive
the correlations between the $\chi_c\to\jpsi\gamma$ and
$\jpsi\to\mu^+\mu^-$ angular distributions. For an illustrative
example to our derived results in the paper, we also compute the
tree-level yields and polarizations of the $\chico$ and $\chict$
inclusive productions at the LHC for a center-of-mass energy of $8$
TeV. Our complete NLO NRQCD predictions of $\chico$ and $\chict$
including yields and polarizations are given in Ref.~\cite{Shao:2014fca}.

The organization of the paper is as follows. In Sec. \ref{sec:2},
we introduce the basic kinematics and conventions used in the paper.
In the next three sections, we derive the angular distributions of
the $\jpsi$ and the $\mu^+$ from the decays of $\chico$ and
$\chict$. In Sec. \ref{sec:7}, we generalize the
rotation-invariant relations from the vector boson to arbitrary
integer-spin particles. In Sec. \ref{sec:8}, we estimate the
impact of the feed-down contributions from $\chi_c$ and $\psip$ on
the prompt-$\jpsi$ polarization. In Sec. \ref{sec:9}, we present
an example to illustrate our derived formulas. Finally, the
conclusion is drawn in the last section. General expressions of the
decay angular distributions for spin-1 and spin-2 bosons taking
into account higher-order radiation multipoles and allowing for
parity-violating effects, are presented in the Appendixes
\ref{app:a}, \ref{app:b}, \ref{app:c}. A reweighting method is also
proposed to extract the complete set of the SDMEs of the $\chict$ in
Appendix \ref{app:d}.

\section{Kinematics and conventions\label{sec:2}}
In this section, we introduce the conventions and kinematics for our
derivations performed in the following sections. Apart from $\chi_c\to \jpsi+\gamma$, we also consider the
subsequent $\jpsi \to \mu^+ \mu^-$. The spin quantization axis
$\overrightarrow{s}$ can be chosen arbitrarily in the rest frame of
the decaying particle. Generally, the polarization vectors for a
massive spin-1 particle are
\begin{eqnarray}
\epsilon^{\mu}_{0}&=&(|\overrightarrow{k}|,E\sin{\th}\cos{\phi},E\sin{\th}\sin{\phi},E\cos{\th})/m,\nonumber\\
\epsilon^{\mu}_{\pm}&=&\frac{e^{\mp
i\gamma}}{\sqrt{2}}(0,\mp\cos{\th}\cos{\phi}+i\sin{\phi},\nonumber\\&&\mp\cos{\th}\sin{\phi}-i\cos{\phi},\pm\sin{\th}),
\end{eqnarray}
where $\th$ and $\phi$ are the polar and azimuthal decay angles with
respect to $\overrightarrow{s}$ and a chosen plane,\footnote{The
plane is an important component to define the polarization frames.
At the end of this section, we will fix our chosen polarization axis
and corresponding plane.} and the symbol $\gamma$ can be chosen as
an arbitrary real number. $E$, $\overrightarrow{k}$, and $m$ are the
particle's energy, momentum, and mass. We set $\gamma=-\phi$ here.
For a spin-2 tensor particle, its spin wave functions can be
constructed from the spin-1 polarization four-vectors as
\begin{eqnarray}
\epsilon^{\mu\nu}_{\lambda}&=&\sum^{1}_{\lambda_1,\lambda_2=-1}{\langle1,\lambda_1;1,\lambda_2|2,\lambda\rangle\epsilon^{\mu}_{\lambda_1}\epsilon^{\nu}_{\lambda_2}},
\end{eqnarray}
where $\langle1,\lambda_1;1,\lambda_2|2,\lambda\rangle$ are the
Clebsch-Gordan coefficients, and $\lambda,\lambda_1,\lambda_2$ denote
the angular distribution components along the spin-quantization axis
$\overrightarrow{s}$. Thus, we have the identities
$p_{\mu}\epsilon^{\mu}_{\lambda}=p_{\mu}\epsilon^{\mu\nu}_{\lambda}=p_{\nu}\epsilon^{\mu\nu}_{\lambda}=(\epsilon_{\lambda})^{\mu}_{\mu}=0$
and $\epsilon^{\mu\nu}_{\lambda}=\epsilon^{\nu\mu}_{\lambda}$.

The $\chicj\rightarrow\jpsi \gamma$ angular distribution can be
written in terms of the $\chicj$ production SDMEs
$\rho_{\lambda\lambda^{\prime}}$ and of the decay SDMEs
$D_{\lambda\lambda^{\prime}}$,
\begin{eqnarray}
\mathcal{W}(\th,\phi)&=&\sum^{J}_{\lambda,\lambda^{\prime}=-J}{\rho_{\lambda\lambda^{\prime}}D_{\lambda\lambda^{\prime}}(\th,\phi)},\label{eq:WW}
\end{eqnarray}
where $\th$ and $\phi$ are the angles parameterizing the $\jpsi$
direction in the $\chicj$ rest frame. Here,
$\rho_{\lambda\lambda^{\prime}}$ and $D_{\lambda\lambda^{\prime}}$
represent the production and decay amplitudes of the $\chicj$ with
angular momentum projector component $\lambda$ along
$\overrightarrow{s}$ multiplied by the corresponding complex
conjugate amplitudes with component $\lambda^{\prime}$.

Several polarization frame definitions have been used in the
literature to fully describe the polarization of heavy
quarkonium~\cite{Beneke:1998re}, i.e. the HX (recoil or s-channel
helicity) frame, the Collins-Soper frame, the Gottfried-Jackson
frame, and the target frame.\footnote{Another useful frame is the``perpendicular helicity frame"~\cite{Braaten:2008mz,Braaten:2008xg}.
It has been used in the $\Upsilon$ polarization
measurement~\cite{Chatrchyan:2012woa} by the CMS Collaboration.} In the HX frame,
$\overrightarrow{s}$ is chosen as the flight direction of the
decaying quarkonium. In the Collins-Soper frame,
\begin{eqnarray}
\overrightarrow{s}=\left(\overrightarrow{p_1}/|\overrightarrow{p_1}|-\overrightarrow{p_2}/|\overrightarrow{p_2}|\right)/|\overrightarrow{p_1}/|\overrightarrow{p_1}|-\overrightarrow{p_2}/|\overrightarrow{p_2}||,
\end{eqnarray}
where $\overrightarrow{p_1}$ and $\overrightarrow{p_2}$ denote the
momenta of the two initial state colliding particles in the rest
frame of the decaying quarkonium. In the Gottfried-Jackson frame,
$\overrightarrow{s}=\frac{\overrightarrow{p_1}}{|\overrightarrow{p_1}|}$,
and in the target frame,
$\overrightarrow{s}=-\frac{\overrightarrow{p_2}}{|\overrightarrow{p_2}|}$.
All the definitions of
the X, Y, Z coordinates can be found in Ref.~\cite{Beneke:1998re}.
In particular, the Y coordinate points in the direction of
$\overrightarrow{p_1}\times(-\overrightarrow{p_2})$ in the $\chi_c$
rest frame.

\section{Angular distribution of $\chico \rightarrow \jpsi\gamma$\label{sec:3}}
The general vertex function for the decay of a vector or axial
vector particle into two vector particles can be expressed as
\begin{eqnarray}
\mathcal{M}(V_{0}\rightarrow
V_{1}V_{2})&=&f_1~(\epsilon_{V_{0}}\cdot\epsilon^*_{V_{1}})[\epsilon^*_{V_{2}}\cdot(-p_{V_{0}}-p_{V_{1}})]\\
&+&f_2~(\epsilon_{V_{0}}\cdot\epsilon^*_{V_{2}})[\epsilon^*_{V_{1}}\cdot(p_{V_{0}}+p_{V_{2}})]\nonumber\\
&+&f_3~(\epsilon^*_{V_{1}}\cdot\epsilon^*_{V_{2}})[\epsilon_{V_{0}}\cdot(p_{V_{1}}-p_{V_{2}})]\nonumber\\
&+&f_4~[\epsilon_{V_{0}}\cdot(p_{V_{1}}-p_{V_{2}})]\nonumber\\&&[\epsilon^*_{V_{1}}\cdot(p_{V_{0}}+p_{V_{2}})]
[\epsilon^*_{V_{2}}\cdot(p_{V_{0}}+p_{V_{1}})]\nonumber\\&+&f_5~i\varepsilon_{\epsilon_{V_{0}}\epsilon^{*}_{V_{1}}\epsilon^{*}_{V_{2}}p_{V_{0}}}\nonumber\\
&+&f_6~[\epsilon^*_{V_{1}}\cdot(p_{V_{0}}+p_{V_{2}})]i\varepsilon_{\epsilon_{V_{0}}\epsilon^{*}_{V_{2}}p_{V_{0}}p_{V_{2}}}\nonumber\\
&+&f_7~[\epsilon^*_{V_{2}}\cdot(p_{V_{0}}+p_{V_{1}})]i\varepsilon_{\epsilon_{V_{0}}\epsilon^{*}_{V_{1}}p_{V_{0}}p_{V_{1}}},\nonumber
\end{eqnarray}
where
\begin{eqnarray*}
p_{V_{0}}&=&p_{V_{1}}+p_{V_{2}}.
\end{eqnarray*}
and $\varepsilon_{\mu\nu\rho\sigma}$ is the antisymmetric
Levi-Civita tensor. A special notation about the vector contracting
with the Levi-Civita tensor is used, for example,
$\varepsilon_{\mu\nu\rho
k}\equiv\varepsilon_{\mu\nu\rho\sigma}k^{\sigma},\varepsilon_{\mu
q\nu
k}\equiv\varepsilon_{\mu\rho\nu\sigma}q^{\rho}k^{\sigma}$. Specifically,
in the case of $\chico
(1^{++})\rightarrow\jpsi(1^{--})\gamma(1^{--})$, the
$f_1,f_2,f_3,f_4$ terms and the $f_7$ term can be dropped because
of parity conservation in QED and the absence of a longitudinal
polarization component for the photon. If we just consider the
electric dipole (E1) transition, which is the dominant contribution
according to the velocity scaling rule in NRQCD, the $f_6$ term can
also be neglected. Therefore, we calculate the helicity amplitudes
$\mathcal{M}_{\lambda_{\chico}\lambda_{\jpsi}\lambda_{\gamma}}$ for
$\chico\rightarrow\jpsi\gamma$ as
\begin{eqnarray}
\mathcal{M}_{+++}&=&\mathcal{M}^*_{---}=\frac{m_{\chico}e^{-i\phi}\sin{\th}}{\sqrt{2}},\nonumber\\
\mathcal{M}_{+--}&=&\mathcal{M}^*_{-++}=-\frac{m_{\chico}e^{3i\phi}\sin{\th}}{\sqrt{2}},\nonumber\\
\mathcal{M}_{+0+}&=&-\mathcal{M}_{-0-}=-\frac{(m_{\chico}^2+m_{\jpsi}^2)\sin^2{\frac{\th}{2}}}{2m_{\jpsi}},\nonumber\\
\mathcal{M}_{+0-}&=&-\mathcal{M}^*_{-0+}=\frac{(m_{\chico}^2+m_{\jpsi}^2)e^{2i\phi}\cos^2{\frac{\th}{2}}}{2m_{\jpsi}},\nonumber\\
\mathcal{M}_{0++}&=&-\mathcal{M}^*_{0--}=-m_{\chico}e^{-2i\phi}\cos{\th},\nonumber\\
\mathcal{M}_{00+}&=&\mathcal{M}^*_{00-}=\frac{(m_{\chico}^2+m_{\jpsi}^2)e^{-i\phi}\sin{\th}}{2\sqrt{2}m_{\jpsi}},\label{eq:amp1}
\end{eqnarray}
where a factor $f_5 (m_{\chico}^2-m_{\jpsi}^2)$ common to all
amplitudes has been omitted. The decay SDMEs are obtained as
$D_{\lambda\lambda^{\prime}}=\sum_{\lambda_{\jpsi},\lambda_{\gamma}}{\mathcal{M}_{\lambda\lambda_{\jpsi}\lambda_{\gamma}}\mathcal{M}^*_{\lambda^{\prime}\lambda_{\jpsi}\lambda_{\gamma}}}$.
Using these ingredients [i.e. Eqs.(\ref{eq:WW}) and (\ref{eq:amp1})] and
assuming $m_{\chico}=m_{\jpsi}$, we can work out the general form of
the angular distribution of $\chico\to\jpsi\gamma$:
\begin{eqnarray}
\mathcal{W}^{\chico\rightarrow \jpsi
\gamma}(\th,\phi)&\propto&\frac{N_{\chico\rightarrow \jpsi
\gamma}}{3+\lambda_{\th}}\left(1+\lambda_{\th}\cos^2{\th}\right.\label{eq:wchic1}\\
&+&\lambda_{\phi}\sin^2{\th}\cos{2\phi}+\lambda_{\th\phi}\sin{2\th}\cos{\phi}\nonumber\\
&+&\left.\lambda^{\perp}_{\phi}\sin^2{\th}\sin{2\phi}+\lambda^{\perp}_{\th\phi}\sin{2\th}\sin{\phi}\right),\nonumber
\end{eqnarray}
with
\begin{eqnarray}
\lambda_{\th}&=&\frac{3\rho_{0,0}-N_{\chico}}{3N_{\chico}-\rho_{0,0}},\lambda_{\phi}=-\frac{2\Re{\rho_{1,-1}}}{3N_{\chico}-\rho_{0,0}},\nonumber\\
\lambda_{\th\phi}&=&-\frac{\sqrt{2}(\Re{\rho_{1,0}}-\Re{\rho_{-1,0}})}{3N_{\chico}-\rho_{0,0}},\nonumber\\
\lambda^{\perp}_{\phi}&=&\frac{2\Im{\rho_{1,-1}}}{3N_{\chico}-\rho_{0,0}},\lambda^{\perp}_{\th\phi}=\frac{\sqrt{2}(\Im{\rho_{1,0}}+\Im{\rho_{-1,0}})}{3N_{\chico}-\rho_{0,0}},
\label{eq:chic1}
\end{eqnarray}
where $\rho_{i,j}$ are SDMEs for the $\chico$ yields and
$N_{\chico}=\rho_{1,1}+\rho_{0,0}+\rho_{-1,-1}$.

\section{Angular distribution of $\chict \rightarrow \jpsi \gamma$\label{sec:4}}
Similarly, in the $\chict$ case, we can write down the general vertex
function for a spin-2 tensor particle $T$ decaying into two vector particles
\begin{eqnarray}
\mathcal{M}(T\rightarrow
V_{1}V_{2})&=&g_1~\epsilon^*_{V_2}\cdot\epsilon_{T}\cdot\epsilon^*_{V_1}\nonumber\\
&+&g_2~[(p_{V_1}-p_{V_2})\cdot\epsilon_{T}\cdot\epsilon^*_{V_1}]
[\epsilon^*_{V_2}\cdot(-p_{T}-p_{V_1})]\nonumber\\
&+&g_3~[(p_{V_1}-p_{V_2})\cdot\epsilon_{T}\cdot\epsilon^*_{V_2}]
[\epsilon^*_{V_1}\cdot(p_{T}+p_{V_2})]\nonumber\\
&+&g_4~[(p_{V_1}-p_{V_2})\cdot\epsilon_{T}\cdot(p_{V_1}-p_{V_2})]
(\epsilon^*_{V_1}\cdot\epsilon^*_{V_2})\nonumber\\
&+&g_5~[(p_{V_1}-p_{V_2})\cdot\epsilon_{T}\cdot(p_{V_1}-p_{V_2})]\nonumber\\
&&[\epsilon^*_{V_1}\cdot(p_{T}+p_{V_2})]
[\epsilon^*_{V_2}\cdot(p_{T}+p_{V_1})]\nonumber\\
&+&{\small\text{Levi-Civita~terms}},
\end{eqnarray}
where $p_{T}=p_{V_{1}}+p_{V_{2}}$. Because of parity conservation, we
drop the Levi-Civita terms in
$\chict(2^{++})\rightarrow\jpsi(1^{--}) \gamma(1^{--})$. The
$g_2,g_3,g_4,g_5$ terms can also be ignored in consideration of the
fact that we only include the leading-order contribution, i.e., the
E1 transition, and these terms are suppressed by
$(m_{\chict}^2-m_{\jpsi}^2)^2$ as compared to the $g_1$
term. Moreover, some of these terms vanish exactly
when the photon is transversely polarized. Thus, the helicity
amplitudes
$\mathcal{M}_{\lambda_{\chict}\lambda_{\jpsi}\lambda_{\gamma}}$ for
$\chict\rightarrow \jpsi\gamma$ become
\begin{eqnarray}
\mathcal{M}_{2++}&=&\mathcal{M}_{-2--}=\frac{\sin^2{\th}}{4},\nonumber\\
\mathcal{M}_{2+-}&=&\mathcal{M}^*_{-2-+}=e^{2i\phi}\cos^4{\frac{\th}{2}},\nonumber\\
\mathcal{M}_{2--}&=&\mathcal{M}^*_{-2++}=\frac{e^{4i\phi}\sin^2{\th}}{4},\nonumber\\
\mathcal{M}_{2-+}&=&\mathcal{M}^*_{-2+-}=e^{2i\phi}\sin^4{\frac{\th}{2}},\nonumber\\
\mathcal{M}_{20+}&=&-\mathcal{M}^*_{-20-}=-\frac{m_{\chict}^2+m_{\jpsi}^2}{\sqrt{2}m_{\chict}m_{\jpsi}}e^{i\phi}\sin^3{\frac{\th}{2}}\cos{\frac{\th}{2}},\nonumber\\
\mathcal{M}_{20-}&=&-\mathcal{M}^*_{-20+}=-\frac{m_{\chict}^2+m_{\jpsi}^2}{\sqrt{2}m_{\chict}m_{\jpsi}}e^{3i\phi}\cos^3{\frac{\th}{2}}\sin{\frac{\th}{2}},\nonumber\\
\mathcal{M}_{1++}&=&-\mathcal{M}^*_{-1--}=-\frac{e^{-i\phi}\sin{2\th}}{4},\nonumber\\
\mathcal{M}_{1+-}&=&-\mathcal{M}^*_{-1-+}=2e^{i\phi}\cos^3{\frac{\th}{2}}\sin{\frac{\th}{2}},\nonumber\\
\mathcal{M}_{1--}&=&-\mathcal{M}^*_{-1++}=-\frac{e^{3i\phi}\sin{2\th}}{4},\nonumber\\
\mathcal{M}_{1-+}&=&-\mathcal{M}^*_{-1+-}=-2e^{i\phi}\sin^3{\frac{\th}{2}}\cos{\frac{\th}{2}},\nonumber\\
\mathcal{M}_{10+}&=&\mathcal{M}_{-10-}=\frac{m_{\chict}^2+m_{\jpsi}^2}{2\sqrt{2}m_{\chict}m_{\jpsi}}\sin^2{\frac{\th}{2}}(1+2\cos{\th}),\nonumber\\
\mathcal{M}_{10-}&=&\mathcal{M}^*_{-10+}=\frac{m_{\chict}^2+m_{\jpsi}^2}{2\sqrt{2}m_{\chict}m_{\jpsi}}e^{2i\phi}\cos^2{\frac{\th}{2}}(2\cos{\th}-1),\nonumber\\
\mathcal{M}_{0++}&=&\mathcal{M}^*_{0--}=\frac{e^{-2i\phi}(1+3\cos{2\th})}{4\sqrt{6}},\nonumber\\
\mathcal{M}_{0+-}&=&\mathcal{M}_{0-+}=\frac{3\sin^2{\th}}{2\sqrt{6}},\nonumber\\
\mathcal{M}_{00+}&=&-\mathcal{M}^*_{00-}=-\frac{\sqrt{3}(m_{\chict}^2+m_{\jpsi}^2)}{8m_{\chict}m_{\jpsi}}e^{-i\phi}\sin{2\th}.\label{eq:hel}
\end{eqnarray}
The angular distribution of the $\chict\to\jpsi\gamma$ decay has, in the E1 approximation and assuming $m_{\chict}=m_{\jpsi}$, the same general
expression of the $\chico\to\jpsi\gamma$ case:
\begin{eqnarray}
\mathcal{W}^{\chict\rightarrow \jpsi
\gamma}(\th,\phi)&\propto&\frac{N_{\chict\rightarrow \jpsi
\gamma}}{3+\lambda_{\th}}\left(1+\lambda_{\th}\cos^2{\th}\right.\label{eq:wchic2}\\
&+&\lambda_{\phi}\sin^2{\th}\cos{2\phi}+\lambda_{\th\phi}\sin{2\th}\cos{\phi}\nonumber\\
&+&\left.\lambda^{\perp}_{\phi}\sin^2{\th}\sin{2\phi}+\lambda^{\perp}_{\th\phi}\sin{2\th}\sin{\phi}\right)\nonumber
\end{eqnarray}
with
\begin{eqnarray}
\lambda_{\th}&=&\frac{6N_{\chict}-9(\rho_{1,1}+\rho_{-1,-1})-12\rho_{0,0}}{6N_{\chict}+3(\rho_{1,1}+\rho_{-1,-1})+4\rho_{0,0}},\nonumber\\
\lambda_{\phi}&=&\frac{2\sqrt{6}(\Re{\rho_{2,0}}+\Re{\rho_{-2,0}})+6\Re{\rho_{1,-1}}}{6N_{\chict}+3(\rho_{1,1}+\rho_{-1,-1})+4\rho_{0,0}},\nonumber\\
\lambda_{\th\phi}&=&\frac{6(\Re{\rho_{2,1}}-\Re{\rho_{-2,-1}})+\sqrt{6}(\Re{\rho_{1,0}}-\Re{\rho_{-1,0}})}{6N_{\chict}+3(\rho_{1,1}+\rho_{-1,-1})+4\rho_{0,0}},\nonumber\\
\lambda^{\perp}_{\phi}&=&-\frac{2\sqrt{6}(\Im{\rho_{2,0}}-\Im{\rho_{-2,0}})+6\Im{\rho_{1,-1}}}{6N_{\chict}+3(\rho_{1,1}+\rho_{-1,-1})+4\rho_{0,0}},\nonumber\\
\lambda^{\perp}_{\th\phi}&=&-\frac{6(\Im{\rho_{2,1}}+\Im{\rho_{-2,-1}})+\sqrt{6}(\Im{\rho_{1,0}}+\Im{\rho_{-1,0}})}{6N_{\chict}+3(\rho_{1,1}+\rho_{-1,-1})+4\rho_{0,0}},\nonumber\\
N_{\chict}&=&\rho_{2,2}+\rho_{1,1}+\rho_{0,0}+\rho_{-1,-1}+\rho_{-2,-2}.
\label{eq:chic2}
\end{eqnarray}

If the magnetic quadrupole
(M2) and electric octupole (E3) contributions are also taken into account by keeping the relevant terms in the vertex functions,
the expression of the angular distribution acquires further terms in the $\chict$ case while, both
for $\chico$ and $\chict$, the existing terms are modified. The modifications depend on
one additional coefficient (expressing the fractional M2 amplitude contribution) in the $\chico$ case
and on two additional coefficients (M2 and E3 contributions) in the $\chict$ case.
If these coefficients are not $< \mathcal{O}(1\%)$, they can modify the angular distributions
significantly~\cite{Faccioli:2011be}. However, inconsistencies in their current experimental determinations
exist~\cite{Artuso:2009aa,Oreglia:1981fx,Armstrong:1993fk,Ambrogiani:2001jw}.
The complete formulas for the angular distributions including the higher-order multipole effects are presented
in the appendixes, while only the E1 transition is considered, for simplicity, throughout the body of the paper.

Our results [Eqs.(\ref{eq:chic1}) and (\ref{eq:chic2})] are exactly
the same as those given in
Refs.~\cite{Kniehl:2003pc,Faccioli:2011be}. Actually, our
derivations, which are based on the general effective decay vertex
functions, are equivalent to those obtained there using the angular
momentum conservation, since the effective amplitudes written by us
are also originated from general considerations on the spins of the
involved particles.

\section{Lepton distribution in $\chicj\rightarrow\jpsi \gamma\rightarrow\mu^+\mu^-\gamma$\label{sec:5}}
We are now in a position to investigate the $\mu^+$ angular
distributions from the cascade decay $\chicj\rightarrow\jpsi
\gamma\rightarrow\mu^+\mu^-\gamma$.

We start from a general formalism to study it. We denote with
$\overrightarrow{s_1}$ and $\overrightarrow{s_2}$ the quantization
axes of the $\chicj$ and of the $\jpsi$, respectively. The general
form of the angular distribution of the $\mu^+$ is
\begin{eqnarray}
&&\mathcal{W}^{\chicj\rightarrow\jpsi
\gamma\rightarrow\mu^+\mu^-\gamma}(\th^{\prime},\phi^{\prime})\nonumber\\&=&\int{\rm{d}^2\Omega[\th,\phi]\rho^{\chicj}_{J_z,J_z^{\prime}}
\mathcal{M}^{\chicj\rightarrow\jpsi\gamma}_{J_zs_z\lambda_{\gamma}}(\th,\phi)\mathcal{M}^{\jpsi\rightarrow\mu^+\mu^-}_{s_z\lambda_{\mu^+}\lambda_{\mu^-}}}(\th^{\prime},\phi^{\prime})\nonumber\\
&&\left(\mathcal{M}^{\chicj\rightarrow\jpsi\gamma}_{J_z^{\prime}s_z^{\prime}\lambda_{\gamma}}(\th,\phi)\mathcal{M}^{\jpsi\rightarrow\mu^+\mu^-}_{s_z^{\prime}\lambda_{\mu^+}\lambda_{\mu^-}}(\th^{\prime},\phi^{\prime})\right)^*,
\label{eq:WFF}
\end{eqnarray}
where the $\rho^{\chicj}_{J_z,J_z^{\prime}}$ coefficients are the
production SDMEs of the $\chicj$,
$\mathcal{M}^{\chicj\rightarrow\jpsi\gamma}_{J_zs_z\lambda_{\gamma}},\mathcal{M}^{\jpsi\rightarrow\mu^+\mu^-}_{s_z\lambda_{\mu^+}\lambda_{\mu^-}}$
are the amplitudes of the two successive decays,{\footnote{We use the
general vector current amplitudes for the $\jpsi$ decay into a muon
pair.}} $J_z$ is the $\chicj$ angular momentum projection with
respect to $\overrightarrow{s_1}$, $s_z$ the $\jpsi$ angular
momentum projection with respect to $\overrightarrow{s_2}$,
$\lambda_{\gamma},\lambda_{\mu^+},\lambda_{\mu^-}$ the photon and
lepton helicities. The angles $\th$ and $\phi$ define the $\jpsi$
direction in the $\chicj$ rest frame with respect to
$\overrightarrow{s_1}$. $\th^{\prime}$ and $\phi^{\prime}$ determine the
$\mu^+$ direction in the $\jpsi$ rest frame with respect to
$\overrightarrow{s_2}$. Indices appearing twice imply a summation,
with $J_z,J_z^{\prime}=\pm
J,\pm(J-1),\ldots,0$,$~s_z,s_z^{\prime}=\pm1,0$, and
$\lambda_{\gamma},\lambda_{\mu^+},\lambda_{\mu^-}=\pm1$.

We will consider two different definitions of
$\overrightarrow{s_2}$. In the first option, $\overrightarrow{s_2}$
is the flight direction of the $\jpsi$ in the rest frame of the
$\chicj$. The $\jpsi\to\mu^+\mu^-$ angular disribution can be
parametrized in the same form of
Eqs.(\ref{eq:wchic1}) and (\ref{eq:wchic2}), with five observable
coefficients depending on the $\chicj$ SDMEs
$\rho^{\chicj}_{J_z,J_z^{\prime}}$:
\begin{eqnarray}
\lambda^{\chico}_{\th^{\prime}}&=&-\frac{1}{3},\lambda^{\chico}_{\phi^{\prime}}=\lambda^{\perp\chico}_{\phi^{\prime}}=0,\label{eq:mu1}\\
\lambda^{\chico}_{\th^{\prime}\phi^{\prime}}&=&\frac{\sqrt{2}(\Re(\rho^{\chico}_{1,0})-\Re(\rho^{\chico}_{-1,0}))}{12
N_{\chico}},\nonumber\\
\lambda^{\perp\chico}_{\th^{\prime}\phi^{\prime}}&=&-\frac{\sqrt{2}(\Im(\rho^{\chico}_{1,0})+\Im(\rho^{\chico}_{-1,0}))}{12
N_{\chico}},\nonumber\\
\lambda^{\chict}_{\th^{\prime}}&=&\frac{1}{13},\nonumber\\
\lambda^{\chict}_{\phi^{\prime}}&=&\frac{7\sqrt{6}(\Re{\rho^{\chict}_{0,2}}+\Re{\rho^{\chict}_{0,-2}})+12\Re{\rho^{\chict}_{1,-1}}}{78N_{\chict}},\nonumber\\
\lambda^{\chict}_{\th^{\prime}\phi^{\prime}}&=&\frac{\sqrt{6}(\Re{\rho^{\chict}_{-1,0}}-\Re{\rho^{\chict}_{1,0}})+24(\Re{\rho^{\chict}_{-2,-1}}-\Re{\rho^{\chict}_{2,1}})}{156N_{\chict}},\nonumber\\
\lambda^{\perp\chict}_{\phi^{\prime}}&=&\frac{7\sqrt{6}(\Im{\rho^{\chict}_{0,2}}-\Im{\rho^{\chict}_{0,-2}})-12\Im{\rho^{\chict}_{1,-1}}}{78N_{\chict}},\nonumber\\
\lambda^{\perp\chict}_{\th^{\prime}\phi^{\prime}}&=&\frac{\sqrt{6}(\Im{\rho^{\chict}_{-1,0}}+\Im{\rho^{\chict}_{1,0}})+24(\Im{\rho^{\chict}_{-2,-1}}+\Im{\rho^{\chict}_{2,1}})}{156N_{\chict}},\nonumber
\end{eqnarray}
with
$N_{\chicj}=\sum^{J}_{\lambda=-J}{\rho^{\chicj}_{\lambda\lambda}}$.
We see that the same results can be obtained in another
formalism using the language of angular momentum theory. The spin
correlations between the $\chicj$ production SDMEs
$\rho^{\chicj}_{J_z,J_z^{\prime}}$, referred to the quantization
axis $\overrightarrow{s_1}$ and the SDMEs of the $\jpsi$ coming
from $\chicj$, $\rho^{\chicj\rightarrow
\jpsi\gamma}_{s_z,s_z^{\prime}}$ referred to
$\overrightarrow{s_2}$, can be expressed as
\begin{eqnarray}
&&\rho^{\chicj\rightarrow
\jpsi\gamma}_{s_z,s_z^{\prime}}=\frac{3}{8\pi}\int{\rm{d}\Omega[\th,\phi]\rho^{\chicj}_{J_z,J_z^{\prime}}\mathcal{D}^{J*}_{J_z,I_{z}}\mathcal{D}^{J}_{J_z^{\prime},I_{z}^{\prime}}}\\
&&\langle1,\lambda_{\gamma};1,s_z|J,I_{z}\rangle \langle
J,I_{z}^{\prime}|1,\lambda_{\gamma};1,s_z^{\prime}\rangle
\rm{Br}(\chicj\rightarrow\jpsi\gamma),\nonumber
\end{eqnarray}
where implicit summations run over
$J_z,J_z^{\prime},I_{z},I_{z}^{\prime}=\pm J,\pm (J-1),\ldots,0$
and over $\lambda_{\gamma}=\pm1$,
$\rm{Br}(\chicj\rightarrow\jpsi\gamma)$ is the branching ratio of
the radiative decay and
$\mathcal{D}^{J}_{J_z,J_z^{\prime}}\equiv\mathcal{D}^{J}_{J_z,J_z^{\prime}}(-\phi,\th,\phi)=e^{i\phi(J_z-J_z^{\prime})}d^{J}_{J_z,J_z^{\prime}}(\th)$,
$d^{J}_{J_z,J_z^{\prime}}(\th)$ being the well-known Wigner
$d$ function
\begin{eqnarray}
&&d^{J}_{J_z,J_z^{\prime}}(\th)=\sum^{\min(J+J_z^{\prime},J-J_z)}_{k=\max(0,J_z^{\prime}-J_z)}{(-)^{k-J_z^{\prime}+J_z}}\\
&&\times\frac{\sqrt{(J+J_z^{\prime})!(J-J_z^{\prime})!(J+J_z)!(J-J_z)!}}{(J+J_z^{\prime}-k)!k!(J-J_z-k)!(k-J_z^{\prime}+J_z)!}\nonumber\\
&&\times\left(\cos{\frac{\th}{2}}\right)^{2J-2k+J_z^{\prime}-J_z}\left(\sin{\frac{\th}{2}}\right)^{2k-J_z^{\prime}+J_z}.\nonumber
\end{eqnarray}
One can easily verify that after substituting
$\rho^{\chicj\rightarrow \jpsi\gamma}_{s_z,s_z^{\prime}}$ calculated
from the above equation into the well-known expression of the
angular distribution of the muon from $\jpsi\rightarrow\mu^+\mu^-$,
\begin{eqnarray}
\lambda^{\jpsi}_{\th^{\prime}}&=&\frac{N_{\jpsi}-3\rho^{\jpsi}_{0,0}}{N_{\jpsi}+\rho^{\jpsi}_{0,0}},\nonumber\\
\lambda^{\jpsi}_{\phi^{\prime}}&=&\frac{2\Re{\rho^{\jpsi}_{1,-1}}}{N_{\jpsi}+\rho^{\jpsi}_{0,0}},\nonumber\\
\lambda^{\jpsi}_{\th^{\prime}\phi^{\prime}}&=&\frac{\sqrt{2}(\Re{\rho^{\jpsi}_{1,0}}-\Re{\rho^{\jpsi}_{-1,0}})}{N_{\jpsi}+\rho^{\jpsi}_{0,0}},\nonumber\\
\lambda^{\perp\jpsi}_{\phi^{\prime}}&=&-\frac{2\Im{\rho^{\jpsi}_{1,-1}}}{N_{\jpsi}+\rho^{\jpsi}_{0,0}},\nonumber\\
\lambda^{\perp\jpsi}_{\th^{\prime}\phi^{\prime}}&=&-\frac{\sqrt{2}(\Im{\rho^{\jpsi}_{1,0}}+\Im{\rho^{\jpsi}_{-1,0}})}{N_{\jpsi}+\rho^{\jpsi}_{0,0}},
\label{eq:jpsi}
\end{eqnarray}
the expressions in Eq.(\ref{eq:mu1}) are recovered.

As a second option, $\overrightarrow{s_2}$ is chosen as coinciding
with $\overrightarrow{s_1}$. The $\jpsi$ spin state $|1,s_z\rangle$
with respect to $\overrightarrow{s_2}$ is no longer its helicity
state, like in the first case. This option is actually a much``easier" choice for the experiment, at least at not very low $\jpsi$
momentum, because it does not require the use of the photon momentum,
and it actually coincides with the usual set of reference frames
adopted in the study of prompt $\jpsi$~\cite{Faccioli:2011be}. With
a direct calculation following Eq.(\ref{eq:WFF}), we obtain
\begin{eqnarray}
\lambda^{\chico}_{\th^{\prime}}&=&\frac{-N_{\chico}+3\rho^{\chico}_{0,0}}{3N_{\chico}-\rho^{\chico}_{0,0}},\label{eq:mu2}\\
\lambda^{\chico}_{\phi^{\prime}}&=&-\frac{2\Re{\rho^{\chico}_{1,-1}}}{3N_{\chico}-\rho^{\chico}_{0,0}},\nonumber\\
\lambda^{\chico}_{\th^{\prime}\phi^{\prime}}&=&-\frac{\sqrt{2}(\Re{\rho^{\chico}_{1,0}}-\Re{\rho^{\chico}_{-1,0}})}{3N_{\chico}-\rho^{\chico}_{0,0}},\nonumber\\
\lambda^{\perp\chico}_{\phi^{\prime}}&=&\frac{2\Im{\rho^{\chico}_{1,-1}}}{3N_{\chico}-\rho^{\chico}_{0,0}},\nonumber\\
\lambda^{\perp\chico}_{\th^{\prime}\phi^{\prime}}&=&\frac{\sqrt{2}(\Im{\rho^{\chico}_{1,0}}+\Im{\rho^{\chico}_{-1,0}})}{3N_{\chico}-\rho^{\chico}_{0,0}},\nonumber
\end{eqnarray}
\begin{eqnarray*}
\lambda^{\chict}_{\th^{\prime}}&=&\frac{6N_{\chict}-9(\rho^{\chict}_{1,1}+\rho^{\chict}_{-1,-1})-12\rho^{\chict}_{0,0}}{6N_{\chict}+3(\rho^{\chict}_{1,1}+\rho^{\chict}_{-1,-1})+4\rho^{\chict}_{0,0}},\nonumber\\
\lambda^{\chict}_{\phi^{\prime}}&=&\frac{2\sqrt{6}(\Re{\rho^{\chict}_{2,0}}+\Re{\rho^{\chict}_{-2,0}})+6\Re{\rho^{\chict}_{1,-1}}}{6N_{\chict}+3(\rho^{\chict}_{1,1}+\rho^{\chict}_{-1,-1})+4\rho^{\chict}_{0,0}},\nonumber\\
\lambda^{\chict}_{\th^{\prime}\phi^{\prime}}&=&\frac{6(\Re{\rho^{\chict}_{2,1}}-\Re{\rho^{\chict}_{-2,-1}})+\sqrt{6}(\Re{\rho^{\chict}_{1,0}}-\Re{\rho^{\chict}_{-1,0}})}
{6N_{\chict}+3(\rho^{\chict}_{1,1}+\rho^{\chict}_{-1,-1})+4\rho^{\chict}_{0,0}},\nonumber\\
\lambda^{\perp\chict}_{\phi^{\prime}}&=&\frac{2\sqrt{6}(\Im{\rho^{\chict}_{0,2}}-\Im{\rho^{\chict}_{0,-2}})-6\Im{\rho^{\chict}_{1,-1}}}{6N_{\chict}+3(\rho^{\chict}_{1,1}+\rho^{\chict}_{-1,-1})+4\rho^{\chict}_{0,0}},\nonumber\\
\lambda^{\perp\chict}_{\th^{\prime}\phi^{\prime}}&=&\frac{6(\Im{\rho^{\chict}_{1,2}}+\Im{\rho^{\chict}_{-1,-2}})+\sqrt{6}(\Im{\rho^{\chict}_{0,1}}+\Im{\rho^{\chict}_{0,-1}})}
{6N_{\chict}+3(\rho^{\chict}_{1,1}+\rho^{\chict}_{-1,-1})+4\rho^{\chict}_{0,0}}.
\end{eqnarray*}
One can also derive these expressions by combining
\begin{eqnarray}
&&\rho^{\chicj\rightarrow\jpsi\gamma}_{s_z,s_z^{\prime}}\propto\sum_{l_z=\pm1,0}{\sum_{J_z,J_z^{\prime}}{\langle1,l_z;1,s_z|J,J_z\rangle}}\nonumber\\
&&\langle J,J_z^{\prime}|1,l_z;1,s_z^{\prime}\rangle
\rho^{\chicj}_{J_z,J_z^{\prime}}\rm{Br}(\chicj\rightarrow\jpsi\gamma)\label{eq:chicjpsi}
\end{eqnarray}
and Eq.(\ref{eq:jpsi}). It is interesting to note that the
expressions for the $\jpsi\to\mu^+\mu^-$ distributions obtained with
this option for $\overrightarrow{s_2}$ are exactly identical (having
the same dependence on the $\chicj$ production SDMEs) to the
expressions obtained before for the $\chicj\to\jpsi\gamma$
distributions. This is only rigorously true when only the E1
transitions are considered. As is discussed quantitatively in
Ref.~\cite{Faccioli:2011be} [whose results are reproduced by
Eq.(\ref{eq:mu2})], the $\jpsi\to\mu^+\mu^-$ distributions are only
mildly corrected by the additional M2 and E3 contributions, while
the $\chicj\to\jpsi\gamma$ distributions are quite sensitive to
them.

\section{Rotation Invariant Relations for Arbitrary Integer-Spin Particles\label{sec:7}}
The partonic Drell-Yan process in perturbative QCD obeys the
well-know Lam-Tung identity~\cite{Lam:1978pu}, which states that the
coefficients $\lambda_{\th}$ and $\lambda_{\phi}$ of the lepton
angular distribution from the Drell-Yan process satisfy
$\lambda_{\th}+4\lambda_{\phi}=1$. Its theoretical relevance is that
the relation remains unchanged up to $\mathcal{O}(\alpha_s^2)$
corrections~\cite{Lam:1980uc} and receives relatively small
corrections even by resummation~\cite{Berger:2007jw}. The
distinctive feature of the identity is that it is independent of the
chosen orientation of the spin axis. It was later pointed out by the
authors of Ref.~\cite{Faccioli:2010ej} that the rotation invariance
of the Lam-Tung relation is a general consequence of the rotational
covariance of $J=1$ angular momentum eigenstates. They presented an
expression formally analogous to the Lam-Tung identity for a $J=1$
boson decaying into a fermion pair with the only assumption that the
spin-quantization axis $\overrightarrow{s}$ should be set in the
production plane, i.e.,
$F_1=\frac{1+\lambda_{\th}+2\lambda_{\phi}}{3+\lambda_{\th}}$. The
new observable $F_1$ is rotation invariant in the production plane.
The condition that the spin-quantization axis is in the production
plane is, indeed, fulfilled in the HX, Collins-Soper,
Gottfried-Jackson, and target frames. From
Ref.~\cite{Faccioli:2010ej}, we know that the rotation-invariant
property of $F_1$ is guaranteed from a relation of the Wigner
functions $d^1_{1,M}(\th)+d^1_{-1,M}(\th)=\delta_{|M|,1}$. In the
section, we want to generalize the relation to the arbitrary spin-$n$
($n$ is an integer) particles. We straightforwardly write down the
linear identities for the Wigner functions:
\begin{eqnarray}
&&\sum^k_{m=-k}{\langle k,m;k,m|2k,2m\rangle
d^{2k}_{2m,M}(\th)}\nonumber\\&=&\langle k,\frac{M}{2};
k,\frac{M}{2}|2k,M\rangle \delta_{\rm{mod}(M,2),0},~~~ n=2k,\nonumber\\
&&\sum^k_{m=0}{\langle 2k+1-m,0;m,0|2k+1,0
\rangle}\nonumber\\&&(d^{2k+1}_{2k+1-m,M}(\th)+d^{2k+1}_{m-2k-1,M}(\th))\nonumber\\
&=&\langle\frac{|M|+1}{2}+k,0;\frac{1-|M|}{2}+k,0|2k+1,0\rangle\nonumber\\
&&\delta_{\rm{mod}(M,2),1},~~~~~~~~~~~~~~~~~~~~~n=2k+1,\label{eq:wigrel}
\end{eqnarray}
where $k$ is a non-negative integer. The amplitudes with respect to
a chosen polarization axis can be symbolically denoted as
$|n\rangle=\sum^n_{m=-n}{a_{m}|n,m\rangle}$, where $|n,m\rangle$ is
a $J_z$ eigenstate with the eigenvalues $m=-n,-n+1,...,n$, and $a_m$
is the production amplitude, i.e., $\rho_{J_z,J_z^{\prime}}\equiv
\langle a_{J_z}a^*_{J_z^{\prime}}\rangle$(average over the events,
assuming that for each event the particle can be produced in a
different angular momentum state). From Eq.(\ref{eq:wigrel}), we can
immediately draw a conclusion that the linear combinations of
amplitudes
\begin{eqnarray*}
b_{2k}\equiv\sum^k_{m=-k}{\langle k,m;k,m|2k,2m\rangle a_{2m}},{\rm
when} ~~n=2k,
\end{eqnarray*}
and
\begin{eqnarray*}
b_{2k+1}\equiv\sum^k_{m=0}{\langle 2k+1-m,0;m,0|2k+1,0
\rangle}\nonumber\\(a_{2k+1-m}+a_{m-1-2k}),{\rm when} ~~n=2k+1,
\end{eqnarray*}
are invariant under the rotation in the production plane. Therefore,
the observables like $F_{n}$ defined as
\begin{eqnarray}
F_{n}&\equiv&\frac{1}{B_{n}}\frac{\langle|b_{n}|^2\rangle}{N_n},\nonumber\\
N_n&\equiv&\sum^n_{m=-n}{\langle|a_m|^2\rangle}\equiv\sum^n_{m=-n}{\rho_{m,m}},\label{eq:Fn}
\end{eqnarray}
are rotation-invariant, where $B_{n}$ is a normalization factor to
ensure $0\leq F_{n}\leq1$. In a more extended sense, any function of
$F_n$ is rotation-invariant. $F_n$ can be expressed in terms of the
coefficients of the decay angular distribution
(e.g., $\lambda_{\th}$,$\lambda_{\phi}$, etc). Specifically, for the
spin-1 particles, the observable is
\begin{eqnarray}
F_1\equiv\frac{1}{2}\frac{\langle|a_{1}+a_{-1}|^2\rangle}{\langle|a_{1}|^2+|a_{0}|^2+|a_{-1}|^2\rangle},\label{eq:F1}
\end{eqnarray}
while for the spin-2 particles, its expression is
\begin{eqnarray}
F_2\equiv\frac{1}{3}\frac{\langle|a_{2}+\sqrt{\frac{2}{3}}a_{0}+a_{-2}|^2\rangle}{\langle|a_{2}|^2+|a_{1}|^2+|a_{0}|^2+|a_{-1}|^2+|a_{-2}|^2\rangle}.\label{eq:F2}
\end{eqnarray}
As examples of $F_1$, we consider the $\jpsi$ decay into two muons
and the $\chico$ decay into a $\jpsi$ and a photon. For the
$\jpsi$,\footnote{Since the invariants are only defined to be
invariant with respect to rotations in the production plane, the invariance of $F_1^{\jpsi\to\mu^+\mu^-}$ is satisfied when $\jpsi$ is
directly produced or from $\chi_c$ decay in the second option but
not in the first option. However, in the first option, one can still
define invariants with respect to rotations in the $\chi_c$ decay
plane.}
\begin{eqnarray}
F_{1}^{\jpsi\to\mu^+\mu^-}=\frac{1+\lambda_{\th^{\prime}}+2\lambda_{\phi^{\prime}}}{3+\lambda_{\th^{\prime}}},
\end{eqnarray}
which has been presented in Ref.~\cite{Faccioli:2010ej}, while for
the $\chico$, one can derive
\begin{eqnarray}
F_{1}^{\chico\to\jpsi\gamma}=\frac{1-\lambda_{\th}-4\lambda_{\phi}}{3+\lambda_{\th}}
\end{eqnarray}
from Eqs.(\ref{eq:chic1}) and (\ref{eq:F1}).\footnote{Note that the
expressions of $F_{1}^{\jpsi\to\mu^+\mu^-}$ and
$F_{1}^{\chico\to\jpsi\gamma}$ as functions of the polarization observables
$\lambda$'s can also be written in one common form being
$F_{1}^{\chico\to\jpsi\gamma}=
1-2F_{1}^{\jpsi\to\mu^+\mu^-}$. This fact can also be expected from
the rotation relations of $\lambda$'s given in
Ref.~\cite{Faccioli:2010kd}.} The $\chict$ provides an example of the
spin-2 particles. In fact, the complete angular distribution of the
$\chict$'s decay product $\jpsi$ is Eq.(\ref{eq:angd}) in stead of
Eq.(\ref{eq:wchic2}). However, the terms absent in
Eq.(\ref{eq:wchic2}) are suppressed, as mentioned above. Hence, the
spin information in Eq.(\ref{eq:chic2}) is not sufficient. In
Appendix \ref{app:b}, we have included the E1,M2, and E3 effects into
the angular distribution of $\jpsi$ in $\chict\to\jpsi\gamma$ and
derived the $F_{2}^{\chict\to\jpsi\gamma}$ there [see
Eq.(\ref{eq:F2F2})]. We suggest that the reader who is interested in this
part to refer to Appendix \ref{app:b}.
These frame-invariant relations can be extended to the study of
other bosons or mesons. The experimentalists can measure these
observables to make a cross-check of their extractions of the angular
distribution coefficients in different frames.

\section{Uncertainty of $\jpsi$ Polarization from Feed-down\label{sec:8}}
The CDF data for the  prompt-$\jpsi$ production include not only
direct-$\jpsi$ production but also the feed-down contributions from
$\chicj$ and  $\psip$. However, the recent NLO calculations of
$\jpsi$ polarization in Refs.~\cite{Chao:2012iv,Butenschoen:2012px} are devoid of the feed-down
contributions. Though the LO NRQCD prediction of the feed down to
the $\jpsi$ polarization in Ref.~\cite{Kniehl:2000nn} was found to
have a minor impact on the final LO result, one may still doubt
whether the NLO feed-down effect on
$\lambda_{\th^{\prime}}$\footnote{Note that, to be consistent
throughout the context, the polarization observables of the $\jpsi$
or the angular distributions of the muon are all denoted by an extra
prime.} of the $\jpsi$ can be neglected, because the NLO correction to
the P wave is large~\cite{Ma:2010vd}. In this section, we will
estimate the possible uncertainty of the $\jpsi$ polarization
$\lambda_{\th^{\prime}}$ arising from the feed down of the $\chi_c$
and $\psip$ decays.

The calculation of the prompt-$\jpsi$ polarization is complex. In
general, prompt data are composed of four parts, i.e., the direct
production of the $\jpsi$, the single-cascade decays of the $\chi_c$
and of the $\psip$, and the double-cascade decay
$\psip\to\chi_c\gamma\to\jpsi\gamma\gamma$. The direct production of
the $\jpsi$ has been studied, e.g., in
Refs.~\cite{Chao:2012iv,Butenschoen:2012px}. The relation between the
production SDMEs of the $\chi_c$ and the SDMEs of the $\jpsi$ from
$\chi_c$ decays is given by Eq.(\ref{eq:chicjpsi}), whereas the
relation between the production SDMEs of the $\psip$ and the SDMEs
of the $\chi_c$ coming from $\psip$ decay is
\begin{eqnarray}
&&\rho^{\psip\to\chicj}_{J_z,J_z^{\prime}}\propto\\&&\sum_{l_z,s_z,s_z^{\prime}=\pm1,0}{\langle1,l_z;1,s_z|J,J_z\rangle\langle1,l_z;1,s_z^{\prime}|J,J_z^{\prime}\rangle\rho^{\psip}_{s_z,s_z^{\prime}}}.\nonumber\label{eq:psipchic}
\end{eqnarray}
By combining Eqs.(\ref{eq:psipchic}) and (\ref{eq:chicjpsi}), the
double-cascade decay component can also be calculated. However, in
the following uncertainties estimation, we will neglect this
contribution, because of the small branching ratio and small cross
section  ratio between $\psip$ and $\jpsi$~\cite{Ma:2010yw}.
Finally, the single-cascade decay $\psip \to \jpsi \pi \pi$ can be
treated in analogy with the double chromoelectric dipole transition $\so
\to \jpsi$~\cite{Kniehl:2000nn}. This part will also not be included
in the uncertainties because of the small cross section ratio
between $\psip$ and $\jpsi$~\cite{Ma:2010yw} and the spin orientation
conserved in $\psip\rightarrow \jpsi \pi
\pi$~\cite{Faccioli:2012kp}.

\begin{figure}
\includegraphics[width=8.5cm]{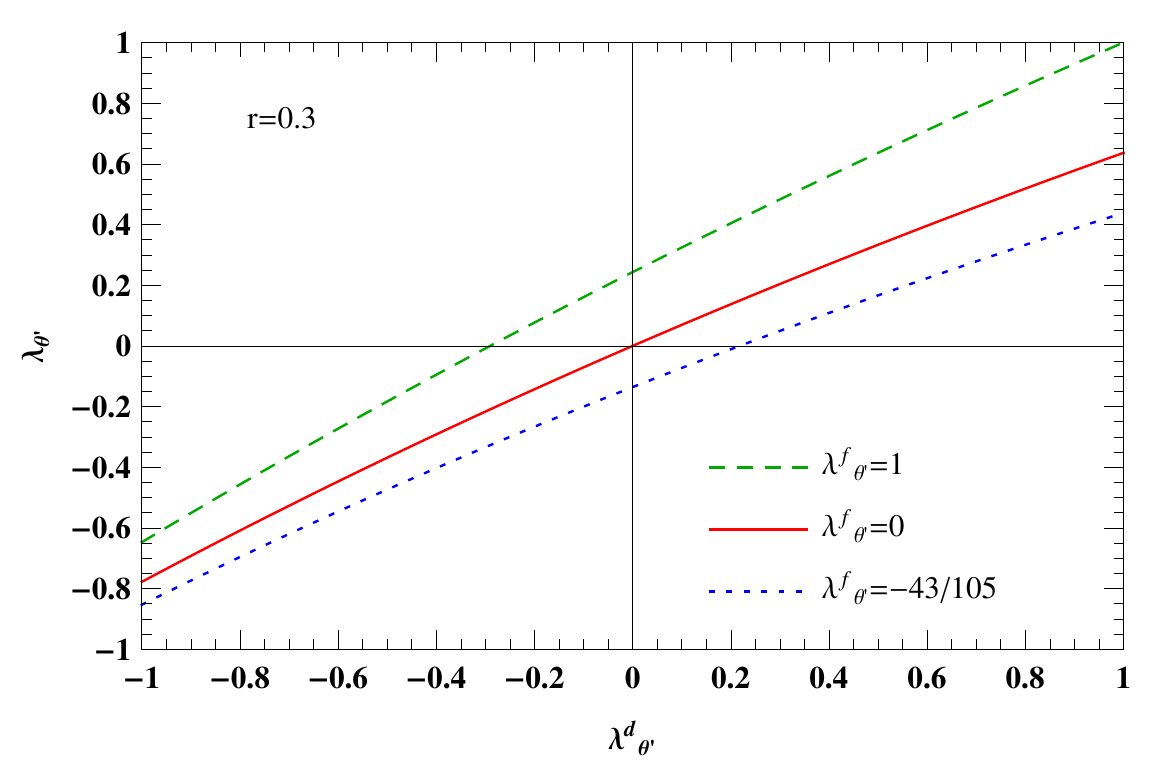}
\caption{\label{fig:uncer}(color online). Possible impact of the feed down on the
prompt-$\jpsi$ polar anisotropy $\lambda_{\th^{\prime}}$ as a
function of the direct-$\jpsi$ polarization
$\lambda^{d}_{\th^{\prime}}$. Here, $r=0.3$ is assumed, and  the
curves correspond to  $\jpsi$ polarizations from the feed down
$\lambda^{f}_{\th^{\prime}}=+1,0,-\frac{43}{105}$.}
\end{figure}

\begin{figure}
\includegraphics[width=8.5cm]{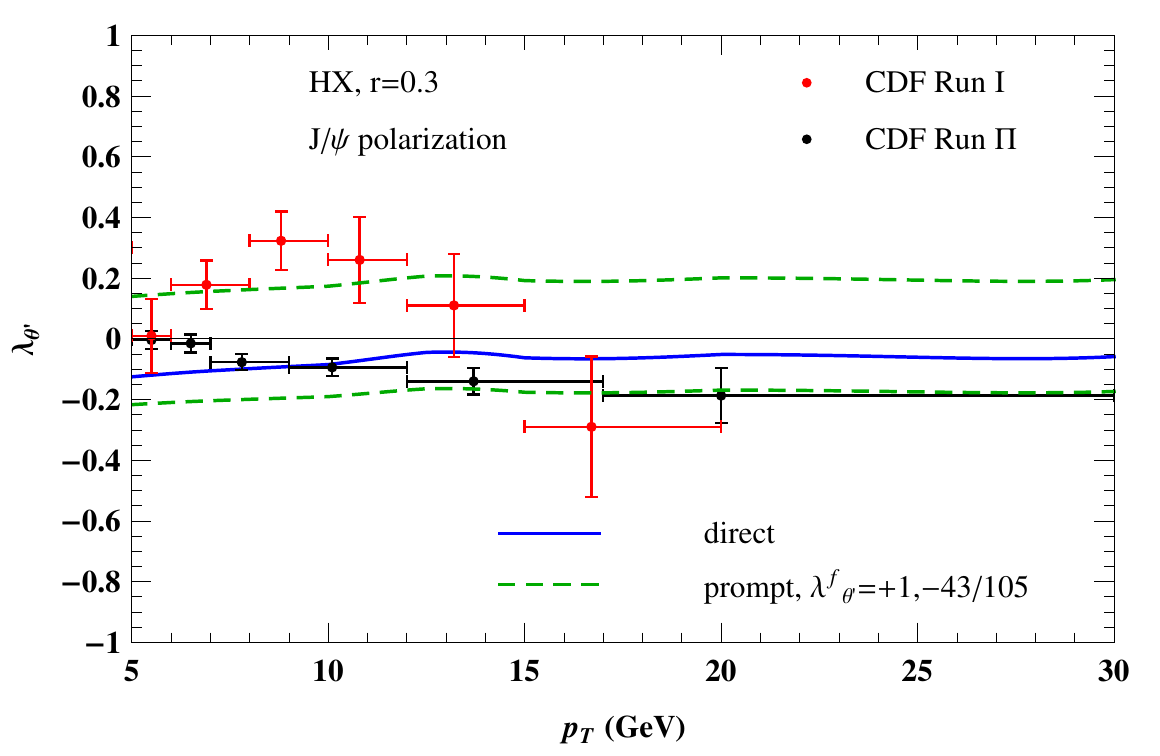}
\caption{\label{fig:poljpsi}(color online). The direct-$\jpsi$ polarization
$\lambda_{\th^{\prime}}$ in the HX frame at the Tevatron calculated
with the LDMEs of Table~\ref{tab:ldmes}, and the corresponding upper
and lower limits for the prompt-$\jpsi$ polarization assuming the
$\chi_c$ feed-down contribution $r=0.3$. CDF data are taken from
Refs.~\cite{Affolder:2000nn,Abulencia:2007us}.}
\end{figure}

In this way, the only contribution to be considered is the feed down
from $\chi_c$. We consider the total prompt-$\jpsi$ yield $\rho$
decomposed in the ``direct" part, $\rho^{d}$, already calculated in
NRQCD at the NLO level, and the ``feed-down" part $\rho^{f}$, with
their corresponding polarization observables
$\lambda^{d}_{\th^{\prime}}$ and $\lambda^{f}_{\th^{\prime}}$, polar
anisotropies of the dilepton decay disributions. The fraction of
the $\jpsi$ yield from feed down with respect to the total prompt yield
is denoted as $r$, i.e.,
$r\equiv\frac{2\rho^{f}_{1,1}+\rho^{f}_{0,0}}{2\rho_{1,1}+\rho_{0,0}}$\footnote{Note
that we use the symmetry property
$\rho^H_{-\lambda,-\lambda^{\prime}}=(-)^{\lambda-\lambda^{\prime}}\rho^H_{\lambda,\lambda^{\prime}}$
which is guaranteed in hadroproduction by parity
invariance~\cite{Chung:1974fq}.} with
$\rho_{s_z,s^{\prime}_z}\equiv\rho^{f}_{s_z,s^{\prime}_z}+\rho^{d}_{s_z,s^{\prime}_z}$
and $\rho\equiv2\rho_{1,1}+\rho_{0,0}$. Hence, the prompt-$\jpsi$
decay polar anisotropy is
\begin{eqnarray}
\rho^{f}_{0,0}&=&r
\frac{1-\lambda^{f}_{\th^{\prime}}}{3+\lambda^{f}_{\th^{\prime}}}\rho,~~
\rho^{f}_{1,1}=r\frac{1+\lambda^{f}_{\th^{\prime}}}{3+\lambda^{f}_{\th^{\prime}}}\rho,\\
\rho^{d}_{0,0}&=&(1-r)
\frac{1-\lambda^{d}_{\th^{\prime}}}{3+\lambda^{d}_{\th^{\prime}}}\rho,~~
\rho^{d}_{1,1}=(1-r)\frac{1+\lambda^{d}_{\th^{\prime}}}{3+\lambda^{d}_{\th^{\prime}}}\rho,\nonumber
\end{eqnarray}
and
\begin{eqnarray}
\lambda_{\th^{\prime}}=\frac{\frac{r\lambda^{f}_{\th^{\prime}}}{3+\lambda^{f}_{\th^{\prime}}}+\frac{(1-r)\lambda^{d}_{\th^{\prime}}}{3+\lambda^{d}_{\th^{\prime}}}}{\frac{r}
{3+\lambda^{f}_{\th^{\prime}}}+\frac{1-r}{3+\lambda^{d}_{\th^{\prime}}}}.\label{eq:uncer}
\end{eqnarray}
All these considerations are valid for any polarization frame. From
Eq.(\ref{eq:mu2}), we know that the $\jpsi$ from $\chico$ and
$\chict$ can have
$-\frac{1}{3}\leq\lambda^{\chico}_{\th^{\prime}}\leq1$ and
$-\frac{3}{5}\leq\lambda^{\chict}_{\th^{\prime}}\leq1$. Therefore,
we take the $-\frac{43}{105}\leq\lambda^{f}_{\th^{\prime}}\leq1$,
which is weighted by the relative contributions of $\chico$ and
$\chict$ to prompt $\jpsi$, i.e.,
$\sigma_{\chico}\mathcal{B}(\chico\to\jpsi\gamma)/\sigma_{\chict}\mathcal{B}(\chict\to\jpsi\gamma)\cong5:2$
, as measured by CDF~\cite{Abulencia:2007bra}. For example, let us
compare the cases $\lambda^{d}_{\th^{\prime}}=0$ and
$\lambda^{d}_{\th^{\prime}}=1$, approximated values of the
direct-$\jpsi$ polarization predictions in the HX frame for
$p_T>10\rm{GeV}$  at the Tevatron according to
Refs.~\cite{Chao:2012iv,Butenschoen:2012px},
respectively. The allowed prompt-$\jpsi$ polarization ranges in the
two cases (with $\lambda^{f}_{\th^{\prime}}$ varying from
$-\frac{43}{105}$ to $1$) are
$-\frac{129r}{272+43r}\leq\lambda_{\th^{\prime}}\leq\frac{3r}{4-r}$
and $\frac{68-111r}{68+37r}\leq\lambda_{\th^{\prime}}\leq1$. If we
fix $r=0.3$ (the central value of the average $\chi_c$ feed-down
fraction to prompt $\jpsi$ measured at the
Tevatron~\cite{Abe:1997yz}), the feed-down contribution may change
$\lambda_{\th^{\prime}}$ from $0.24$ to $-0.14$ when
$\lambda^{d}_{\th^{\prime}}=0$ and from $1$ to $0.44$ when the
polarization of the direct $\jpsi$ is fully transverse.  More
generally, Fig.~\ref{fig:uncer} shows curves for the
prompt-$\jpsi$ polarization $\lambda_{\th^{\prime}}$ as a function
of the direct-$\jpsi$ polarization $\lambda^{d}_{\th^{\prime}}$, for
$\lambda^{f}_{\th^{\prime}}=+1,0,-\frac{43}{105}$, and $r=0.3$. The
upper and lower curves represent physical bounds for
$\lambda_{\th^{\prime}}$. Figure \ref{fig:poljpsi} shows the maximum
possible impact of the feed down from the $\chi_c$ decays on the
prompt-$\jpsi$ polarization predicted in the HX frame for the
Tevatron with $\sqrt{s}=1.96 \rm{TeV}$ and $|y_{\jpsi}|<0.6$. The
LDMEs values (see Table \ref{tab:ldmes}) used for the direct-$\jpsi$
prediction are those obtained in Ref.~\cite{Chao:2012iv} by fitting
the NLO NRQCD calculation to the Tevatron data. Only the central
value of the direct-$\jpsi$ prediction is shown. In particular, the
prediction of an almost unpolarized $\jpsi$ production obtained with
the LDMEs determined in Ref.~\cite{Chao:2012iv} is not drastically
affected by the neglected impact of the $\chi_c$ feed-down
contribution.

\begin{table}[h]
\caption{\label{tab:ldmes} CO LDMEs for $\jpsi$ from
Ref.~\cite{Chao:2012iv} obtained by fitting the differential cross
section and polarization of prompt $\jpsi$ simultaneously at the
Tevatron~\cite{Abulencia:2007us}. The CS LDME is calculated with the
B-T potential model in Ref.~\cite{Eichten:1995ch}.}
\begin{tabular}{c*{3}{c}}
\hline\hline \itshape ~$\langle\mathcal{O}^{\jpsi}(\ss)\rangle$~ &
~$\langle\mathcal{O}^{\jpsi}(\sps)\rangle$~ &
~$\langle\mathcal{O}^{\jpsi}(\so)\rangle$~ &
~$\langle\mathcal{O}^{\jpsi}(\p0)\rangle/m_{c}^2$~
\\
\itshape $\rm{GeV}^3$ & $10^{-2}\rm{GeV}^3$ & $10^{-2}\rm{GeV}^3$ &
$10^{-2}\rm{GeV}^3$
\\\hline ~1.16~& $8.9$ & $0.30$ & $0.56$
\\\hline\hline
\end{tabular}
\end{table}

\section{An example of $\chi_{c1}$ and $\chi_{c2}$ polarization\label{sec:9}}
In this section, we are in a position to give an example for
the inclusive $\chico$ and $\chict$ production at the LHC with
$\sqrt{s}=8~\rm{TeV}$.\footnote{To avoid possible misunderstanding, we want to
remind the readers that we only give LO NRQCD results here as a
simple application of the formulas presented in the paper. We do
not intend to give phenomenological predictions of $\chi_c$
polarization. As pointed out in Ref.~\cite{Ma:2010vd}, there is a
kinematical enhancement of P wave at NLO, which results in a large
cancellation between the P wave and S wave. Our more reliable NLO
phenomenological results for $\chi_c$ are presented in an independent
paper~\cite{Shao:2014fca}.} We also present the contributions of the individual Fock
states to the SDMEs, for the convenience of the readers who want to
use our results with different LDME values.

We use our automatic matrix element and event generator
HELAC-Onia~\cite{Shao:2012iz} to
calculate all the SDMEs under the relevant conditions. The generator
is built on rewritten versions of the published
HELAC~\cite{Kanaki:2000ey,Papadopoulos:2006mh},
PHEGAS~\cite{Papadopoulos:2000tt,Cafarella:2007pc},
RAMBO~\cite{Kleiss:1985gy}, and VEGAS~\cite{Lepage:1977sw} codes. The
program has been extensively tested.

The input parameters in our calculations are
\begin{itemize}
\item[(1)]$m_c=1.5~\rm{GeV}$,$m_{\chi_c}=2m_c=3\rm{GeV}$,
\footnote{We use the approximation $m_{\chi_c}=m_{\jpsi}=2m_c$
here. The mass dependences of the result is mild when
$p_T^{\chi_c}\simeq p_T^{\jpsi}> 2m_c$.}
\item[(2)]$\sqrt{s}=8 ~\rm{TeV},|y_{\chi_c}|<2.4$,
\item[(3)]the CTEQ6L1~\cite{Pumplin:2002vw} set of parton distribution functions,
\item[(4)]renormalization and factorization scales $\mu_r=\mu_f=\sqrt{(2m_c)^2+p_T^2}$,
\item[(5)]CS LDMEs $\langle\mathcal{O}^{\chicj}(\pjs)\rangle=\frac{3(2J+1)2N_c}{4\pi}|R^{\prime}_P(0)|^2$,
with $|R^{\prime}_P(0)|^2=0.075\rm{GeV}^5$~\cite{Eichten:1995ch}, and
\item[(6)]CO LDMEs $\frac{\langle\mathcal{O}^{\chicj}(\so)\rangle}{2J+1}=2.2\times10^{-3}\rm{GeV}^3$~\cite{Ma:2010vd}.
\end{itemize}

The relations between the SDMEs of $\so$ and those of $\pjs$ are
identical to those in Eq.(\ref{eq:psipchic}) and are specifically
\begin{eqnarray}
&&\rho^{\so\to\chicj}_{J_z,J_z^{\prime}}\propto\\&&\sum_{l_z,s_z,s_z^{\prime}=\pm1,0}{\langle1,l_z;1,s_z|J,J_z\rangle\langle1,l_z;1,s_z^{\prime}|J,J_z^{\prime}\rangle\rho^{\so}_{s_z,s_z^{\prime}}}.\nonumber
\end{eqnarray}

The results of the calculations are organized in the following
groups of figures, where in each case, from top to bottom, three
different polarization frames are considered
(HX, Collins-Soper, Gottfried-Jackson):
Figs.(\ref{fig:sigmapo},\ref{fig:sigmapt},\ref{fig:sigmas}) show the
partial cross sections
($\frac{\rm{d}\sigma_{00}}{\rm{d}p_T},\frac{\rm{d}\sigma_{11}}{\rm{d}p_T},\frac{\rm{d}\sigma_{22}}{\rm{d}p_T}$)
and the total cross sections
($\frac{\rm{d}\sigma_{tot}}{{\rm{d}p_T}}$) as a function of $p_T$
for each individual Fock state contributing to $\chi_c$ production
($\tpos,\tpts$, and $\so$, with
$\frac{\rm{d}\sigma_{tot}}{\rm{d}p_T}=2\frac{\rm{d}\sigma_{11}}{\rm{d}p_T}+\frac{\rm{d}\sigma_{00}}{\rm{d}p_T}$
for $\tpos$ and
$\so$,$\frac{\rm{d}\sigma_{tot}}{\rm{d}p_T}=2\frac{\rm{d}\sigma_{22}}{\rm{d}p_T}+2\frac{\rm{d}\sigma_{11}}{\rm{d}p_T}+\frac{\rm{d}\sigma_{00}}{\rm{d}p_T}$
for $\tpts$). The corresponding nondiagonal SDMEs are shown in
Figs.(\ref{fig:rhopo},\ref{fig:rhopt},\ref{fig:rhos}). The frame-dependent parameters $\lambda_{\th},\lambda_{\phi}$, and
$\lambda_{\th\phi}$ of the radiative $\chi_c$ decays are shown in
Figs.(\ref{fig:th1jpsi},\ref{fig:phi1jpsi},\ref{fig:thphi1jpsi}) for
the $\chico$ and in
Figs.(\ref{fig:th2jpsi},\ref{fig:phi2jpsi},\ref{fig:thphi2jpsi}) for
the $\chict$, with distinct curves for the CS-only result and for
the full CS+CO calculation. We remind that these parameters are also
identical, in the E1-only approximation, to the parameters
$\lambda_{\th^{\prime}},\lambda_{\phi^{\prime}}$, and
$\lambda_{\th^{\prime}\phi^{\prime}}$ of the dilepton distribution
of the $\jpsi$ from $\chico$ decays in the second option, for each
considered polarization frame. Finally,
Figs.(\ref{fig:F1jpsi}) and (\ref{fig:F2jpsi}) show the results for the
corresponding frame-independent polarization parameters $F_1$ and
$F_2$.

Some features should be emphasized:
\begin{itemize}
\item[(1)] From the curves of the total cross sections in
Figs.(\ref{fig:sigmapo},\ref{fig:sigmapt},\ref{fig:sigmas}), we may
conclude that the CS dominates in the low transverse momentum
region,
while the CO may dominate when $p_T$ increases because gluon
fragmentation processes~\cite{Braaten:1994kd} become important. The
CS $p_T$ distribution may receive significant contributions from the
higher order radiative corrections~\cite{Ma:2010vd}. The full NLO
predictions for $\chicj$ polarizations are already presented in Ref.~\cite{Shao:2014fca}.

\item[(2)] The polarization parameters in the Gottfried-Jackson frame tend to
be very similar to those in the HX frame, especially at high $p_T$,
while significant differences exist between these two and the
Collins-Soper frame. The HX and Collins-Soper frames are, therefore,
sufficient for the characterization of the angular distributions. In
particular, from Fig.\ref{fig:sigmas}, we see that the longitudinal
cross section of the $\so$ channel is largest in the Collins-Soper
frame when $p_T\gg2m_c$, in agreement with the statement in
Ref.~\cite{Kniehl:2003pc}.

\item[(3)] We have verified that the results of $\rho_{i,j}$ in the Gottfried-Jackson
frame and the target frame coincide within error bars, aside from a
factor $(-1)^{i-j}$, which is also pointed out in
Ref.~\cite{Kniehl:2003pc}. Therefore, we will not present the results in the target frame here.
For diagonal elements, i.e., $\rho_{i,i}$,
the curves in the Gottfried-Jackson are
similar to those in the helicity frame and different from those
in the Collins-Soper frame. Hence, for the polarization
observable $\lambda_{\th}$, the helicity and the Collins-Soper
frames are enough. Moreover, from Fig.(\ref{fig:sigmas}), we see
that the longitudinal cross section of the $\so$ channel is the
largest in the Collins-Soper frame when $p_T\gg2m_c$, which is
consistent with the statement in Ref.~\cite{Braaten:2008xg}.

\item[(4)] In the HX and the Collins-Soper frames, the SDMEs $\rho_{i,j}$
with $|i-j|$ odd are almost zero. Therefore, they should be measured
in the Gottfried-Jackson frame as well as $\lambda_{\th\phi}$ and
$\lambda_{\th^{\prime}\phi^{\prime}}$ shown in
Figs.(\ref{fig:thphi1jpsi}) and (\ref{fig:thphi2jpsi}). In fact, this is a
consequence of the definition of the Y axis taken always
as $P_1\times(-P_2)$ (in the $\chi_c$ rest frame) both at positive and
negative rapidity. Actually, the values of $\lambda_{\th\phi}$ in
the positive and negative rapidity are opposite. The existence of
these relations between the choice of the Y aixs and the sign
$\lambda_{\th\phi}$ was already pointed out in Ref.~\cite{Faccioli:2010kd}.
~On the other hand, for $\rho_{i,j}$ with $|i-j|$ even, the
measurements in the HX and the Collins-Soper frames are more significant
than in the Gottfried-Jackson frame.


\item[(5)] The frame-independent observables defined in
Eqs.(\ref{eq:F1}) and (\ref{eq:F2}) for the $\chico$ and the $\chict$ are
shown in Fig.\ref{fig:F1jpsi}. The figures show the
frame-independent property of $F^{\chico\to\jpsi\gamma}_1$ and
$F^{\chict\to\jpsi\gamma}_2$ by direct numerical calculation. Hence,
it would be interesting to measure these observables at the LHC. In
particular, it is remarkable that $F^{\chict\to\jpsi\gamma}_2$ is
practically independent of $p_T$, contrary to $\lambda_{\th}$ and
$\lambda_{\phi}$ in all frames considered (see
Figs.(\ref{fig:th2jpsi}) and (\ref{fig:phi2jpsi})).
\end{itemize}
\section{Summary \label{sec:10}}

Finally, we draw our conclusion. The upgrade of the integrated
luminosity at the LHC will not only allow us to measure the
polarizations of $S$-wave quarkonium states like $\jpsi$ and
$\psi'$ but also the angular distributions of decay products from
the $P$-wave states $\chicj$. This opens new opportunities to
further test  NRQCD factorization and the quarkonium production
mechanisms, in general.  We have presented a general calculation
framework based on the shape of the decay vertex functions to
investigate the polarizations of the $\chicj$ states and their
impact in the observed prompt-$\jpsi$ polarization. We have derived
general expressions for the polar and azimuthal angle distributions
of the $\chico$ and $\chict$ decays into $\jpsi\gamma$ and for the
subsequent $\jpsi$ decay into muons. The coefficients of the angular
distributions have been calculated as a function of the $\chi_c$
production SDMEs
[Eqs.(\ref{eq:chic1}), (\ref{eq:chic2}), (\ref{eq:mu1}), (\ref{eq:mu2}), and (\ref{eq:reweight})].
We have derived rotation-invariant relations for arbitrary
integer-spin particles [Eq.(\ref{eq:Fn})] and in the specific cases
of $\chico$ and $\chict$ decays. As an example of an application of our
calculation framework, in the NRQCD factorization, we have
calculated the tree-level angular distributions of the $\chico$ and
$\chict$ decays at the LHC considering several polarization frames
and also frame-independent quantities. Moreover, we have also estimated
the impact of the $\chi_c$ feed down in the polarization of
prompt $\jpsi$. We found that our previous direct-$\jpsi$ polarization
results~\cite{Chao:2012iv} will not change much after including this
part. A more detailed phenomenological analysis of the yields and
polarizations of $\chi_c$ in hadroproduction are performed in Ref.~\cite{Shao:2014fca}, based on our complete NLO NRQCD calculations.

{\it Note added:} While this paper was prepared, a new preprint
~\cite{Gong:2012ug} for polarizations of the prompt-$\jpsi$ and
$\psip$ production at the LHC and Tevatron appeared. The authors
extracted a set of CO LDMEs for the $\jpsi$ other than those in
Refs.~\cite{Butenschoen:2012px,Chao:2012iv} by including the $\chicj$
and $\psip$ feed-down contributions.   Their LDMEs result in two
combinations of LDMEs that agree with those extracted in
Refs.~\cite{Chao:2012iv,Ma:2010yw}, and their prompt polarization
result is around the upper limit in Fig.\ref{fig:poljpsi} of this
paper.

\begin{acknowledgments}
We thank J.-P. Lansberg, Q.-H. Cao and Y.-Q. Ma for helpful
discussions. This work was supported in part by the National Natural
Science Foundation of China (Grants No.11021092 and No.11075002) and the
Ministry of Science and Technology of China (Grant No.2009CB825200).
\end{acknowledgments}

\begin{appendix}
\section{Decay angular distribution of spin-1 bosons\label{app:a}}

The most general expression for the decay angular distribution of a
vector boson $V$ without assuming parity conservation is
\begin{eqnarray}
\mathcal{W}^V(\th,\phi)&\propto&
1+\lambda_{\th}\cos^2\th+\lambda_{\phi}\sin^2\th\cos\phi\nonumber\\&+&\lambda_{\th\phi}\sin2\th\cos\phi+\lambda^{\perp}_{\phi}\sin^2\th\sin2\phi\nonumber\\
&+&\lambda^{\perp}_{\th\phi}\sin2\th\sin\phi+2\eta_{\th}\cos\th\nonumber\\&+&2\eta_{\th\phi}\sin\th\cos\phi+2\eta^{\perp}_{\th\phi}\sin\th\sin\phi,
\end{eqnarray}
with
\begin{eqnarray}
\lambda_{\th}&=&(1-3\delta)\frac{N_V-3\rho^V_{0,0}}{(1+\delta)N_V+(1-3\delta)\rho^V_{0,0}},\nonumber\\
\lambda_{\phi}&=&(1-3\delta)\frac{2\Re{\rho^V_{1,-1}}}{(1+\delta)N_V+(1-3\delta)\rho^V_{0,0}},\nonumber\\
\lambda_{\th\phi}&=&(1-3\delta)\frac{\sqrt{2}(\Re{\rho^V_{1,0}}-\Re{\rho^V_{-1,0}})}{(1+\delta)N_V+(1-3\delta)\rho^V_{0,0}},\nonumber\\
\lambda^{\perp}_{\phi}&=&-(1-3\delta)\frac{2\Im{\rho^V_{1,-1}}}{(1+\delta)N_V+(1-3\delta)\rho^V_{0,0}},\nonumber\\
\lambda^{\perp}_{\th\phi}&=&-(1-3\delta)\frac{\sqrt{2}(\Im{\rho^V_{1,0}}+\Im{\rho^V_{-1,0}})}{(1+\delta)N_V+(1-3\delta)\rho^V_{0,0}},\nonumber\\
\eta_{\th}&=&\alpha\frac{\rho^V_{1,1}-\rho^V_{-1,-1}}{(1+\delta)N_V+(1-3\delta)\rho^V_{0,0}},\nonumber\\
\eta_{\th\phi}&=&\alpha\frac{\sqrt{2}(\Re{\rho^V_{1,0}}+\Re{\rho^V_{-1,0}})}{(1+\delta)N_V+(1-3\delta)\rho^V_{0,0}},\nonumber\\
\eta^{\perp}_{\th\phi}&=&-\alpha\frac{\sqrt{2}(\Im{\rho^V_{1,0}}-\Im{\rho^V_{-1,0}})}{(1+\delta)N_V+(1-3\delta)\rho^V_{0,0}},\label{eq:appa1}
\end{eqnarray}
where $N_V=\rho^V_{1,1}+\rho^V_{0,0}+\rho^V_{-1,-1}$, and the
parameters $\alpha,\delta$ depend on the identity of $V$ and of its
decay products. In particular, $\alpha$ is induced by the parity-violating interactions in the decay. In other words, it is nonzero
only when the decay is not a parity conservative process. For the
$\jpsi$ decays into dilepton, $\alpha=0,\delta=0$, whereas for the
pure E1 radiative transition
$\chico\to\jpsi\gamma$, $\alpha=0,\delta=\frac{1}{2}$. If one also
wants to include the M2 transition in the $\chico$ decay, $\delta$
should be changed to $\frac{1+2a^{J=1}_1a^{J=1}_{2}}{2}$, where
$a^{J=1}_1, a^{J=1}_{2}$ represent the E1 and M2 amplitudes,
respectively, with the normalization
$(a^{J=1}_1)^2+(a^{J=1}_{2})^2=1$. They have been measured
in Refs.~\cite{Artuso:2009aa,Oreglia:1981fx,Ambrogiani:2001jw}. The
numerical values measured are shown in Table \ref{tab:multi1}. Without
losing generality, the rotation-invariant observable $F_1$ defined
in Eq.(\ref{eq:F1}) can be written as\footnote{One just substitutes Eq.(\ref{eq:appa1}) into Eq.(\ref{eq:F1}) to get the following expression. We want to remind the readers that because the rotation-invariant observable is not uniquely defined, a more general form of the rotation-invariant observable is an arbitary function of $F_1$.}
\begin{eqnarray}
F_1&=&\frac{1-3\delta+(1-\delta)\lambda_{\th}+2\lambda_{\phi}}{(1-3\delta)(3+\lambda_{\th})}.
\end{eqnarray}

\begin{table}[h]
\caption{\label{tab:multi1} The normalized M2 amplitude $a^{J=1}_2$
for $\chico\to\jpsi\gamma$ as measured by different experiments.}
\begin{tabular}{c|{c}}
\hline\hline \itshape ~\rm{Experiment}~ & ~$a^{J=1}_2(10^{-2})$~
\\\hline ~\rm{CLEO}~\cite{Artuso:2009aa}~& $-6.26\pm0.63\pm0.24$\\
~\rm{Crystal Ball}~\cite{Oreglia:1981fx}~& $-0.2^{+0.8}_{-2.0}$\\
~\rm{E835}~\cite{Ambrogiani:2001jw}~ & $0.2\pm3.2\pm0.4$
\\\hline\hline
\end{tabular}
\end{table}

\begin{figure}
\includegraphics[width=8.5cm]{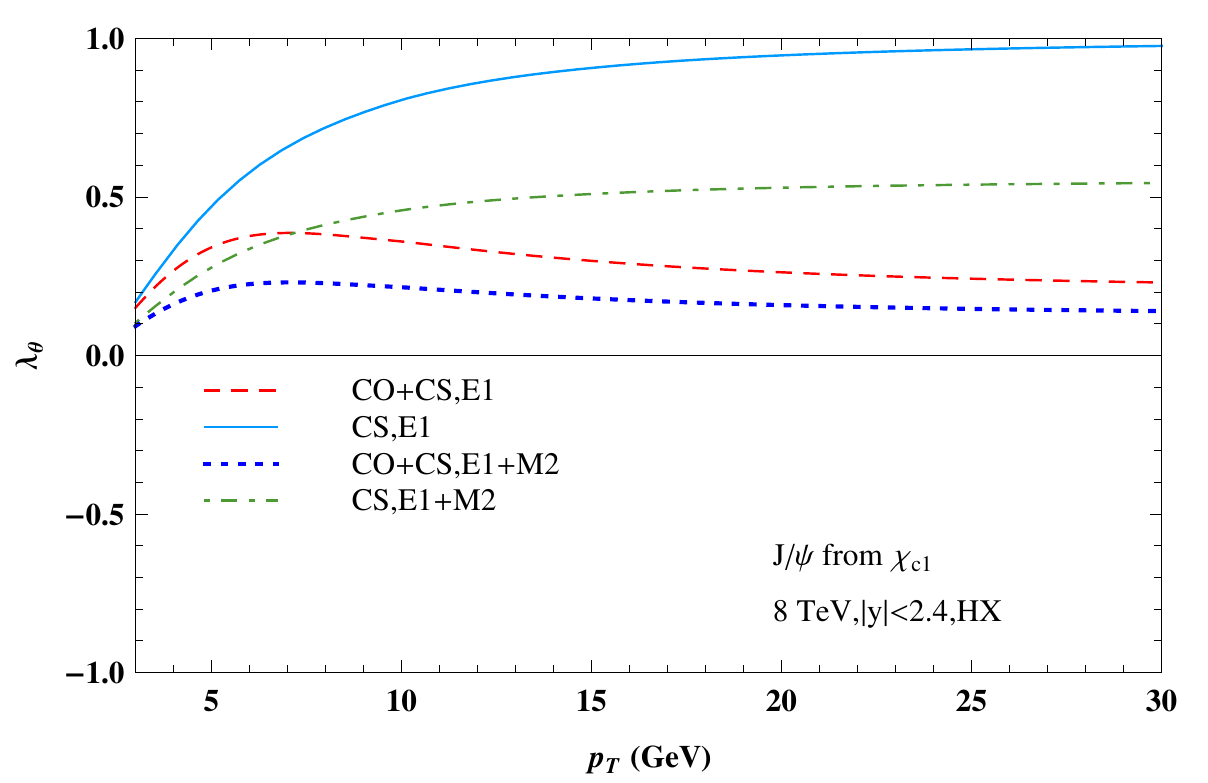}
\caption{\label{fig:chic1m2}(color online). M2 contribution to the $\chico$
polarization $\lambda_{\th}$ in $\chico\to\jpsi\gamma$ in the HX
frame. The CLEO measurement of $a^{J=1}$ listed in Table \ref{tab:multi1} has been used.}
\end{figure}

In order to show the impact of the higher-order multipole M2
contribution to the $\chico$ polarizations, we take the example
illustrated in Sec. \ref{sec:8}. As an illustrative case, only
$\lambda_{\th}$ in the HX frame is shown in Fig.
\ref{fig:chic1m2}, where E1 means pure E1 transtion approximation and
E1+M2 means that we have included the full E1 and M2 transitions using the
CLEO~\cite{Artuso:2009aa} measured $a^{J=1}$ in
Table \ref{tab:multi1}.

\section{Decay angular distribution of spin-2 bosons\label{app:b}}

The general decay angular distributions of spin-2 tensor particles $T$ can have up to 24 observable parameters:
\begin{eqnarray}
\mathcal{W}^{T}(\th,\phi)&\propto&1+\lambda_{\th}\cos^2\th+\lambda_{2\th}\cos^4\th
\nonumber\\&+&\lambda_{\th\phi}\sin2\th\cos\phi+\lambda_{2\th\phi}\sin2\th\sin^2\th\cos\phi
\nonumber\\&+&\lambda^{\perp}_{\th\phi}\sin2\th\sin\phi+\lambda^{\perp}_{2\th\phi}\sin2\th\sin^2\th\sin\phi\nonumber\\
&+&\lambda_{\phi}\sin^2\th\cos2\phi+\lambda_{2\phi}\sin^4\th\cos2\phi\nonumber\\
&+&\lambda^{\perp}_{\phi}\sin^2\th\sin2\phi+\lambda^{\perp}_{2\phi}\sin^4\th\sin2\phi\nonumber\\
&+&\lambda_{3\th\phi}\sin2\th\sin^2\th\cos3\phi\nonumber\\
&+&\lambda^{\perp}_{3\th\phi}\sin2\th\sin^2\th\sin3\phi\nonumber\\&+&\lambda_{4\phi}\sin^4\th\cos4\phi+\lambda^{\perp}_{4\phi}\sin^4\th\sin4\phi\nonumber\\
&+&2\eta_{\th}\cos\th+2\eta_{2\th}\cos^3\th\nonumber\\&+&2\eta_{\th\phi}\sin\th\cos\phi+2\eta_{2\th\phi}\sin^3\th\cos\phi\nonumber\\
&+&2\eta^{\perp}_{\th\phi}\sin\th\sin\phi+2\eta^{\perp}_{2\th\phi}\sin^3\th\sin\phi\nonumber\\
&+&2\eta_{\phi}\sin^2\th\cos\th\cos2\phi+2\eta^{\perp}_{\phi}\sin^2\th\cos\th\sin2\phi
\nonumber\\&+&2\eta_{3\th\phi}\sin^3\th\cos3\phi\nonumber\\
&+&2\eta^{\perp}_{3\th\phi}\sin^3\th\sin3\phi,\label{eq:wspin2}
\end{eqnarray}
where
\begin{eqnarray*}
\lambda_{\th}&=&6[(1-3\delta_0-\delta_1)N_{T}\nonumber\\&-&(1-7\delta_0+\delta_1)(\rho^{T}_{1,1}+\rho^{T}_{-1,-1})\nonumber\\&-&(3-\delta_0-7\delta_1)\rho^{T}_{0,0}]/R,\nonumber\\
\lambda_{2\th}&=&(1+5\delta_0-5\delta_1)[N_{T}-5(\rho^{T}_{1,1}+\rho^{T}_{-1,-1})\nonumber\\&+&5\rho^{T}_{0,0}]/R,\nonumber\\
\lambda_{\th\phi}&=&4[2(1-\delta_0-2\delta_1)(\Re{\rho^{T}_{2,1}}-\Re{\rho^{T}_{-2,-1}})\nonumber\\&-&\sqrt{6}(2\delta_0-\delta_1)(\Re{\rho^{T}_{1,0}}-\Re{\rho^{T}_{-1,0}})]/R,\nonumber\\
\lambda_{2\th\phi}&=&-2(1+5\delta_0-5\delta_1)[(\Re{\rho^{T}_{2,1}}-\Re{\rho^{T}_{-2,-1}})\nonumber\\&-&\sqrt{6}(\Re{\rho^{T}_{1,0}}-\Re{\rho^{T}_{-1,0}})]/R,\nonumber\\
\lambda^{\perp}_{\th\phi}&=&4[-2(1-\delta_0-2\delta_1)(\Im{\rho^{T}_{2,1}}+\Im{\rho^{T}_{-2,-1}})\nonumber\\&+&\sqrt{6}(2\delta_0-\delta_1)(\Im{\rho^{T}_{1,0}}+\Im{\rho^{T}_{-1,0}})]/R,\nonumber\\
\lambda^{\perp}_{2\th\phi}&=&2(1+5\delta_0-5\delta_1)[(\Im{\rho^{T}_{2,1}}+\Im{\rho^{T}_{-2,-1}})\nonumber\\&-&\sqrt{6}(\Im{\rho^{T}_{1,0}}+\Im{\rho^{T}_{-1,0}})]/R,\nonumber\\
\lambda_{\phi}&=&4[\sqrt{6}(1+\delta_0-3\delta_1)(\Re{\rho^{T}_{2,0}}+\Re{\rho^{T}_{-2,0}})\nonumber\\&-&6(2\delta_0-\delta_1)\Re{\rho^{T}_{1,-1}}]/R,\nonumber\\
\lambda_{2\phi}&=&-2(1+5\delta_0-5\delta_1)[\sqrt{6}(\Re{\rho^{T}_{2,0}}+\Re{\rho^{T}_{-2,0}})\nonumber\\&-&4\Re{\rho^{T}_{1,-1}}]/R,\nonumber
\end{eqnarray*}
\begin{eqnarray}
\lambda^{\perp}_{\phi}&=&-4[\sqrt{6}(1+\delta_0-3\delta_1)(\Im{\rho^{T}_{2,0}}-\Im{\rho^{T}_{-2,0}})\nonumber\\&-&6(2\delta_0-\delta_1)\Im{\rho^{T}_{1,-1}}]/R,\nonumber\\
\lambda^{\perp}_{2\phi}&=&2(1+5\delta_0-5\delta_1)[\sqrt{6}(\Im{\rho^{T}_{2,0}}-\Im{\rho^{T}_{-2,0}})\nonumber\\&-&4\Im{\rho^{T}_{1,-1}}]/R,\nonumber\\
\lambda_{3\th\phi}&=&2(1+5\delta_0-5\delta_1)\frac{\Re{\rho^{T}_{2,-1}}-\Re{\rho^{T}_{-2,1}}}{R},\nonumber\\
\lambda^{\perp}_{3\th\phi}&=&-2(1+5\delta_0-5\delta_1)\frac{\Im{\rho^{T}_{2,-1}}+\Im{\rho^{T}_{-2,1}}}{R},\nonumber\\
\lambda_{4\phi}&=&2(1+5\delta_0-5\delta_1)\frac{\Re{\rho^{T}_{2,-2}}}{R},\nonumber\\
\lambda^{\perp}_{4\phi}&=&-2(1+5\delta_0-5\delta_1)\frac{\Im{\rho^{T}_{2,-2}}}{R},
\end{eqnarray}
and
\begin{eqnarray}
\eta_{\th}&=&2[(2\alpha_1+\alpha_2)(\rho^{T}_{2,2}-\rho^{T}_{-2,-2})\nonumber\\&-&2(\alpha_1-\alpha_2)(\rho^{T}_{1,1}-\rho^{T}_{-1,-1})]/R,\nonumber\\
\eta_{2\th}&=&-2(2\alpha_1-\alpha_2)[(\rho^{T}_{2,2}-\rho^{T}_{-2,-2})\nonumber\\&-&2(\rho^{T}_{1,1}-\rho^{T}_{-1,-1})]/R,\nonumber\\
\eta_{\th\phi}&=&-4[2(\alpha_1-\alpha_2)(\Re{\rho^{T}_{2,1}}+\Re{\rho^{T}_{-2,-1}})\nonumber\\&-&\sqrt{6}\alpha_1(\Re{\rho^{T}_{1,0}}+\Re{\rho^{T}_{-1,0}})]/R,\nonumber\\
\eta_{2\th\phi}&=&2(2\alpha_1-\alpha_2)[3(\Re{\rho^{T}_{2,1}}+\Re{\rho^{T}_{-2,-1}})\nonumber\\&-&\sqrt{6}(\Re{\rho^{T}_{1,0}}+\Re{\rho^{T}_{-1,0}})]/R,\nonumber\\
\eta^{\perp}_{\th\phi}&=&4[2(\alpha_1-\alpha_2)(\Im{\rho^{T}_{2,1}}-\Im{\rho^{T}_{-2,-1}})\nonumber\\&-&\sqrt{6}\alpha_1(\Im{\rho^{T}_{1,0}}-\Im{\rho^{T}_{-1,0}})]/R,\nonumber\\
\eta^{\perp}_{2\th\phi}&=&-2(2\alpha_1-\alpha_2)[3(\Im{\rho^{T}_{2,1}}-\Im{\rho^{T}_{-2,-1}})\nonumber\\&-&\sqrt{6}(\Im{\rho^{T}_{1,0}}-\Im{\rho^{T}_{-1,0}})]/R,\nonumber\\
\eta_{\phi}&=&-2\sqrt{6}(2\alpha_1-\alpha_2)\frac{\Re{\rho^{T}_{2,0}}-\Re{\rho^{T}_{-2,0}}}{R},\nonumber\\
\eta^{\perp}_{\phi}&=&2\sqrt{6}(2\alpha_1-\alpha_2)\frac{\Im{\rho^{T}_{2,0}}+\Im{\rho^{T}_{-2,0}}}{R},\nonumber\\
\eta_{3\th\phi}&=&-2(2\alpha_1-\alpha_2)\frac{\Re{\rho^{T}_{2,-1}}+\Re{\rho^{T}_{-2,1}}}{R},\nonumber\\
\eta^{\perp}_{3\th\phi}&=&2(2\alpha_1-\alpha_2)\frac{\Im{\rho^{T}_{2,-1}}-\Im{\rho^{T}_{-2,1}}}{R},\label{eq:eta}
\end{eqnarray}
with
\begin{eqnarray}
N_{T}&=&\rho^{T}_{2,2}+\rho^{T}_{1,1}+\rho^{T}_{0,0}+\rho^{T}_{-1,-1}+\rho^{T}_{-2,-2},\nonumber\\
R&=&(1+5\delta_0+3\delta_1)N_{T}\nonumber\\
&+&3(1-3\delta_0-\delta_1)(\rho^{T}_{1,1}+\rho^{T}_{-1,-1})\nonumber\\
&+&(5-7\delta_0-9\delta_1)\rho^{T}_{0,0}.
\end{eqnarray}
The parameters $\alpha_1$ and $\alpha_2$ vanish when  parity is
conserved, as in the $\chict$ decay. Other two parameters $\delta_0$ and $\delta_1$ can be determined from the specific
processes considered. For the $\chict$ decays into a $\jpsi$ and a
photon, through pure E1 transition, $\delta_0=\frac{1}{10}$ and
$\delta_1=\frac{3}{10}$, while after including the higher-order
multipole amplitudes in the radiative transitions, the coefficients
$\delta_0$ and $\delta_1$ can be expressed as the following
polynomials in the E1, M2, and E3 amplitudes
$a^{J=2}_1,a^{J=2}_2,a^{J=2}_3$:
\begin{eqnarray}
\delta_0&=&[1+2a^{J=2}_1(\sqrt{5}a^{J=2}_2+2a^{J=2}_3)\nonumber\\&+&4a^{J=2}_2(a^{J=2}_2+\sqrt{5}a^{J=2}_3)+3(a^{J=2}_3)^2]/10,\nonumber\\
\delta_1&=&[9+6a^{J=2}_1(\sqrt{5}a^{J=2}_2-4a^{J=2}_3)\nonumber\\&-&4a^{J=2}_2(a^{J=2}_2+2\sqrt{5}a^{J=2}_3)+7(a^{J=2}_3)^2]/30.
\end{eqnarray}
Again, the normalization of
$(a^{J=2}_1)^2+(a^{J=2}_2)^2+(a^{J=2}_3)^2=1$ has been imposed. The
measurements of the multipole amplitudes
Refs.~\cite{Artuso:2009aa,Oreglia:1981fx,Armstrong:1993fk,Ambrogiani:2001jw}
are listed in Table \ref{tab:multi2}. Finally, the expression of the
frame-independent parameter $F_2$ in terms of the coefficients in
Eq.(\ref{eq:wspin2}):
\begin{eqnarray}
F_{2}&=&\frac{n_1+n_2
\lambda_{\th}+n_3\lambda_{2\th}+n_4\lambda_{\phi}+n_5\lambda_{2\phi}+n_6\lambda_{4\phi}}{d_1+d_2\lambda_{\th}+d_3\lambda_{2\th}},\nonumber\\
n_1&=&\frac{1}{6},\nonumber\\
n_2&=&\frac{4-4\delta_0-3\delta_1}{18(2-4\delta_0-3\delta_1)},\nonumber\\
n_3&=&\frac{2+2\delta_0-7\delta_1-4\delta_0^2+\delta_0\delta_1+3\delta_1^2}{6(1+5\delta_0-5\delta_1)(2-4\delta_0-3\delta_1)},\nonumber\\
n_4&=&\frac{1}{3(2-4\delta_0-3\delta_1)},\nonumber\\
n_5&=&\frac{2\delta_0-\delta_1}{(1+5\delta_0-5\delta_1)(2-4\delta_0-3\delta_1)},\nonumber\\
n_6&=&\frac{1}{1+5\delta_0-5\delta_1},\nonumber\\
d_1&=&\frac{15}{16},d_2=\frac{5}{16},d_{3}=\frac{3}{16}.\label{eq:F2F2}
\end{eqnarray}

\begin{table}[h]
\caption{\label{tab:multi2}The normalized M2 and E3 amplitudes
$a^{J=2}_2,a^{J=2}_3$ for $\chict\to\jpsi\gamma$ as measured by
various experiments.}
\begin{tabular}{c|c*{1}{c}}
\hline\hline \itshape ~\rm{Experiment}~  & ~$a^{J=2}_2(10^{-2})$~ &
~$a^{J=2}_3(10^{-2})$~
\\\hline ~\rm{CLEO(Fit 1)}~\cite{Artuso:2009aa}~& $-9.3\pm1.6\pm0.3$ &
$0$\rm{(fixed)}\\~\rm{CLEO(Fit 2)}~\cite{Artuso:2009aa}~&
$-7.9\pm1.9\pm0.3$ &
$1.7\pm1.4\pm0.3$\\
~\rm{Crystal Ball}~\cite{Oreglia:1981fx}~ & $-33.3^{+11.6}_{-29.2}$
&
$0$\rm{(fixed)}\\
~\rm{E760(Fit 1)}~\cite{Armstrong:1993fk}~ & $-14\pm6$ & $0$\rm{(fixed)}\\
~\rm{E760(Fit 2)}~\cite{Armstrong:1993fk}~ & $-14^{+8}_{-7}$ &
$0^{+6}_{-5}$\\~\rm{E835(Fit 1)}~\cite{Ambrogiani:2001jw}~ &
$-9.3^{+3.9}_{-4.1}\pm0.6$ & $0$\rm{(fixed)}\\
~\rm{E835(Fit 2)}~\cite{Ambrogiani:2001jw}~ &
$-7.6^{+5.4}_{-5.0}\pm0.9$ & $2.0^{+5.5}_{-4.4}\pm0.9$
\\\hline\hline
\end{tabular}
\end{table}

\begin{figure}
\includegraphics[width=8.5cm]{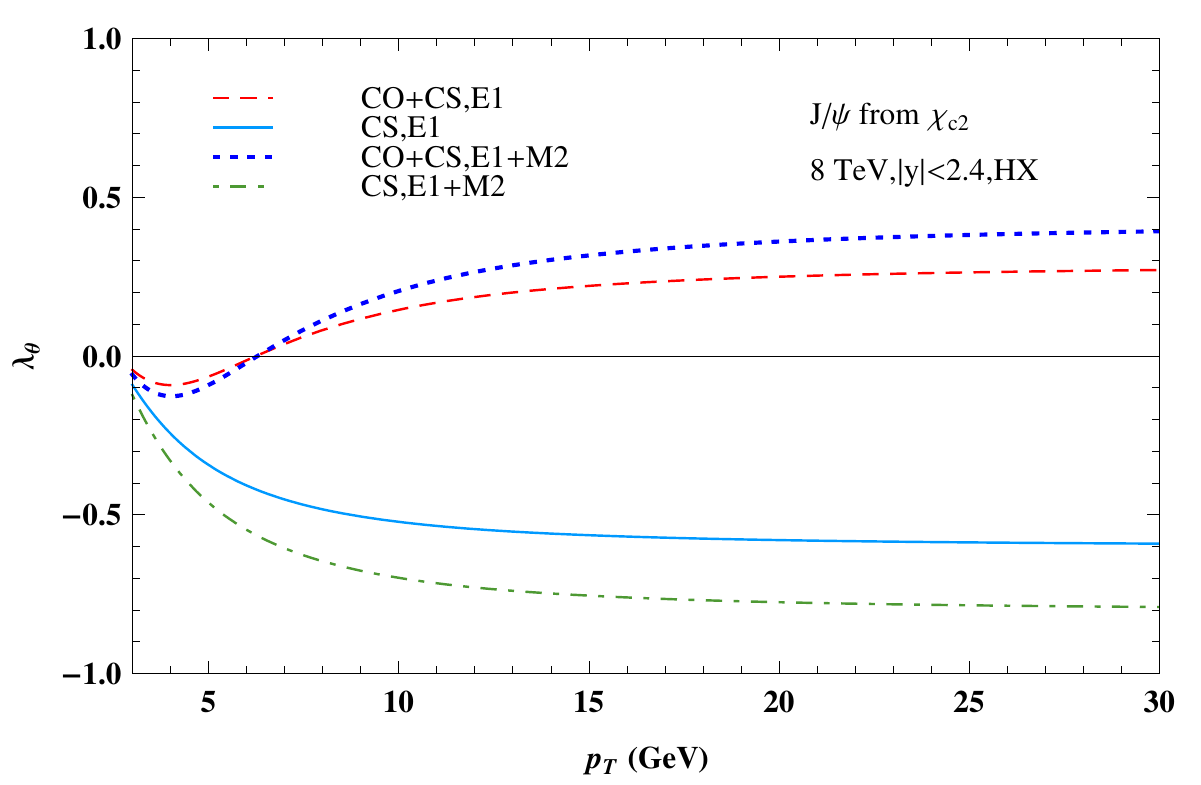}
\caption{\label{fig:chic2m2}(color online). M2 contribution to the $\chict$
polarization $\lambda_{\th}$ in $\chict\to\jpsi\gamma$ in the HX
frame. The CLEO measurement of $a^{J=2}_2$ listed in Table \ref{tab:multi2} has been used.}
\end{figure}

In order to show the impact of the higher-order multipole M2
contribution\footnote{The E3 amplitude for the $\chict$ decay is
zero from the consideration of the single quark radiation
hypothesis.} to the $\chict$ polarizations, we take the example
illustrated in Sec. \ref{sec:8} as well. As an illustrative
example, only $\lambda_{\th}$ in the HX frame is illustrated in Fig.
\ref{fig:chic2m2}, where E1 means pure E1 transition approximation and
E1+M2 means that we have included the full E1 and M2 transitions, with
$a^{J=2}_2$ as measured by the CLEO collaboration~\cite{Artuso:2009aa} (Table \ref{tab:multi2}).

\section{Dilepton angular distribution in $\chi_c$ decay with multipole effects\label{app:c}}
We consider here the general expression of the dilepton angular
distribution in the decays
$\chi_c\to\jpsi\gamma\to\mu^+\mu^-\gamma$, including also the
higher-order multipole transitions neglected in Sec. \ref{sec:5}.

In the first option discussed in Sec. \ref{sec:5} for the
definition of the $\jpsi$ quantization axis, the SDMEs for the
$\jpsi$ from $\chi_c$ decays can be expressed in terms of the
$\chi_c$ production matrix elements as\footnote{If one does not
integrate the angles $\th$ and $\phi$ in the following equation and put
the SDMEs $\rho^{\chicj\rightarrow
J/\psi\gamma}_{s_z,s_z^{\prime}}(\th,\phi)$ into Eq.(\ref{eq:jpsi}),
one obtains the full angular distribution of the decay chain
$\chi_c\rightarrow\jpsi\gamma\rightarrow\mu^+\mu^-\gamma$, including
the correlations between the $\chi_c$ decay angles $\th,\phi$ and
the $\jpsi$ decay angles $\th^{\prime},\phi^{\prime}$, which might
be useful in Monte Carlo simulations of experimental analyses.}
\begin{eqnarray}
&&\rho^{\chicj\rightarrow
\jpsi\gamma}_{s_z,s_z^{\prime}}\nonumber\\&=&\frac{1}{8\pi}\sum^{J+1}_{l=1}\sum_{\lambda_{\gamma}=\pm
l}\int{\rm{d}\Omega[\th,\phi]\rho^{\chicj}_{J_z,J_z^{\prime}}\mathcal{D}^{J*}_{J_z,J_{1z}}\mathcal{D}^{J}_{J_z^{\prime},J_{2z}}}\nonumber\\
&&\langle l,\lambda_{\gamma};1,s_z|J,J_{1z}\rangle \langle
J,J_{2z}|1,\lambda_{\gamma};1,s_z^{\prime}\rangle
\nonumber\\&&(2l+1)(a^{J=J}_l)^2\rm{Br}(\chicj\rightarrow\jpsi\gamma).\label{eq:sdchij}
\end{eqnarray}
The coefficients  are expressed as
\begin{eqnarray*}
\lambda^{\chico}_{\th^{\prime}}&=&-\frac{1-3(a^{J=1}_2)^2}{3-(a^{J=1}_2)^2},\lambda^{\chico}_{\phi^{\prime}}=\lambda^{\perp\chico}_{\phi^{\prime}}=0,\label{eq:mu11}\\
\lambda^{\chico}_{\th^{\prime}\phi^{\prime}}&=&\frac{\sqrt{2}(a^{J=1}_1)^2(\Re(\rho^{\chico}_{1,0})-\Re(\rho^{\chico}_{-1,0}))}{4(3-(a^{J=1}_2)^2)
N_{\chico}},\nonumber\\
\lambda^{\perp\chico}_{\th^{\prime}\phi^{\prime}}&=&-\frac{\sqrt{2}(a^{J=1}_1)^2(\Im(\rho^{\chico}_{1,0})+\Im(\rho^{\chico}_{-1,0}))}{4(3-(a^{J=1}_2)^2)
N_{\chico}},\nonumber\\
\lambda^{\chict}_{\th^{\prime}}&=&\frac{3(1-11(a^{J=2}_2)^2+9(a^{J=2}_3)^2)}{39+11(a^{J=2}_2)^2-9(a^{J=2}_3)^2},\nonumber\\
\lambda^{\chict}_{\phi^{\prime}}&=&\frac{(a^{J=2}_1)^2(7\sqrt{6}(\Re{\rho^{\chict}_{0,2}}+\Re{\rho^{\chict}_{0,-2}})+12\Re{\rho^{\chict}_{1,-1}})}{2(39+11(a^{J=2}_2)^2-9(a^{J=2}_3)^2)N_{\chict}},\nonumber\\
\lambda^{\chict}_{\th^{\prime}\phi^{\prime}}&=&[\sqrt{6}(1-\frac{13}{3}(a^{J=2}_2)^2-(a^{J=2}_3)^2)(\Re{\rho^{\chict}_{-1,0}}-\Re{\rho^{\chict}_{1,0}})\nonumber\\&+&6(4-9(a^{J=2}_2)^2-4(a^{J=2}_3)^2)(\Re{\rho^{\chict}_{-2,-1}}-\Re{\rho^{\chict}_{2,1}})]\nonumber\\
&/&[4(39+11(a^{J=2}_2)^2-9(a^{J=2}_3)^2)N_{\chict}],\nonumber\\
\lambda^{\perp\chict}_{\phi^{\prime}}&=&\frac{(a^{J=2}_1)^2(7\sqrt{6}(\Im{\rho^{\chict}_{0,2}}-\Im{\rho^{\chict}_{0,-2}})-12\Im{\rho^{\chict}_{1,-1}})}{2(39+11(a^{J=2}_2)^2-9(a^{J=2}_3)^2)N_{\chict}},\nonumber
\end{eqnarray*}
\begin{eqnarray}
\lambda^{\perp\chict}_{\th^{\prime}\phi^{\prime}}&=&[\sqrt{6}(1-\frac{13}{3}(a^{J=2}_2)^2-(a^{J=2}_3)^2)(\Im{\rho^{\chict}_{-1,0}}+\Im{\rho^{\chict}_{1,0}})\nonumber\\&+&6(4-9(a^{J=2}_2)^2-4(a^{J=2}_3)^2)
(\Im{\rho^{\chict}_{-2,-1}}+\Im{\rho^{\chict}_{2,1}})]
\nonumber\\&/&[4(39+11(a^{J=2}_2)^2-9(a^{J=2}_3)^2)N_{\chict}].\nonumber
\end{eqnarray}
As remarked previously, some of the polarization observables are
trivial and devoid of spin information. However, these observables
can, in principle, be measured to extract the multipole amplitudes of
the $\chi_c$ decay, for example, in electron-positron collisions.

In the second option, in which the quantization axis for the
$\jpsi\to\mu^+\mu^-$ decay coincides with the $\chi_c$ quantization
axis, the general relation between the SDMEs of the $\chicj$
$\rho^{\chicj}_{J_z,J_z^{\prime}}$ and those of the $\jpsi$ from the
$\chicj$ decay
$\rho^{\chicj\rightarrow\jpsi\gamma}_{s_z,s_z^{\prime}}$ in the full multipole expansion is
\begin{eqnarray}
&&\rho^{\chicj\rightarrow\jpsi\gamma}_{s_z,s_z^{\prime}}\propto\sum^{J+1}_{l=1}\sum^l_{l_z=-l}{\sum_{J_z,J_z^{\prime}}{(a^{J=J}_l)^2\langle l,l_z;1,s_z|J,J_z\rangle}}\nonumber\\
&&\langle J,J_z^{\prime}|l,l_z;1,s_z^{\prime}\rangle
\rho^{\chicj}_{J_z,J_z^{\prime}}\rm{Br}(\chicj\rightarrow\jpsi\gamma).\label{eq:chicjpsi2}
\end{eqnarray}
The coefficients of the $\mu^+$ angular distribution are
\begin{eqnarray}
\lambda^{\chico}_{\th^{\prime}}&=&\frac{-N_{\chico}+3\rho^{\chico}_{0,0}}{R_1},\\
\lambda^{\chico}_{\phi^{\prime}}&=&-\frac{2\Re{\rho^{\chico}_{1,-1}}}{R_1},\nonumber\\
\lambda^{\chico}_{\th^{\prime}\phi^{\prime}}&=&-\frac{\sqrt{2}(\Re{\rho^{\chico}_{1,0}}-\Re{\rho^{\chico}_{-1,0}})}{R_1},\nonumber\\
\lambda^{\perp\chico}_{\phi^{\prime}}&=&\frac{2\Im{\rho^{\chico}_{1,-1}}}{R_1},\nonumber\\
\lambda^{\perp\chico}_{\th^{\prime}\phi^{\prime}}&=&\frac{\sqrt{2}(\Im{\rho^{\chico}_{1,0}}+\Im{\rho^{\chico}_{-1,0}})}{R_1},\nonumber\\
\lambda^{\chict}_{\th^{\prime}}&=&\frac{6N_{\chict}-9(\rho^{\chict}_{1,1}+\rho^{\chict}_{-1,-1})-12\rho^{\chict}_{0,0}}{R_2},\nonumber\\
\lambda^{\chict}_{\phi^{\prime}}&=&\frac{2\sqrt{6}(\Re{\rho^{\chict}_{2,0}}+\Re{\rho^{\chict}_{-2,0}})+6\Re{\rho^{\chict}_{1,-1}}}{R_2},\nonumber\\
\lambda^{\chict}_{\th^{\prime}\phi^{\prime}}&=&\frac{6(\Re{\rho^{\chict}_{2,1}}-\Re{\rho^{\chict}_{-2,-1}})+\sqrt{6}(\Re{\rho^{\chict}_{1,0}}-\Re{\rho^{\chict}_{-1,0}})}
{R_2},\nonumber\\
\lambda^{\perp\chict}_{\phi^{\prime}}&=&\frac{2\sqrt{6}(\Im{\rho^{\chict}_{0,2}}-\Im{\rho^{\chict}_{0,-2}})-6\Im{\rho^{\chict}_{1,-1}}}{R_2},\nonumber\\
\lambda^{\perp\chict}_{\th^{\prime}\phi^{\prime}}&=&\frac{6(\Im{\rho^{\chict}_{1,2}}+\Im{\rho^{\chict}_{-1,-2}})+\sqrt{6}(\Im{\rho^{\chict}_{0,1}}+\Im{\rho^{\chict}_{0,-1}})}
{R_2},\nonumber
\end{eqnarray}
with
\begin{eqnarray*}
R_1&=&[(15-2(a^{J=1}_2)^2)N_{\chico}\nonumber\\&-&(5-6(a^{J=1}_2)^2)\rho^{\chico}_{0,0}]/(5-6(a^{J=1}_2)^2),\nonumber\\
R_2&=&[2(21+14(a^{J=2}_2)^2+5(a^{J=2}_3)^2)N_{\chict}\nonumber\\&+&3(7-14(a^{J=2}_2)^2-5(a^{J=2}_3)^2)(\rho^{\chict}_{1,1}+\rho^{\chict}_{-1,-1})\nonumber\\
&+&4(7-14(a^{J=2}_2)^2-5(a^{J=2}_3)^2)\rho^{\chict}_{0,0}]\nonumber\\&/&(7-14(a^{J=2}_2)^2-5(a^{J=2}_3)^2).
\end{eqnarray*}

\section{A possible way for determining $\rho_{J_z,J_z^{\prime}}(|J_z-J_z^{\prime}|>2)$ of $\chict$ with a reweighting method\label{app:d}}
From Eqs.(\ref{eq:chic2}), (\ref{eq:mu1}), (\ref{eq:mu2}), we see that the
$\chict$ SDMEs $\rho_{J_z,J_z^{\prime}}$ having
$|J_z-J_z^{\prime}|>2$ cannot be measured from the integrated
angular distributions. The fact that the coefficients of these SDMEs
are suppressed by $v^2$ or $(m_{\chict}-m_{\jpsi})$ makes the
measurement of these polarization observables difficult. In this
appendix, we propose a reweighting method to measure these SDMEs.

From Eq.(\ref{eq:hel}), it can be recognized that this fact
originates from the cancellation of the transverse and longitudinal
components of the $\jpsi$ coming from $\chict$.\footnote{It might not be
so straightforward for the reader. It will be clear if the
reader substitutes the Eq.(\ref{eq:hel}) into Eq.(\ref{eq:WW}).} If
the probabilities of the transverse and the longitudinal parts are
made different by reweighting, the suppression of the
$\rho_{J_z,J_z^{\prime}}(|J_z-J_z^{\prime}|>2)$ terms can be
avoided. The only tradeoff is that one should also measure the polar
angle $\th^{\prime}$ of the $\mu^+$ from the subsequent $\jpsi$
decay.\footnote{We are using the first one of the two options in Sec.
\ref{sec:5}.} As is well known, the polar angle distribution of the
$\mu^+$ from the transverse polarized $\jpsi$ decay is
$\frac{(1+\cos^2{\th^{\prime}})}{2}$, while that from the
longitudinal one is $(1-\cos^2{\th^{\prime}})$. Therefore, after
integrating over solid angles of the $\mu^{+}$, the transverse and
longitudinal $\jpsi$ components receive equal weight
$D_{J_z,J_z^{\prime}}=\sum_{\lambda_{\jpsi}=\pm,0,\lambda_{\gamma}=\pm}{\mathcal{M}_{J_z,\lambda_{\jpsi},\lambda_{\gamma}}\mathcal{M}^*_{J_z^{\prime},\lambda_{\jpsi},\lambda_{\gamma}}}$
resulting in the cancellation of the terms containing
$\rho_{J_z,J_z^{\prime}}(|J_z-J_z^{\prime}|>2)$ of the $\chict$. It
is possible to avoid these cancellations by measuring the angle
$\th^{\prime}$ and assigning each reconstructed event the extra
weight $w(\cos^2\th^{\prime})$. For example, if one chooses
$w(\cos^2\th^{\prime})\propto\cos^2\th^{\prime}$, because of
\begin{eqnarray*}
\int{\rm{d}\cos\th^{\prime}~(1-\cos^2\th^{\prime})w(\cos^2\th^{\prime})}&&\\:\int{\rm{d}\cos\th^{\prime}~\frac{(1+\cos^2\th^{\prime})}{2}w(\cos^2\th^{\prime})}&&=1:2,
\end{eqnarray*}
the decay SDMEs become
\begin{eqnarray}
D_{J_z,J_z^{\prime}}&=&2\sum_{\lambda_{\jpsi}=\pm,\lambda_{\gamma}=\pm}{\mathcal{M}_{J_z,\lambda_{\jpsi},\lambda_{\gamma}}\mathcal{M}^*_{J_z^{\prime},\lambda_{\jpsi},\lambda_{\gamma}}}\nonumber\\
&+&\sum_{\lambda_{\gamma}=\pm}{\mathcal{M}_{J_z,0,\lambda_{\gamma}}\mathcal{M}^*_{J_z^{\prime},0,\lambda_{\gamma}}},
\end{eqnarray}
and the cancellations do not occur. Without losing  generality, we
define $r_L$ and $r_T$ such that
\begin{eqnarray*}
\int{\rm{d}\cos\th^{\prime}~(1-\cos^2\th^{\prime})w(\cos^2\th^{\prime})}&&\\:\int{\rm{d}\cos\th^{\prime}~\frac{(1+\cos^2\th^{\prime})}{2}w(\cos^2\th^{\prime})}&&=r_L:r_T,
\end{eqnarray*}
and
\begin{eqnarray}
D_{J_z,J_z^{\prime}}&=&r_T\sum_{\lambda_{\jpsi}=\pm,\lambda_{\gamma}=\pm}{\mathcal{M}_{J_z,\lambda_{\jpsi},\lambda_{\gamma}}\mathcal{M}^*_{J_z^{\prime},\lambda_{\jpsi},\lambda_{\gamma}}}\nonumber\\
&+&r_L\sum_{\lambda_{\gamma}=\pm}{\mathcal{M}_{J_z,0,\lambda_{\gamma}}\mathcal{M}^*_{J_z^{\prime},0,\lambda_{\gamma}}}.
\end{eqnarray}
The weighted angular distribution of $\chict\to\jpsi\gamma$ is, thus,
\begin{eqnarray}
\mathcal{\tilde{W}}^{\chict\to\jpsi\gamma}(\th,\phi)&\propto&1+\lambda_{\th}\cos^2\th+\lambda_{2\th}\cos^4\th\nonumber\\
&+&\lambda_{\phi}\sin^2\th\cos2\phi+\lambda^{\perp}_{\phi}\sin^2\th\sin2\phi\nonumber\\
&+&\lambda_{\th\phi}\sin2\th\cos\phi+\lambda^{\perp}_{\th\phi}\sin2\th\sin\phi\nonumber\\
&+&\lambda_{2\phi}\sin^4\th\cos2\phi+\lambda^{\perp}_{2\phi}\sin^4\th\sin2\phi\nonumber\\
&+&\lambda_{2\th\phi}\sin2\th\sin^2\th\cos\phi\nonumber\\
&+&\lambda^{\perp}_{2\th\phi}\sin2\th\sin^2\th\sin\phi\nonumber\\
&+&\lambda_{3\th\phi}\sin2\th\sin^2\th\cos3\phi\nonumber\\
&+&\lambda^{\perp}_{3\th\phi}\sin2\th\sin^2\th\sin3\phi\nonumber\\
&+&\lambda_{4\phi}\sin^4\th\cos4\phi\nonumber\\
&+&\lambda^{\perp}_{4\phi}\sin^4\th\sin4\phi.\label{eq:angd}
\end{eqnarray}
The explicit expressions for the coefficients are {\small
\begin{eqnarray*}
N_{\chict}&=&\rho_{2,2}+\rho_{1,1}+\rho_{0,0}+\rho_{-1,-1}+\rho_{-2,-2},\nonumber\\
R&=&3(r_T+r_L)N_{\chict}+3r_T(\rho_{1,1}+\rho_{-1,-1})+(7r_T-3r_L)\rho_{0,0},\nonumber\\
\lambda_{\th}&=&\frac{6r_T
N_{\chict}-9r_L(\rho_{1,1}+\rho_{-1,-1})-6(5r_T-3r_L)\rho_{0,0}}{R},\nonumber\\
\lambda_{2\th}&=&(r_T-r_L)\frac{3N_{\chict}-15(\rho_{1,1}+\rho_{-1,-1})+15\rho_{0,0}}{R},\nonumber\\
\end{eqnarray*}
\begin{eqnarray}
\lambda_{\phi}&=&\left[2\sqrt{6}(4r_T-3r_L)(\Re{\rho_{2,0}}+\Re{\rho_{-2,0}})\right.\nonumber\\&&\left.-6(2r_T-3r_L)\Re{\rho_{1,-1}}\right]/R,\nonumber\\
\lambda^{\perp}_{\phi}&=&-\left[2\sqrt{6}(4r_T-3r_L)(\Im{\rho_{2,0}}-\Im{\rho_{-2,0}})\right.\nonumber\\&&\left.-6(2r_T-3r_L)\Im{\rho_{1,-1}}\right]/R,\nonumber\\
\lambda_{\th\phi}&=&\left[6(2r_T-r_L)(\Re{\rho_{2,1}}-\Re{\rho_{-2,-1}})\right.\nonumber\\&&\left.-\sqrt{6}(2r_T-3r_L)(\Re{\rho_{1,0}}-\Re{\rho_{-1,0}})\right]/R,\nonumber\\
\lambda^{\perp}_{\th\phi}&=&-\left[6(2r_T-r_L)(\Im{\rho_{2,1}}+\Im{\rho_{-2,-1}})\right.\nonumber\\&&\left.-\sqrt{6}(2r_T-3r_L)(\Im{\rho_{1,0}}+\Im{\rho_{-1,0}})\right]/R,\nonumber\\
\lambda_{2\phi}&=&(r_T-r_L)\frac{24\Re{\rho_{1,-1}}-6\sqrt{6}(\Re{\rho_{2,0}}+\Re{\rho_{-2,0}})}{R},\nonumber\\
\lambda^{\perp}_{2\phi}&=&-(r_T-r_L)\frac{24\Im{\rho_{1,-1}}-6\sqrt{6}(\Im{\rho_{2,0}}-\Im{\rho_{-2,0}})}{R},\nonumber\\
\lambda_{2\th\phi}&=&6(r_T-r_L)\left[-(\Re{\rho_{2,1}}-\Re{\rho_{-2,-1}})\right.\nonumber\\&&\left.+\sqrt{6}(\Re{\rho_{1,0}}-\Re{\rho_{-1,0}})\right]/R,\nonumber\\
\lambda^{\perp}_{2\th\phi}&=&6(r_T-r_L)\left[(\Im{\rho_{2,1}}+\Im{\rho_{-2,-1}})\right.\nonumber\\&&\left.-\sqrt{6}(\Im{\rho_{1,0}}+\Im{\rho_{-1,0}})\right]/R,\nonumber\\
\lambda_{3\th\phi}&=&6(r_T-r_L)\frac{\Re{\rho_{2,-1}}-\Re{\rho_{-2,1}}}{R},\nonumber\\
\lambda^{\perp}_{3\th\phi}&=&-6(r_T-r_L)\frac{\Im{\rho_{2,-1}}+\Im{\rho_{-2,1}}}{R},\nonumber\\
\lambda_{4\phi}&=&(r_T-r_L)\frac{6\Re{\rho_{2,-2}}}{R},\nonumber\\
\lambda^{\perp}_{4\phi}&=&-(r_T-r_L)\frac{6\Im{\rho_{2,-2}}}{R}.\label{eq:reweight}
\end{eqnarray}}
From the above equations, we find that one can measure
$\lambda_{3\th\phi},\lambda^{\perp}_{3\th\phi},\lambda_{4\phi},\lambda^{\perp}_{4\phi}$
to determine the values of
$\rho_{J_z,J_z^{\prime}}(|J_z-J_z^{\prime}|>2)$ if $r_T\neq r_L$.
\end{appendix}
\begin{figure}
\includegraphics[width=8.5cm]{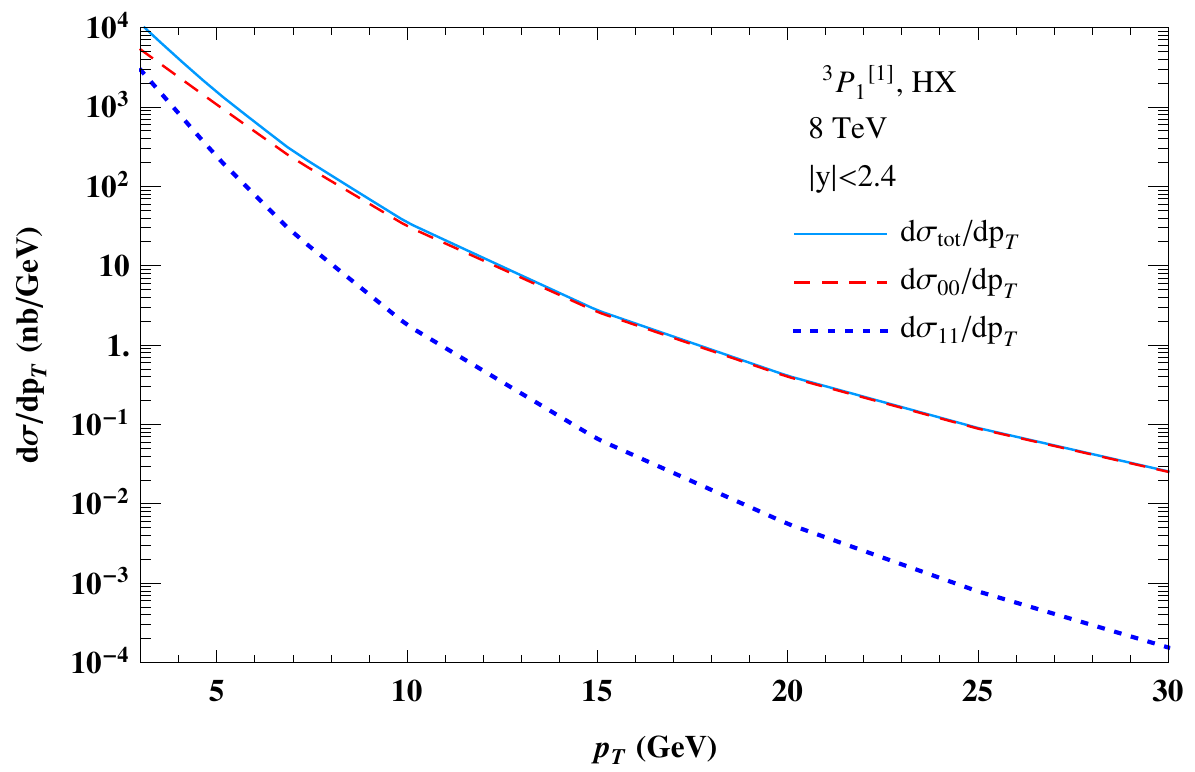}
(a)
\includegraphics[width=8.5cm]{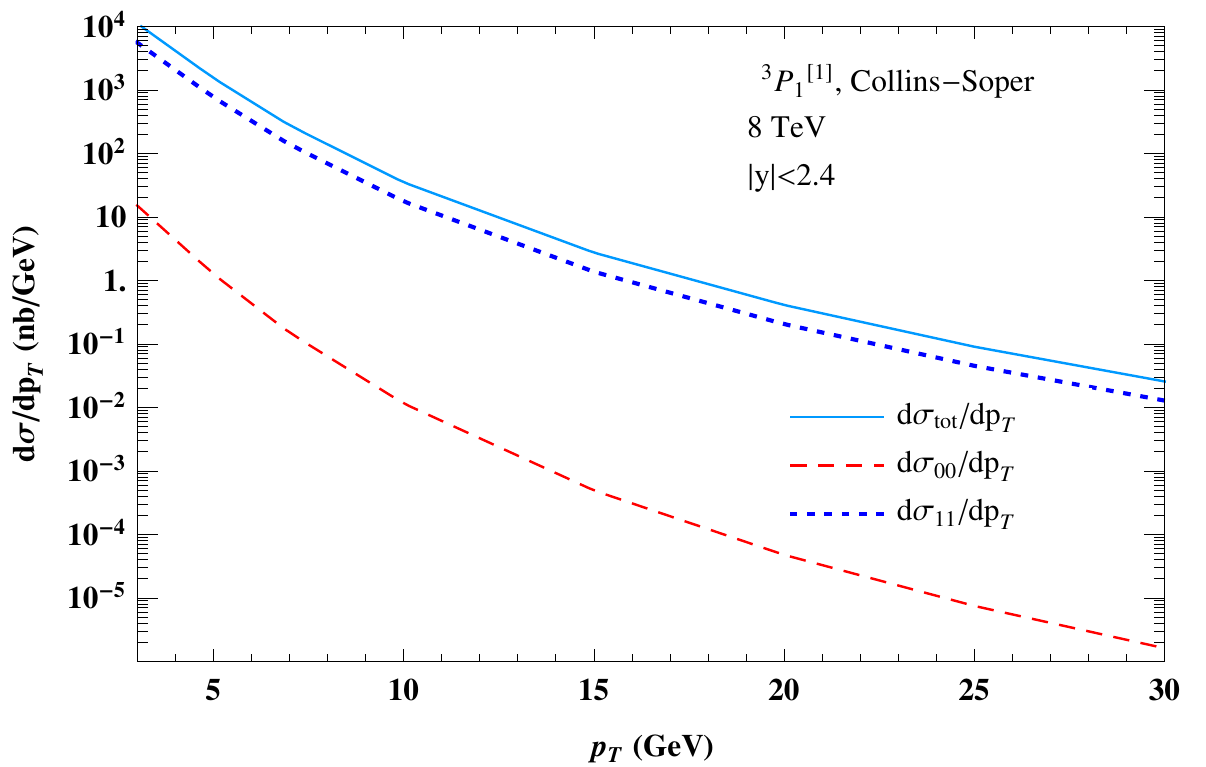}
(b)
\includegraphics[width=8.5cm]{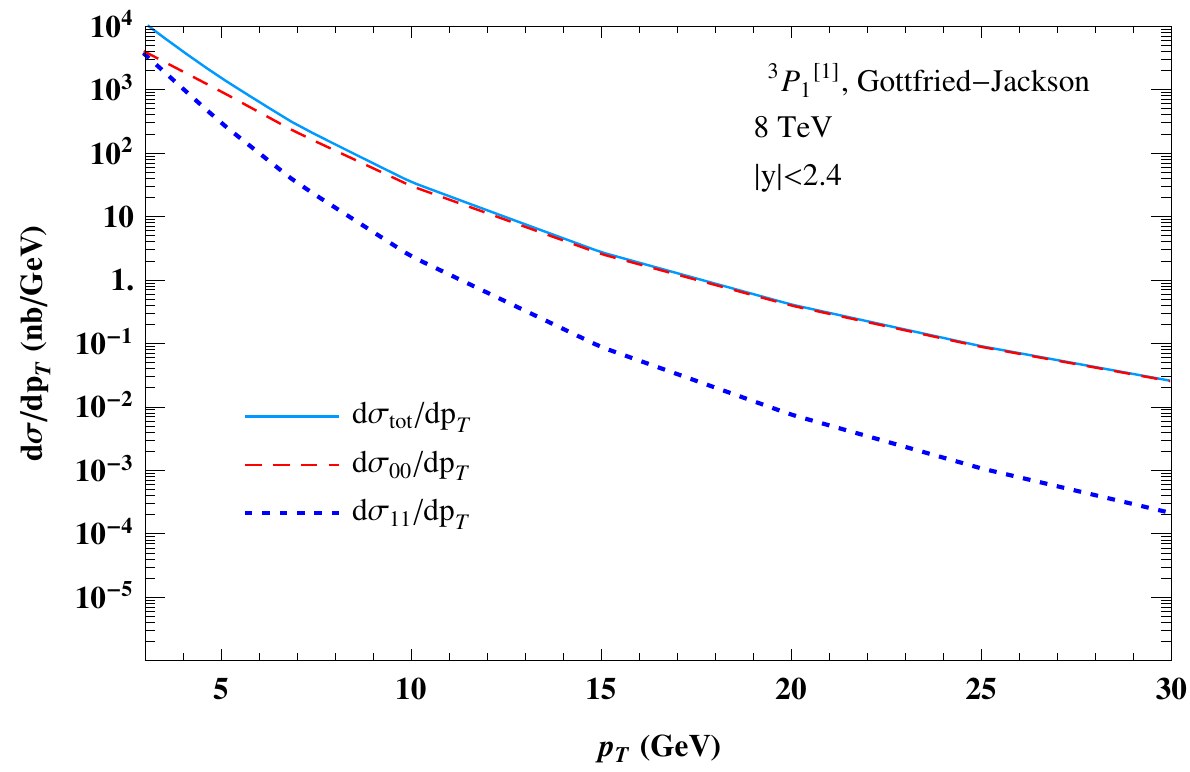}
(c) \caption{\label{fig:sigmapo}(color online). Distributions of
$\frac{\rm{d}\sigma_{00}}{\rm{d}p_T}$,$\frac{\rm{d}\sigma_{11}}{\rm{d}p_T}$,
and
$\frac{\rm{d}\sigma_{tot}}{\rm{d}p_T}=2\frac{\rm{d}\sigma_{11}}{\rm{d}p_T}+\frac{\rm{d}\sigma_{00}}{\rm{d}p_T}$
for $\tpos$ in the (a) HX, (b) Collins-Soper, and (c) Gottfried-Jackson frames.}
\end{figure}
\begin{figure}
\includegraphics[width=8.5cm]{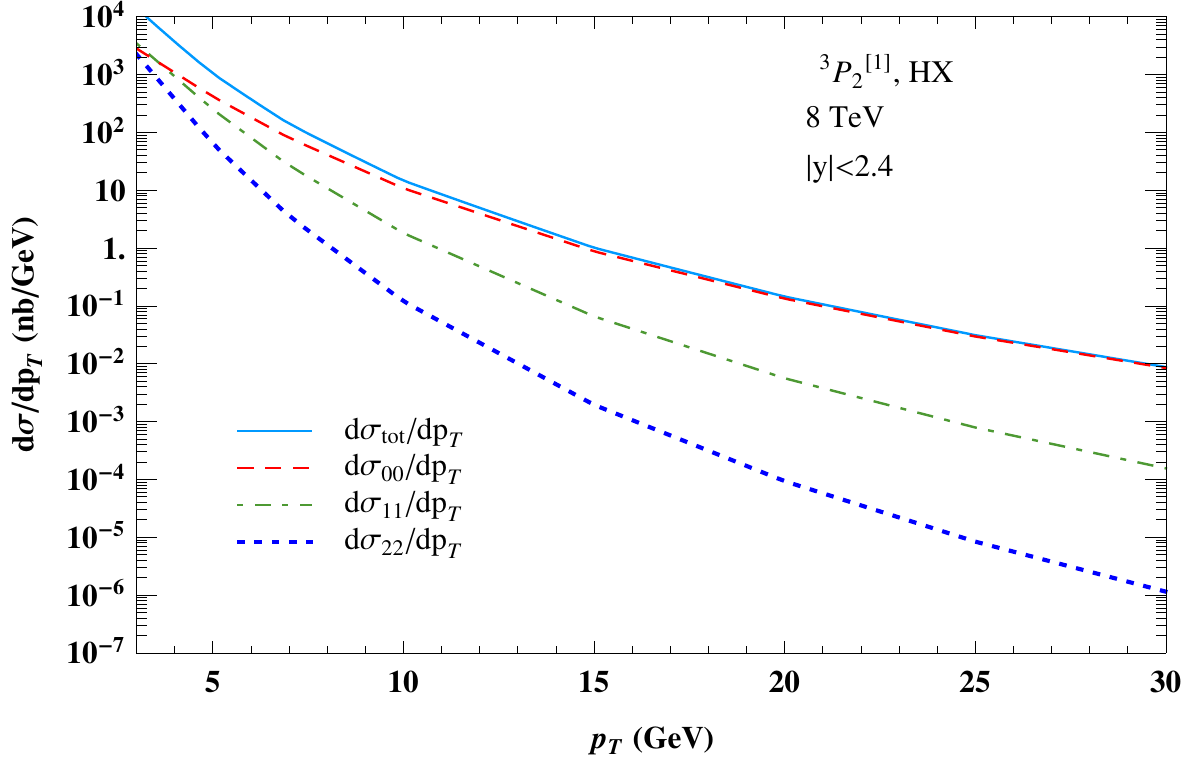}
(a)
\includegraphics[width=8.5cm]{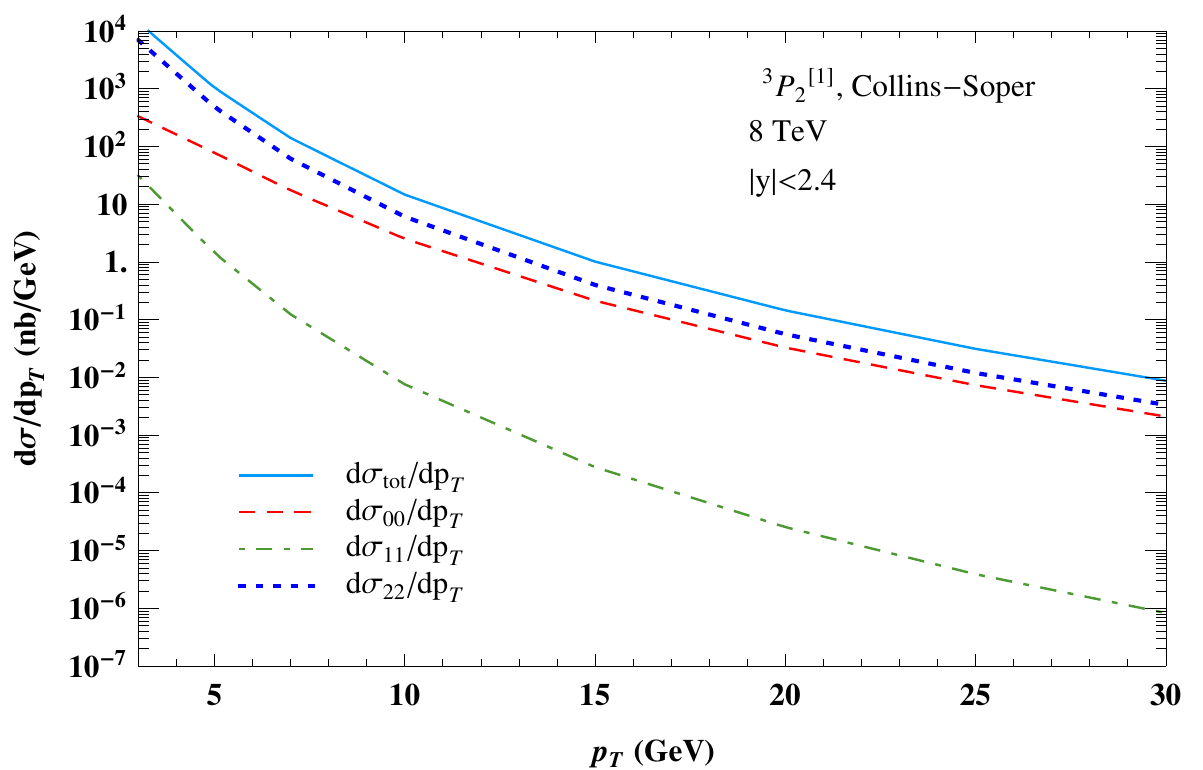}
(b)
\includegraphics[width=8.5cm]{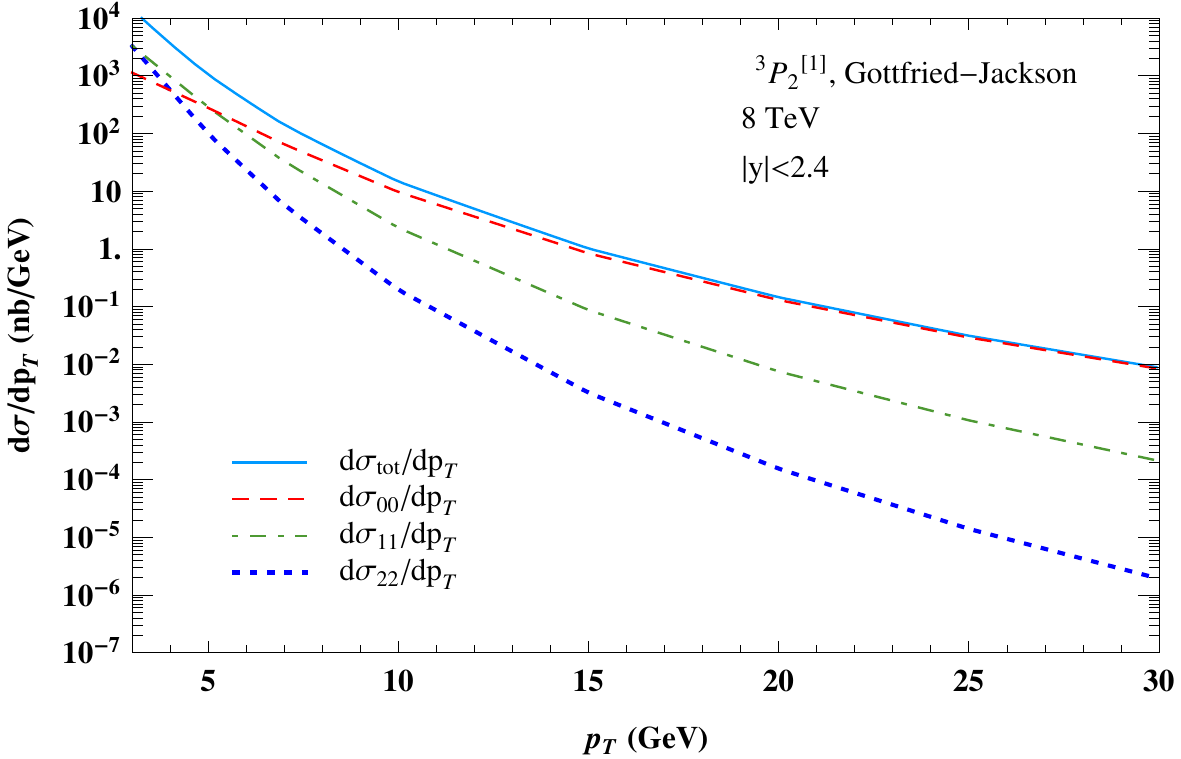}
(c) \caption{\label{fig:sigmapt}(color online). Distributions of
$\frac{\rm{d}\sigma_{00}}{\rm{d}p_T}$, $\frac{\rm{d}\sigma_{11}}{\rm{d}p_T}$, $\frac{\rm{d}\sigma_{22}}{\rm{d}p_T}$,
and
$\frac{\rm{d}\sigma_{tot}}{\rm{d}p_T}=2\frac{\rm{d}\sigma_{22}}{\rm{d}p_T}+2\frac{\rm{d}\sigma_{11}}{\rm{d}p_T}+\frac{\rm{d}\sigma_{00}}{\rm{d}p_T}$
for $\tpts$ in the (a) HX, (b) Collins-Soper, and (c) Gottfried-Jackson frames.}
\end{figure}
\begin{figure}
\includegraphics[width=8.5cm]{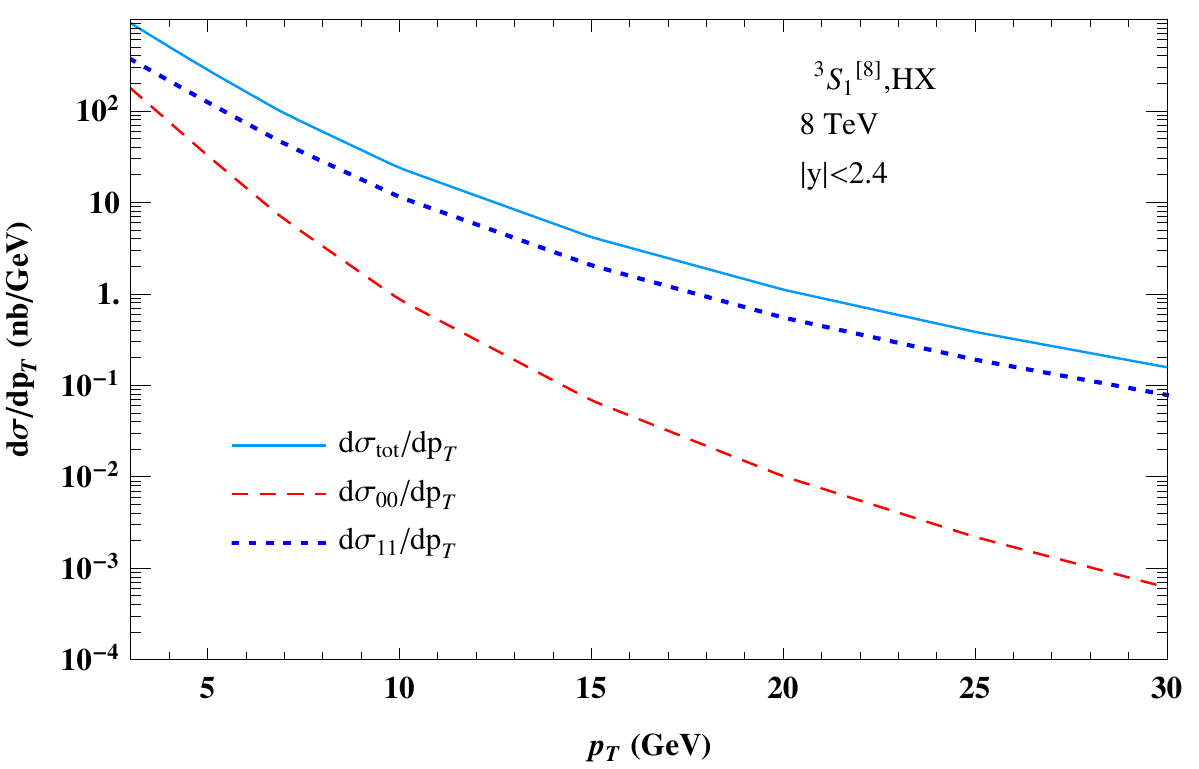}
(a)
\includegraphics[width=8.5cm]{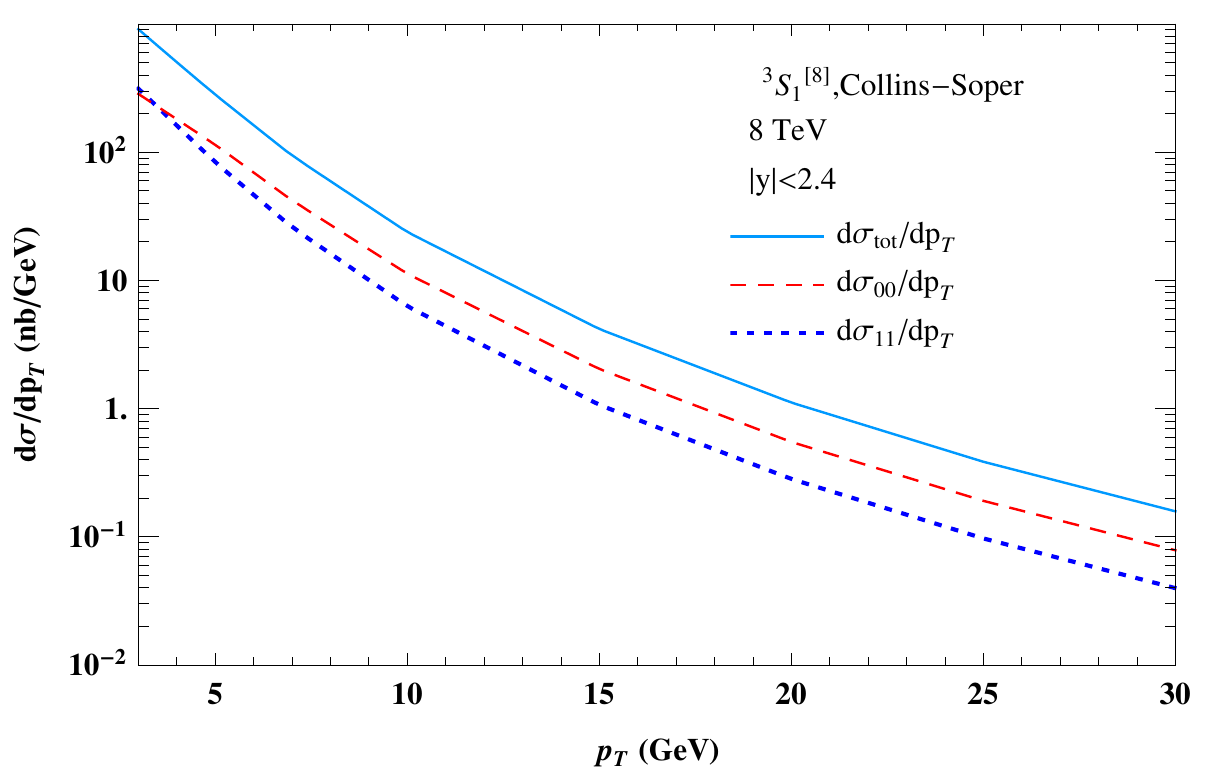}
(b)
\includegraphics[width=8.5cm]{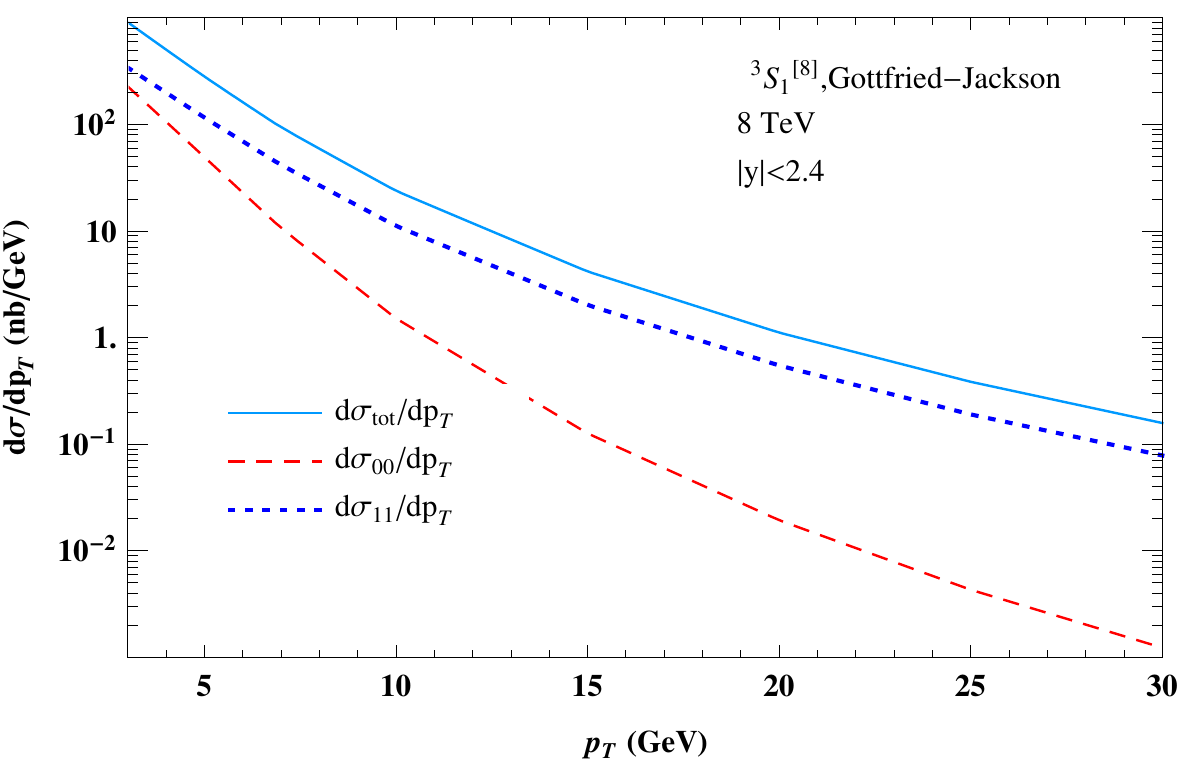}
(c) \caption{\label{fig:sigmas}(color online). Distributions of
$\frac{\rm{d}\sigma_{00}}{\rm{d}p_T}$, $\frac{\rm{d}\sigma_{11}}{\rm{d}p_T}$,
and
$\frac{\rm{d}\sigma_{tot}}{\rm{d}p_T}=2\frac{\rm{d}\sigma_{11}}{\rm{d}p_T}+\frac{\rm{d}\sigma_{00}}{\rm{d}p_T}$
for $\so$ [including only $\langle\mathcal{O}^{\chicz}(\so)\rangle$]
in the (a) HX, (b) Collins-Soper, and (c) Gottfried-Jackson frames.}
\end{figure}
\begin{figure}
\includegraphics[width=8.5cm]{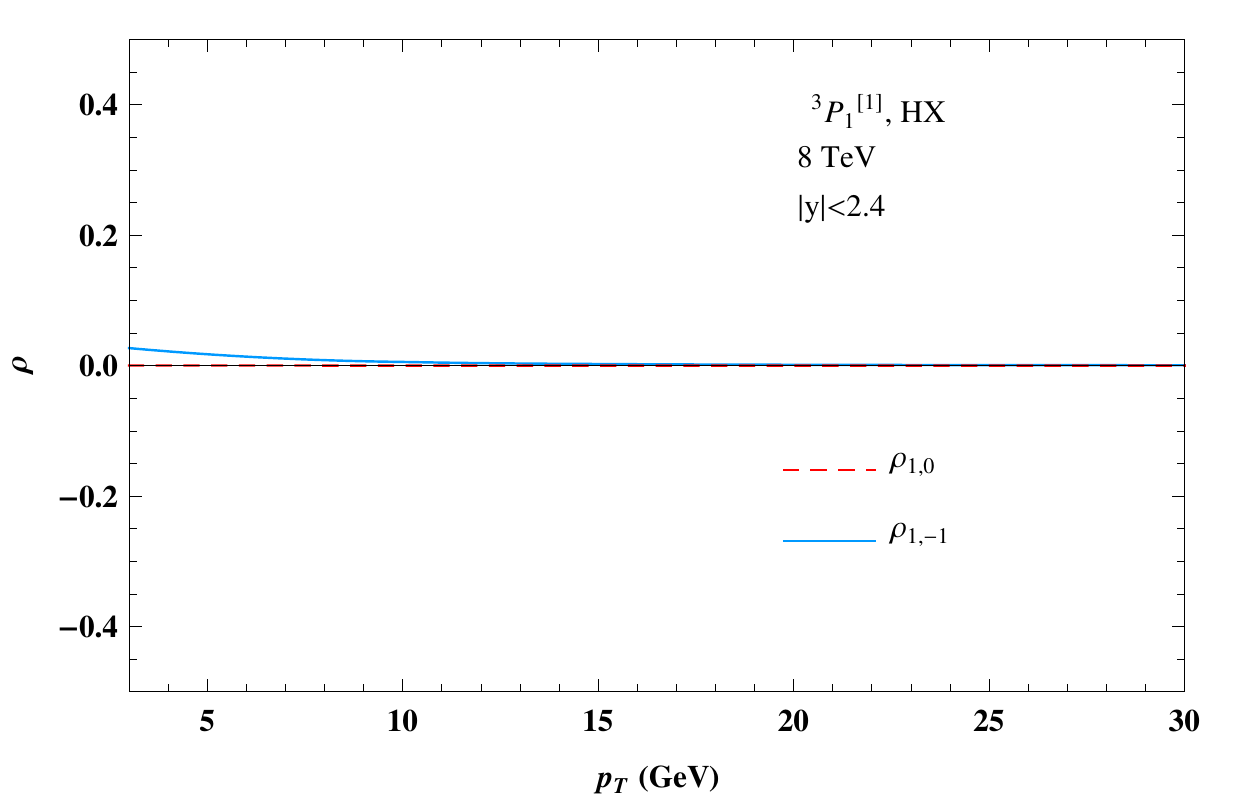}
(a)
\includegraphics[width=8.5cm]{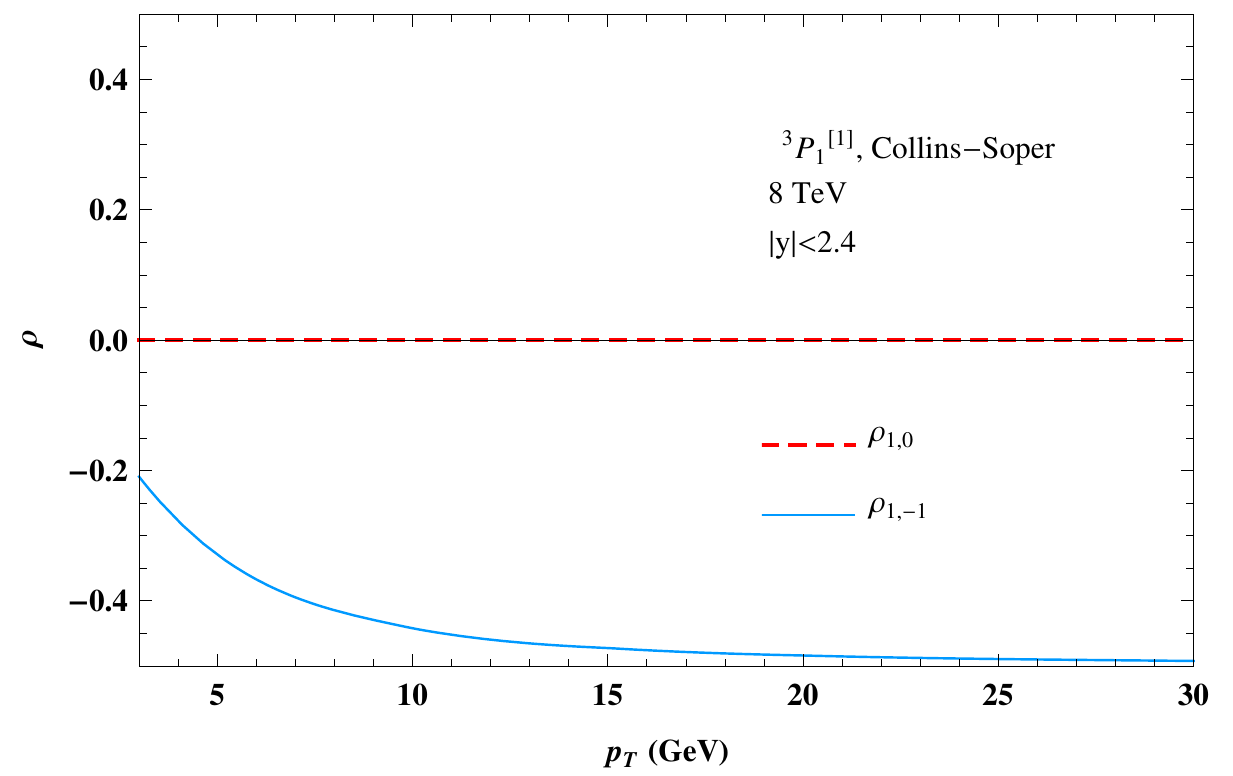}
(b)
\includegraphics[width=8.5cm]{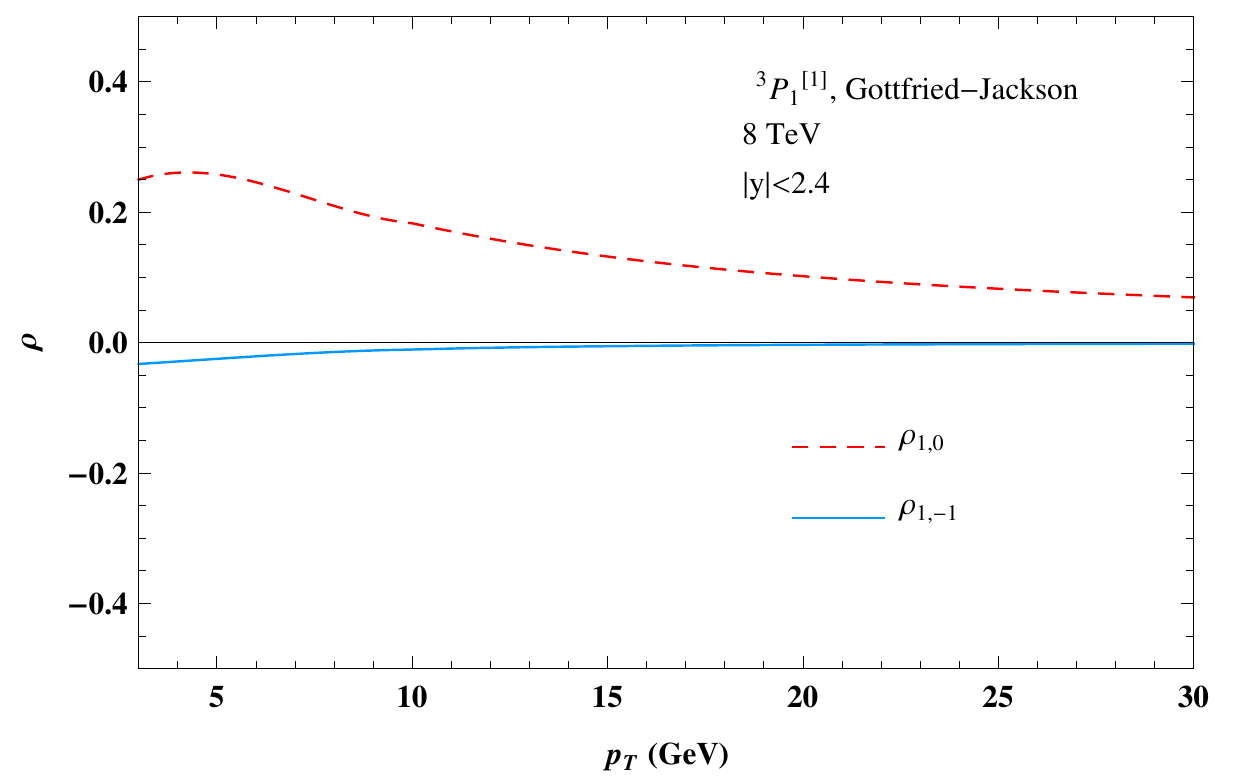}
(c) \caption{\label{fig:rhopo}(color online). Real parts of the nondiagonal SDMEs
(normalized by
$\frac{\rm{d}\sigma_{tot}}{\rm{d}p_T}=2\frac{\rm{d}\sigma_{11}}{\rm{d}p_T}+\frac{\rm{d}\sigma_{00}}{\rm{d}p_T}$)
for $\tpos$ in the (a) HX, (b) Collins-Soper, and (c) Gottfried-Jackson frames.}
\end{figure}
\begin{figure}
\includegraphics[width=8.5cm]{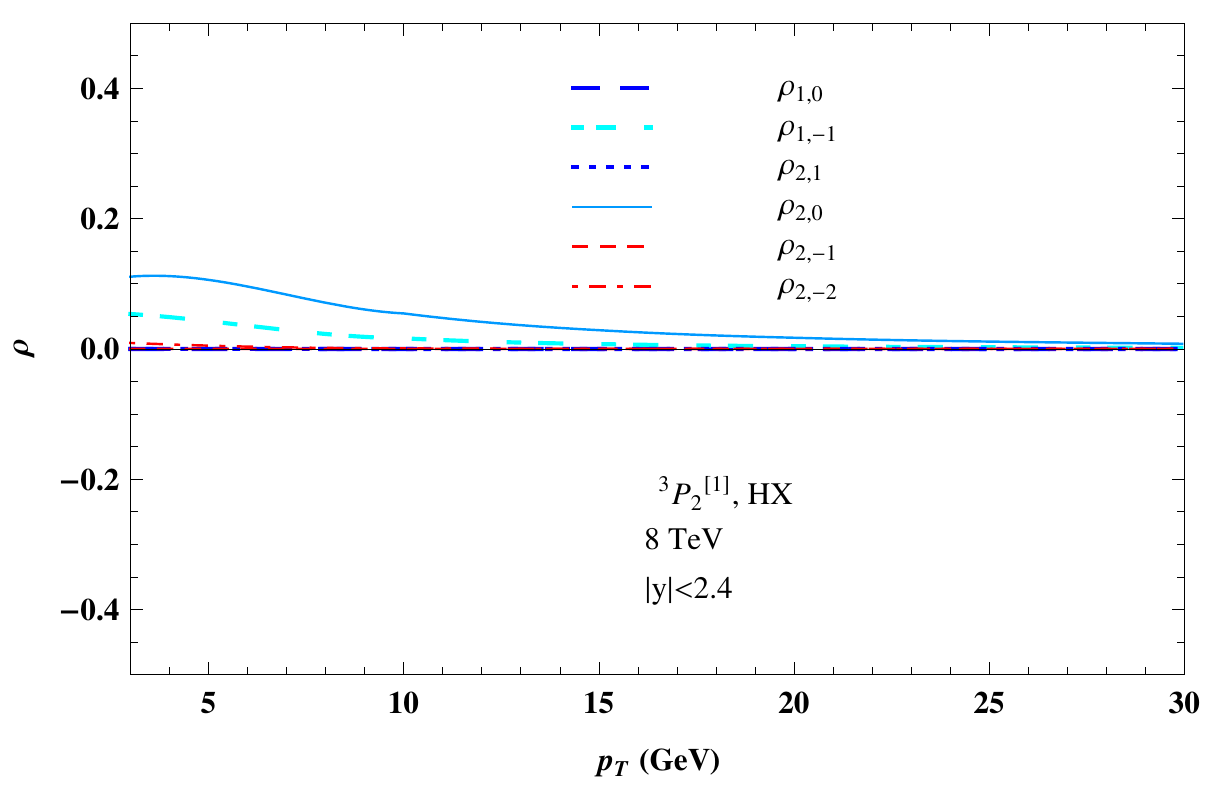}
(a)
\includegraphics[width=8.5cm]{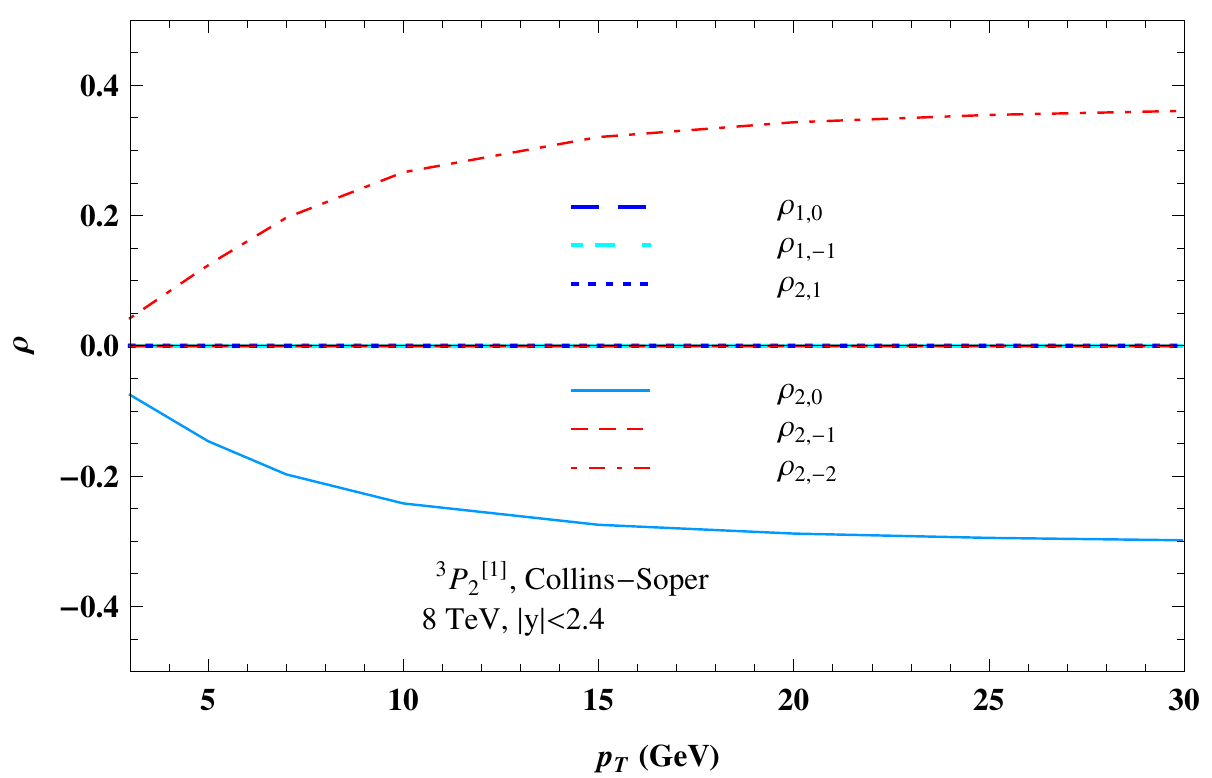}
(b)
\includegraphics[width=8.5cm]{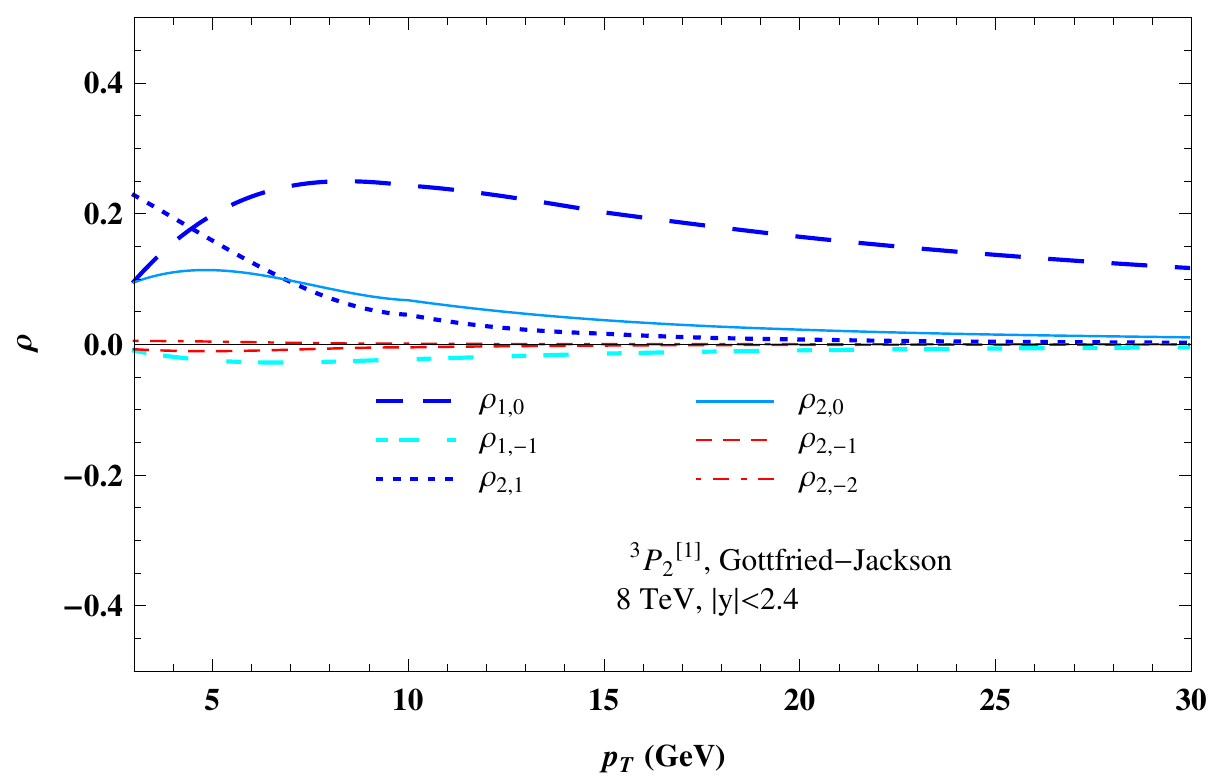}
(c) \caption{\label{fig:rhopt}(color online). Real parts of the nondiagonal SDMEs
(normalized by
$\frac{\rm{d}\sigma_{tot}}{\rm{d}p_T}=2\frac{\rm{d}\sigma_{22}}{\rm{d}p_T}+2\frac{\rm{d}\sigma_{11}}{\rm{d}p_T}+\frac{\rm{d}\sigma_{00}}{\rm{d}p_T}$)
for $\tpts$ in the (a) HX, (b) Collins-Soper, and (c) Gottfried-Jackson frames. In Fig.7(a), $\rho_{1,0}$, $\rho_{2,1}$, and $\rho_{2,-1}$ are almost zero. In Fig.7(b),  $\rho_{1,0}$, $\rho_{1,-1}$, $\rho_{2,1}$, and $\rho_{2,-1}$ vanish.}
\end{figure}
\begin{figure}
\includegraphics[width=8.5cm]{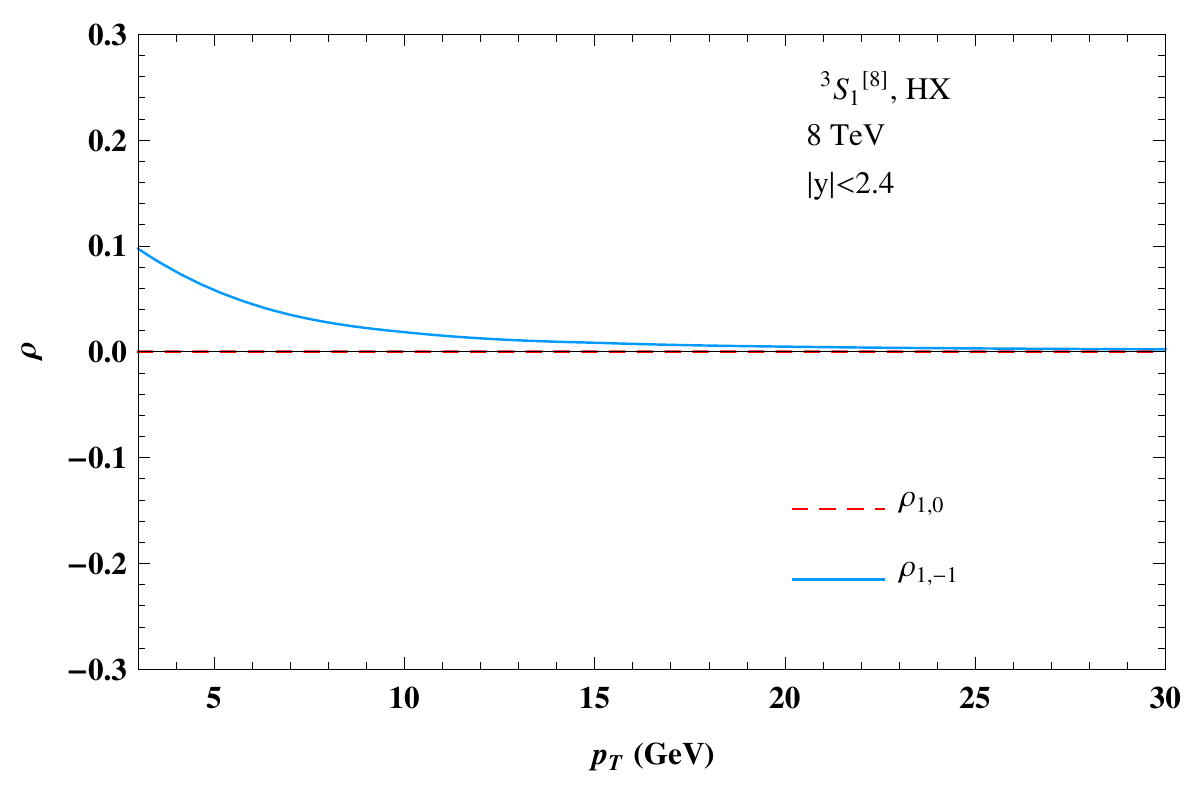}
(a)
\includegraphics[width=8.5cm]{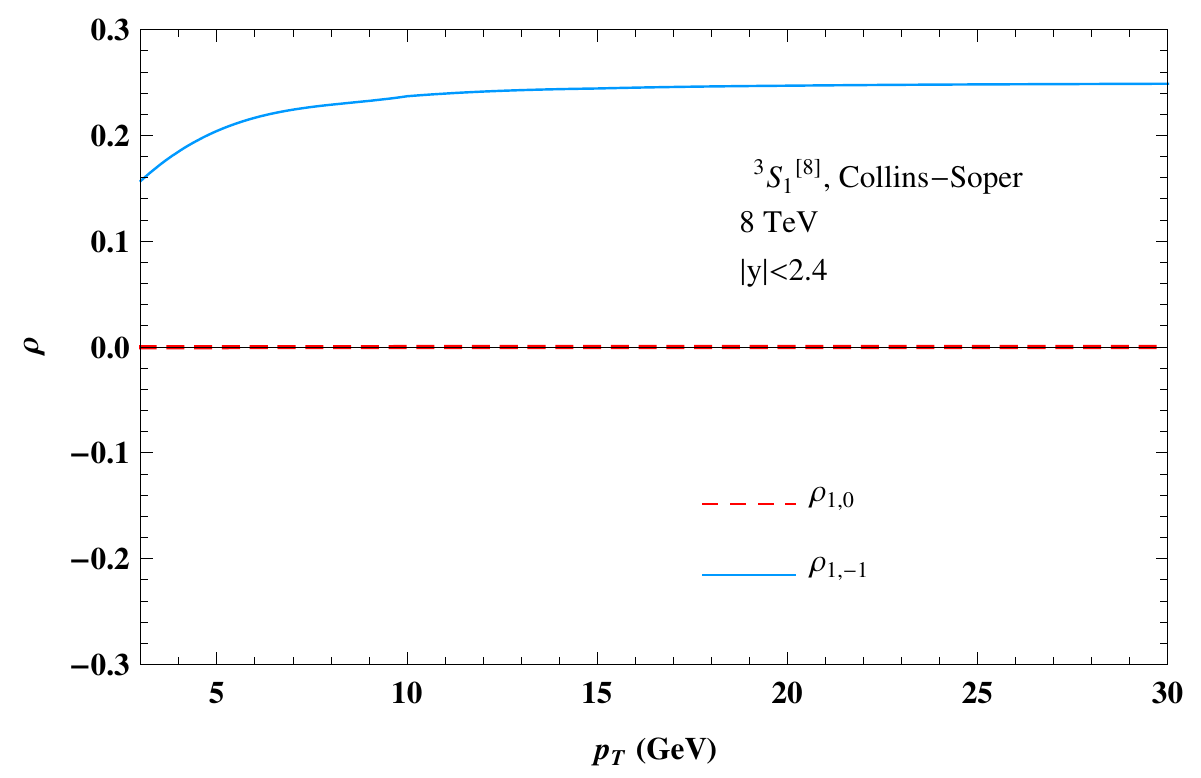}
(b)
\includegraphics[width=8.5cm]{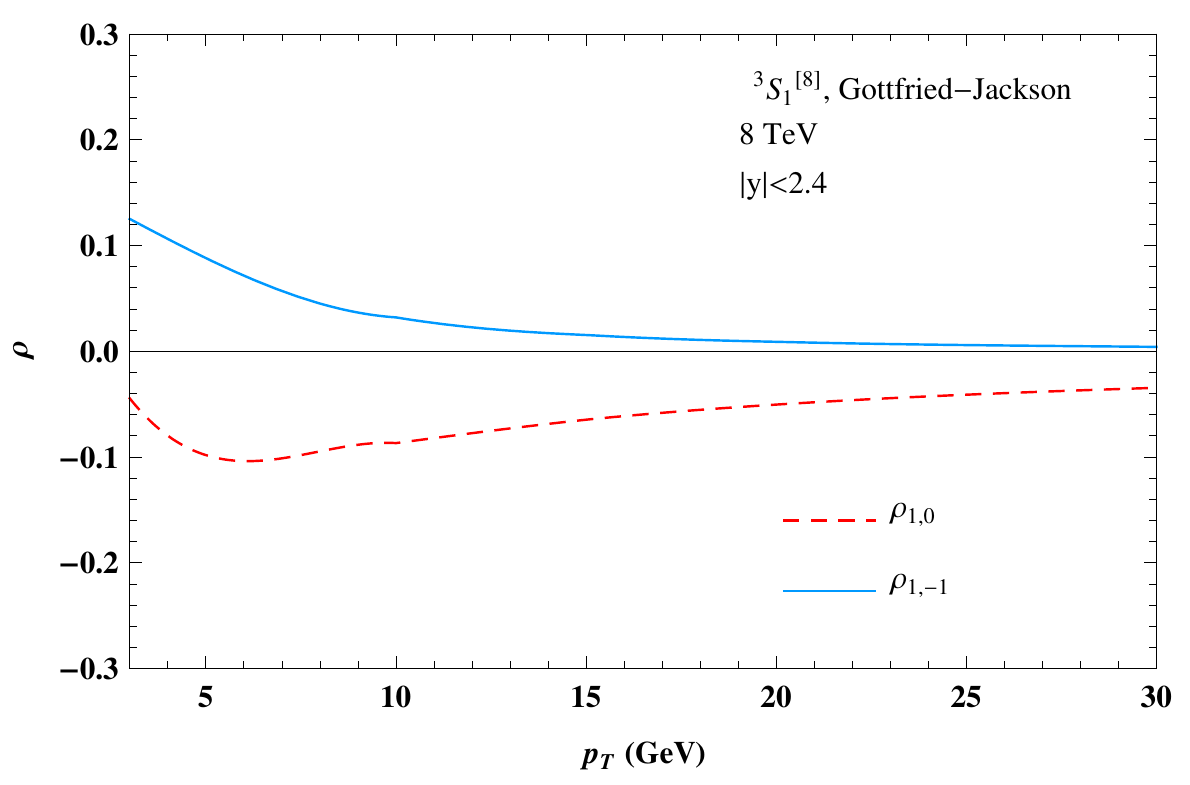}
(c) \caption{\label{fig:rhos}(color online). Real parts of the nondiagonal SDMEs
(normalized by
$\frac{\rm{d}\sigma_{tot}}{\rm{d}p_T}=2\frac{\rm{d}\sigma_{11}}{\rm{d}p_T}+\frac{\rm{d}\sigma_{00}}{\rm{d}p_T}$)
for $\so$ in the (a) HX, (b) Collins-Soper, and (c) Gottfried-Jackson frames.}
\end{figure}
\begin{figure}
\includegraphics[width=8.5cm]{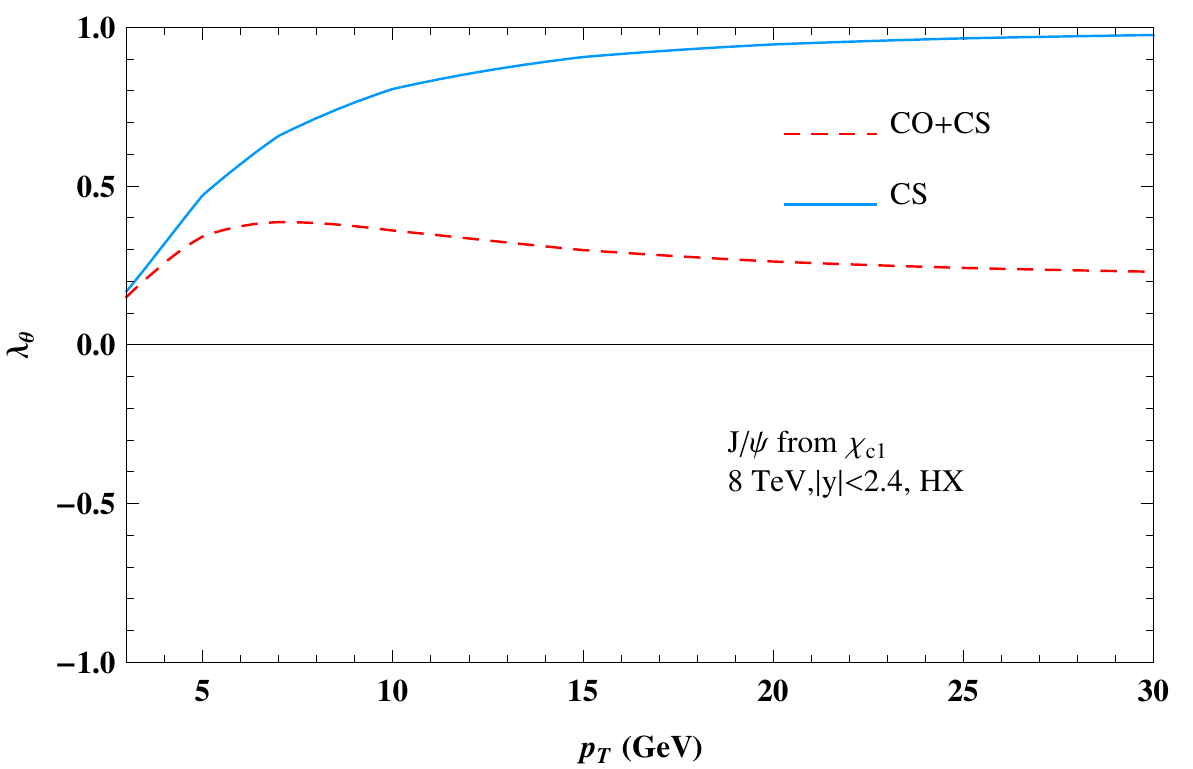}
(a)
\includegraphics[width=8.5cm]{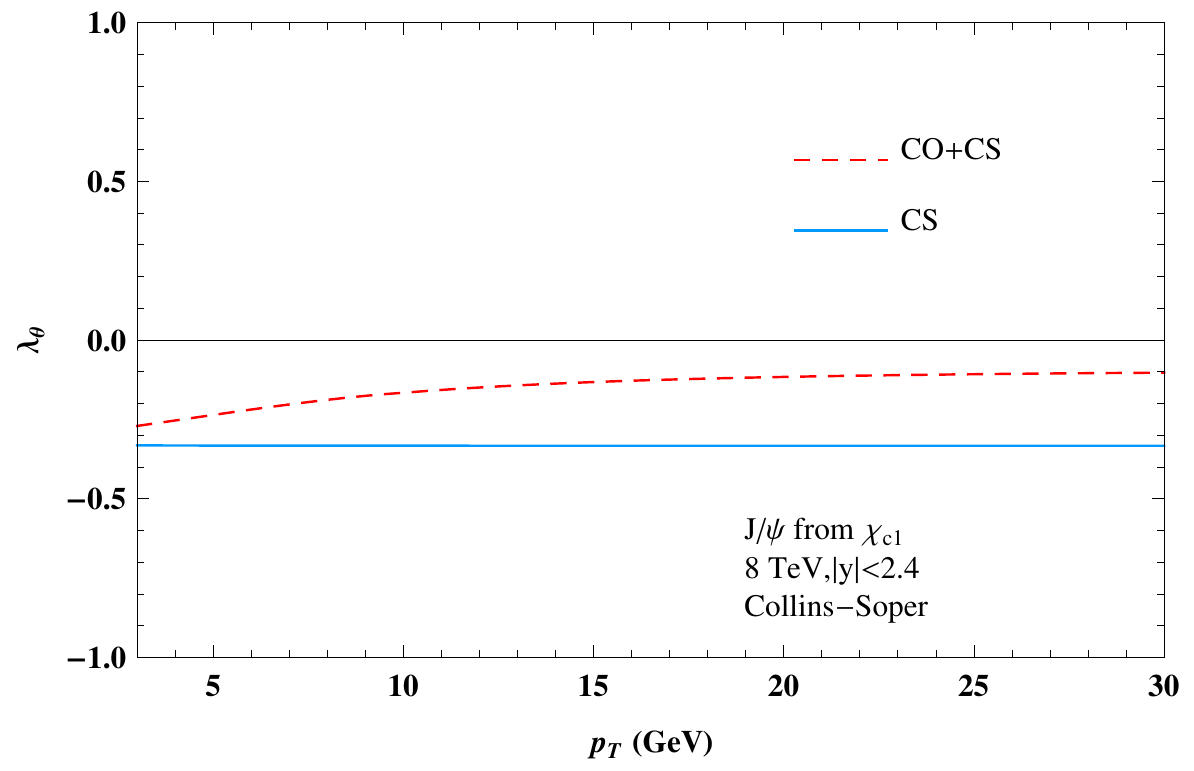}
(b)
\includegraphics[width=8.5cm]{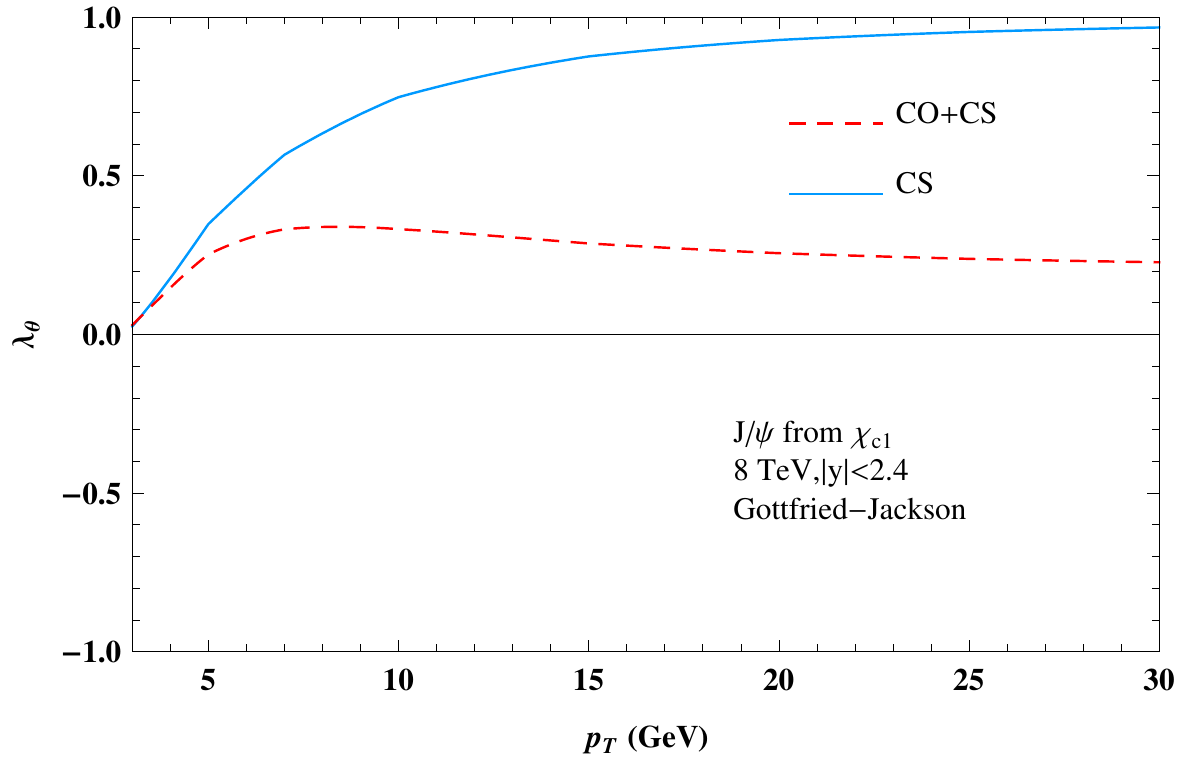}
(c) \caption{\label{fig:th1jpsi}(color online). Transverse momentum dependence of
the parameter $\lambda_{\th}$ of $\chico\to\jpsi\gamma$ angular
distribution, in the (a) HX, (b) Collins-Soper, and (c) Gottfried-Jackson frames.}
\end{figure}
\begin{figure}
\includegraphics[width=8.5cm]{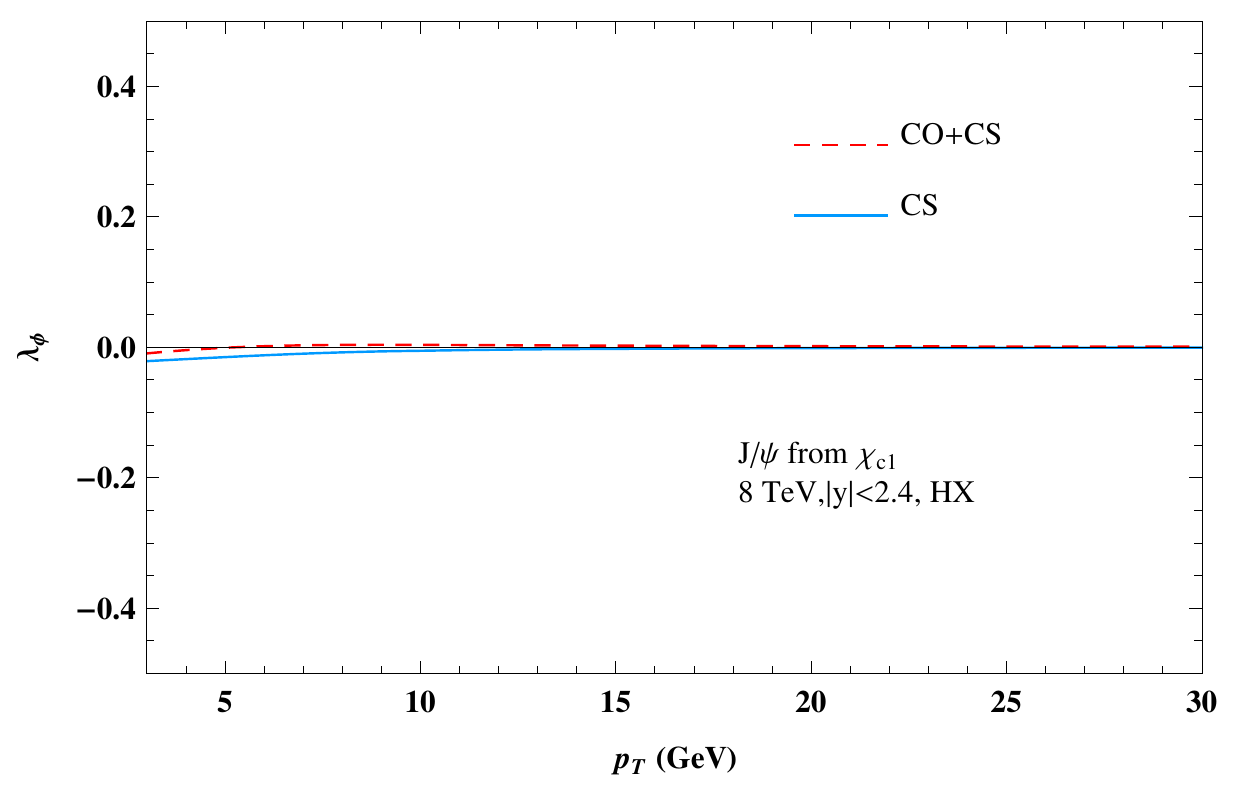}
(a)
\includegraphics[width=8.5cm]{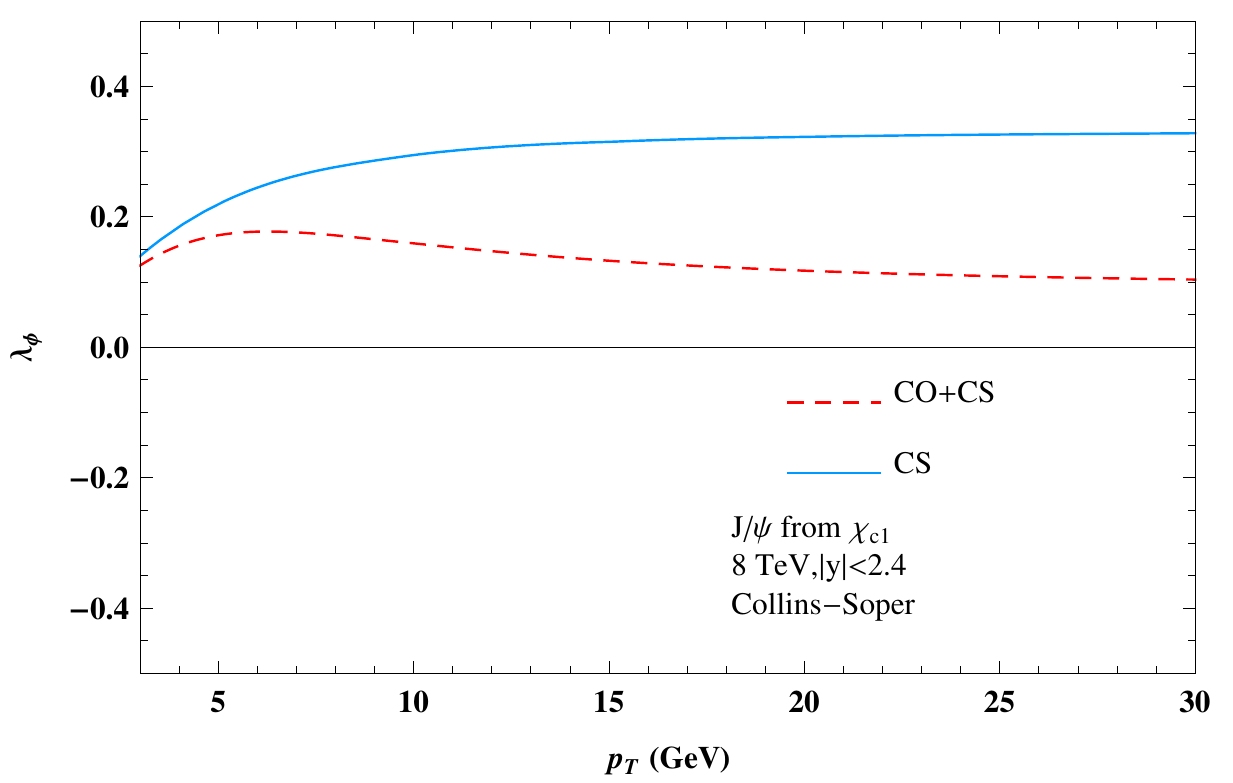}
(b)
\includegraphics[width=8.5cm]{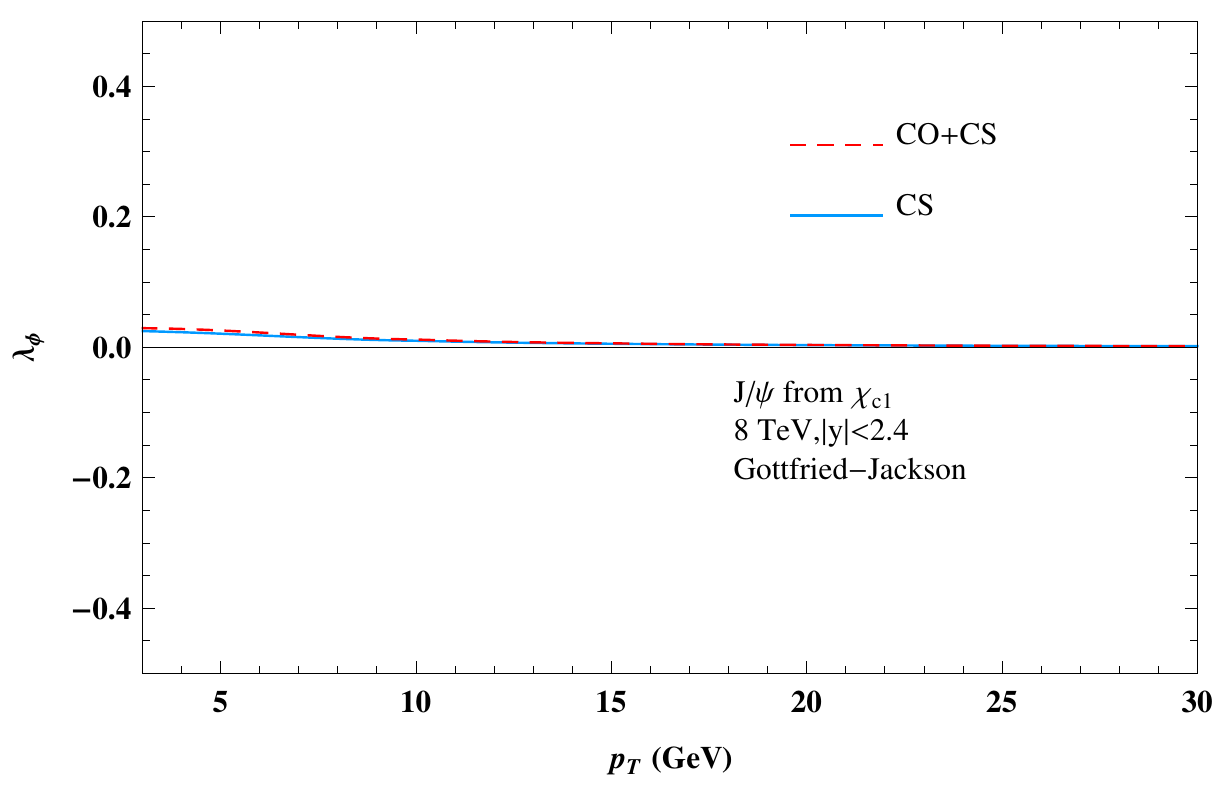}
(c) \caption{\label{fig:phi1jpsi}(color online). Transverse momentum dependence of
$\lambda_{\phi}$ of $\chico\to\jpsi\gamma$ angular distribution, in the (a) HX, (b) Collins-Soper, and (c) Gottfried-Jackson frames.}
\end{figure}
\begin{figure}
\includegraphics[width=8.5cm]{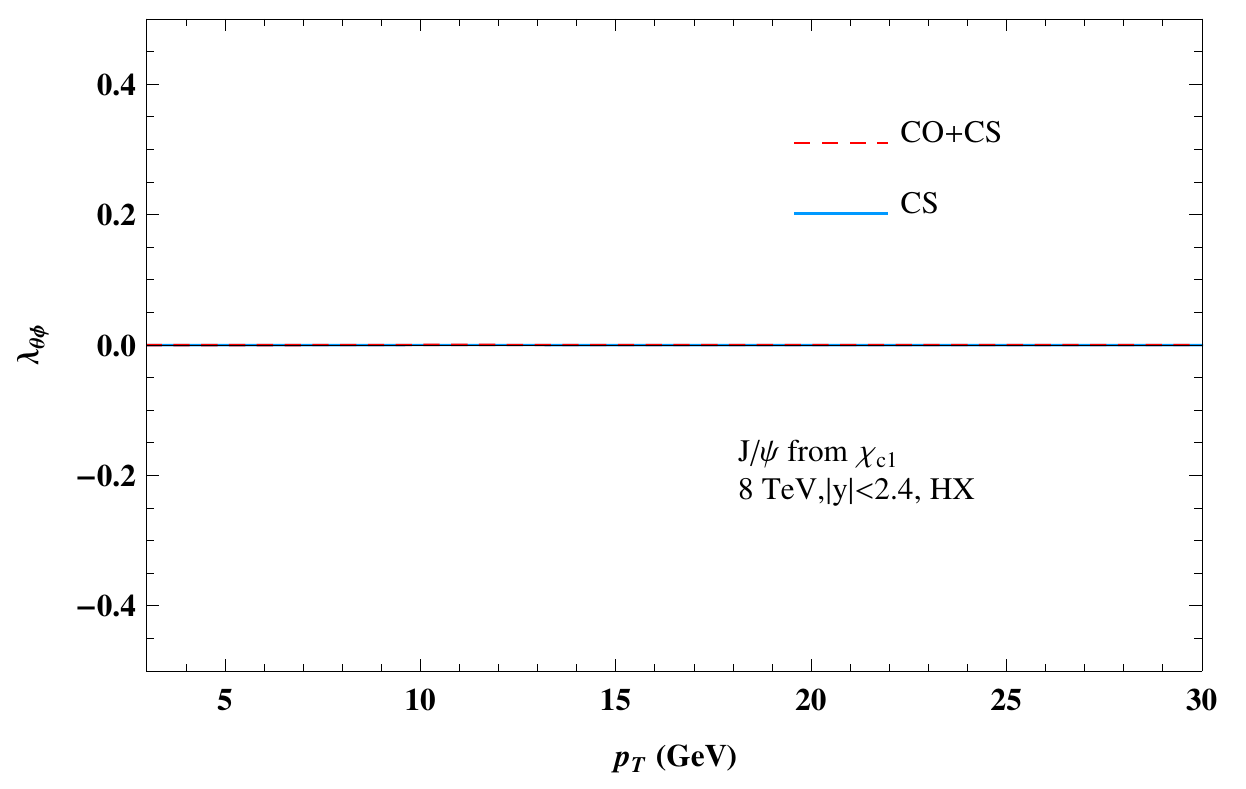}
(a)
\includegraphics[width=8.5cm]{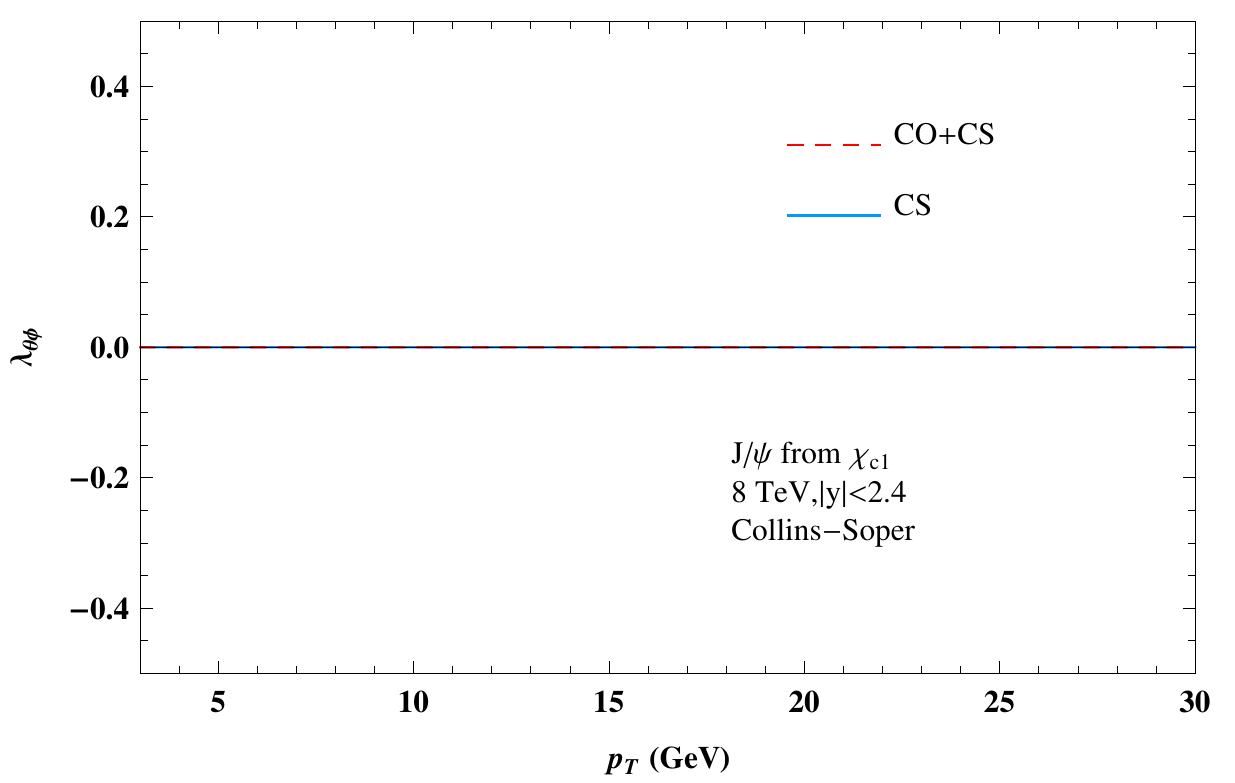}
(b)
\includegraphics[width=8.5cm]{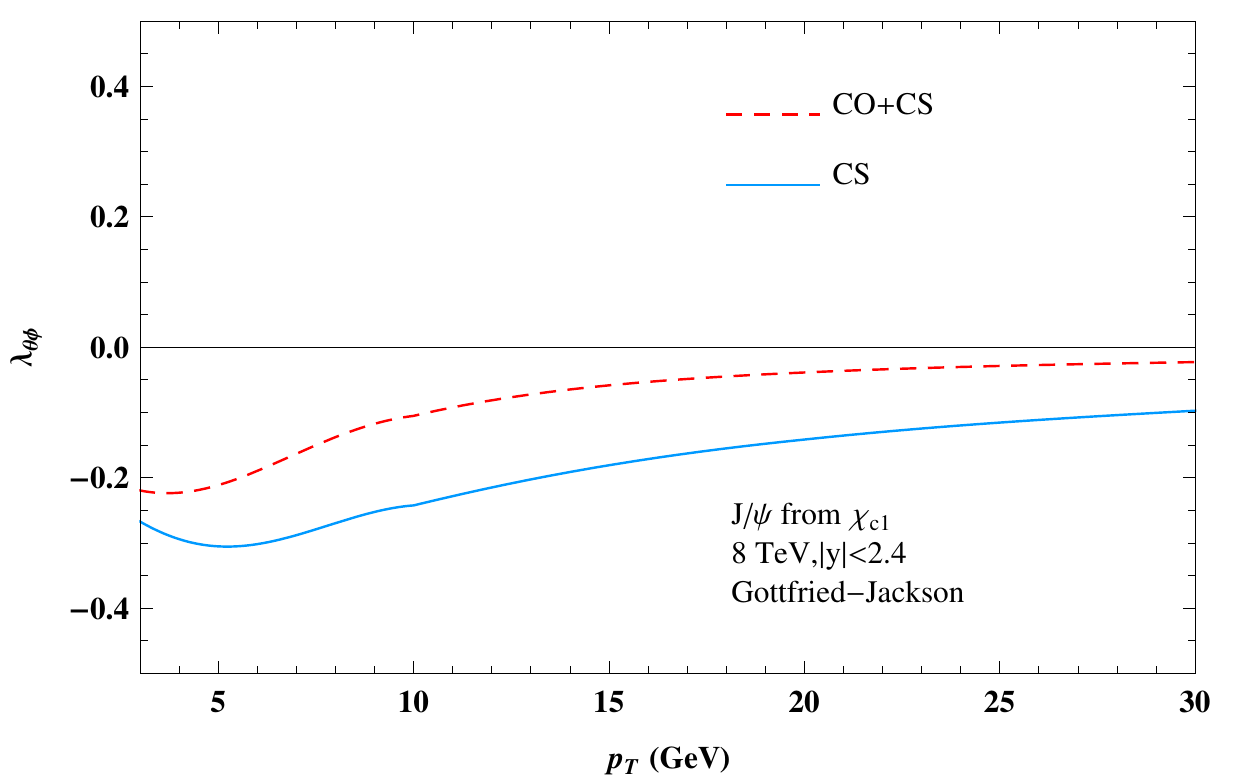}
(c) \caption{\label{fig:thphi1jpsi}(color online). Transverse momentum dependence
of $\lambda_{\th\phi}$ of $\chico\to\jpsi\gamma$ angular
distribution, in the (a) HX, (b) Collins-Soper, and (c) Gottfried-Jackson frames.}
\end{figure}
\begin{figure}
\includegraphics[width=8.5cm]{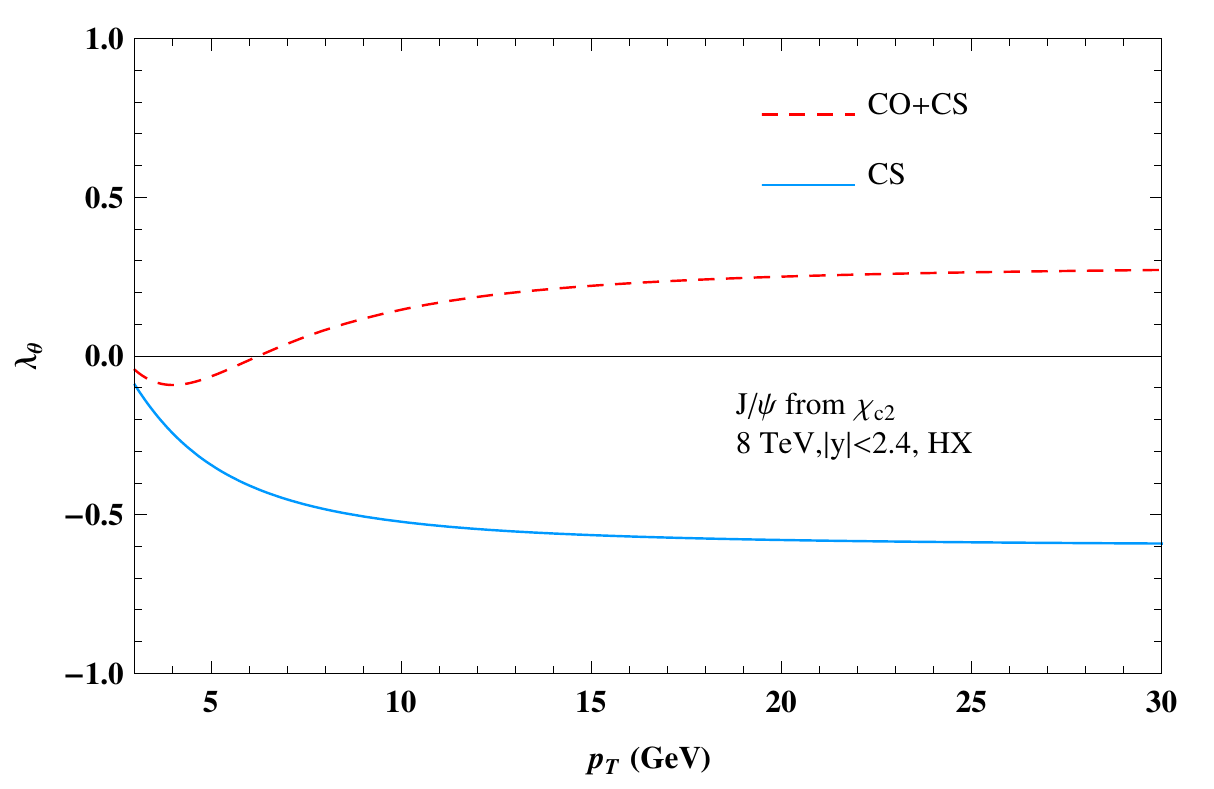}
(a)
\includegraphics[width=8.5cm]{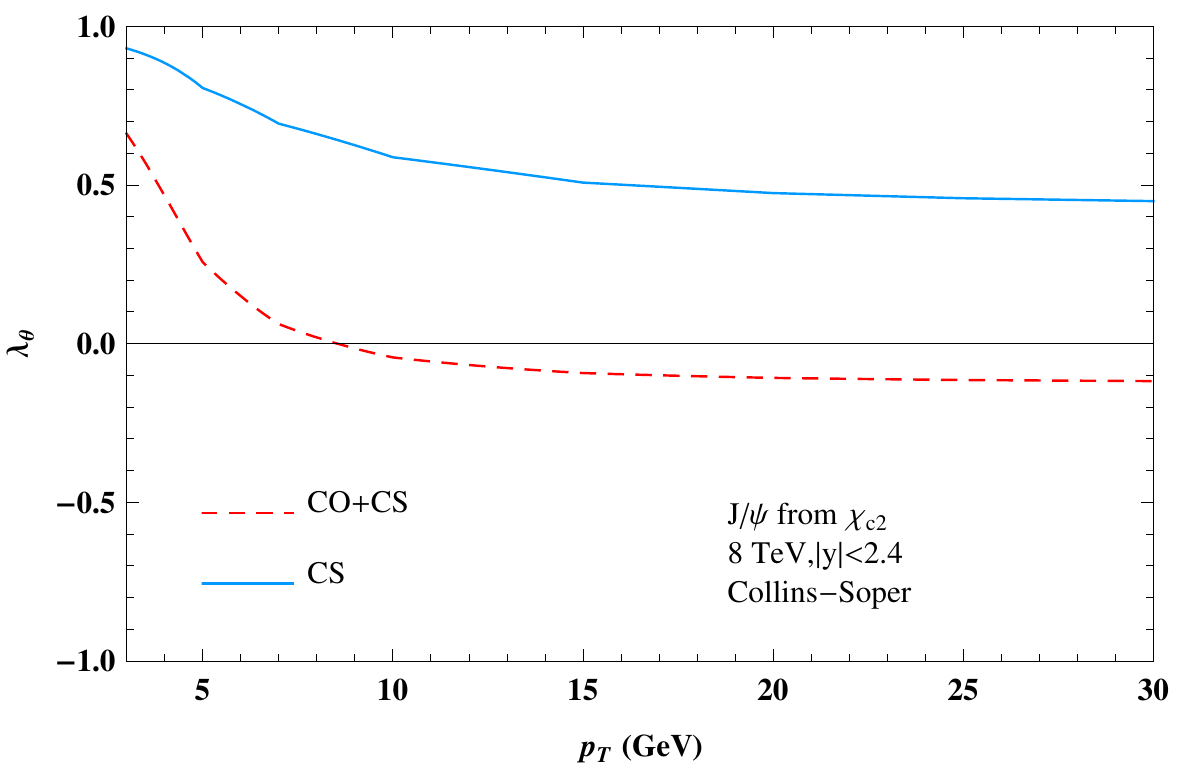}
(b)
\includegraphics[width=8.5cm]{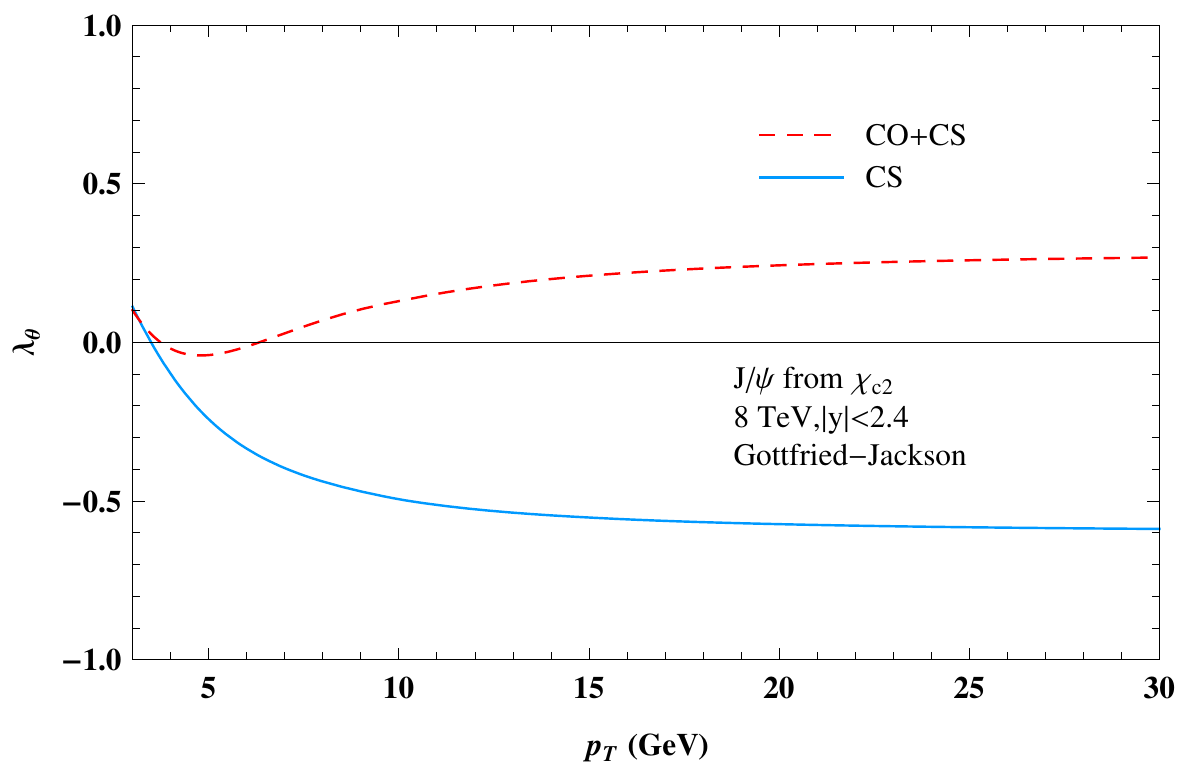}
(c) \caption{\label{fig:th2jpsi}(color online). Transverse momentum dependence of
$\lambda_{\th}$ of $\chict\to\jpsi\gamma$ angular distribution, in
the three polarization frames.}
\end{figure}
\begin{figure}
\includegraphics[width=8.5cm]{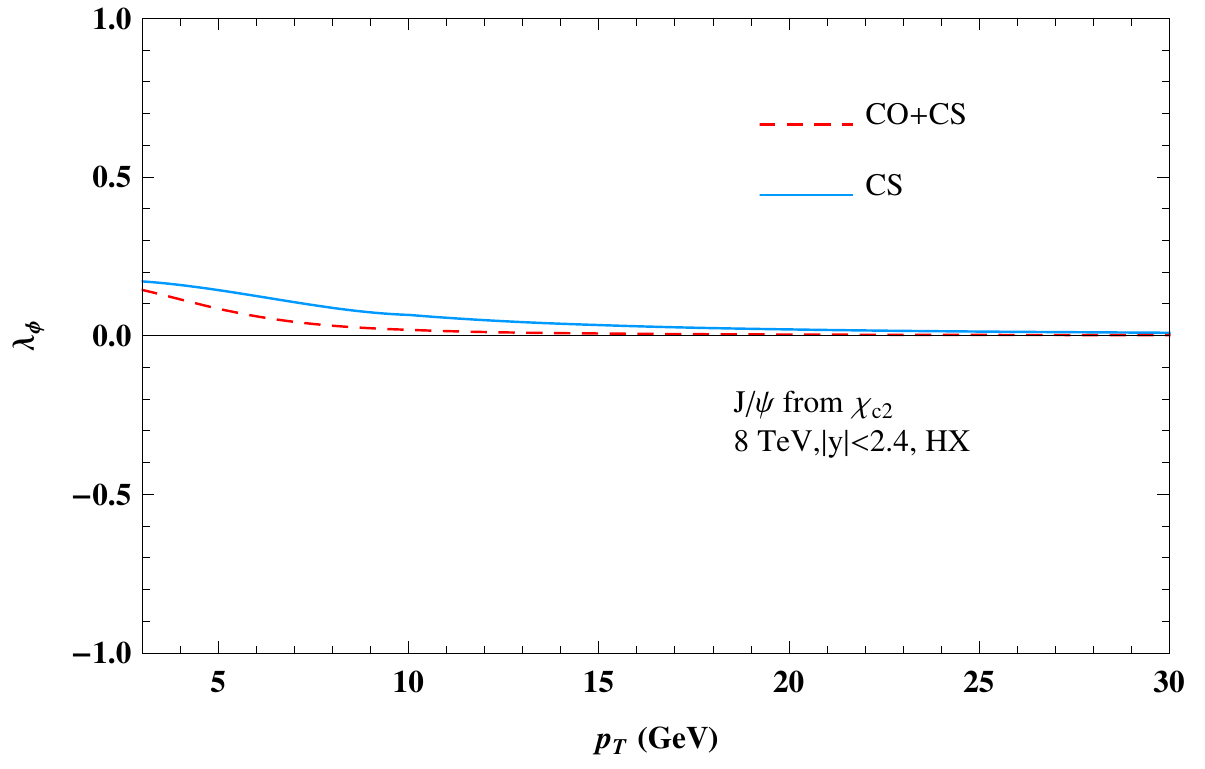}
(a)
\includegraphics[width=8.5cm]{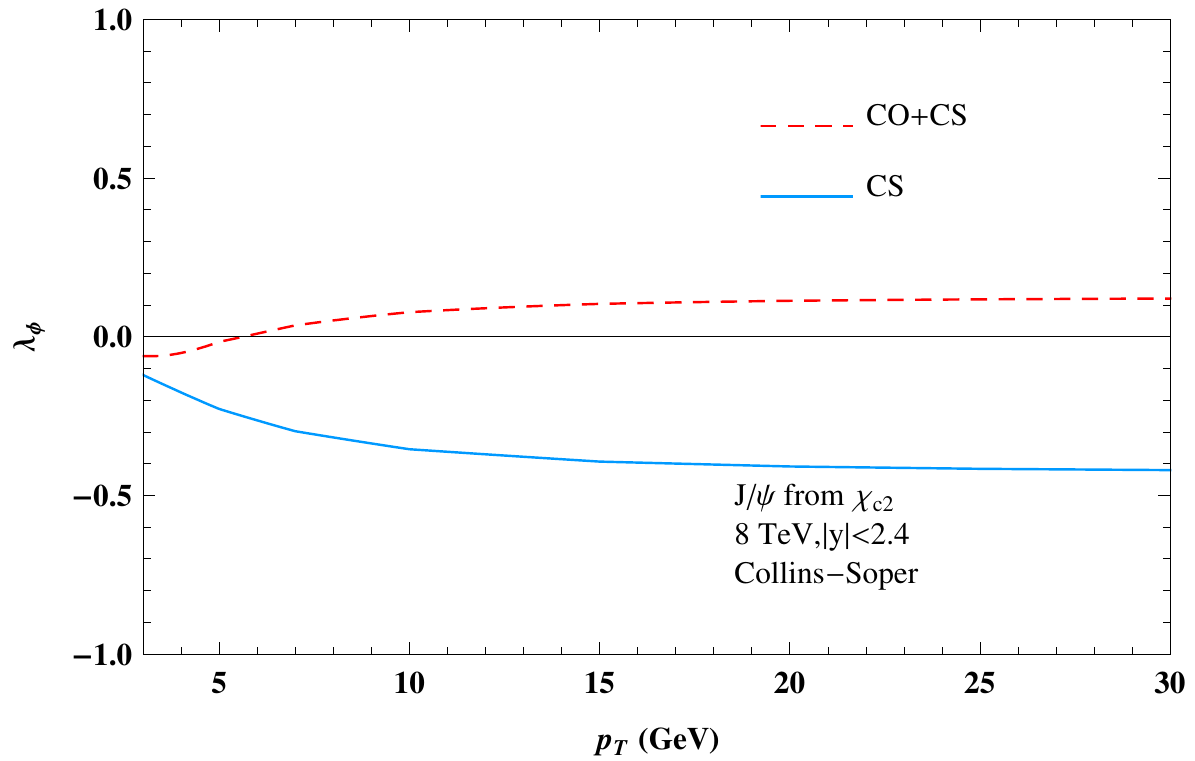}
(b)
\includegraphics[width=8.5cm]{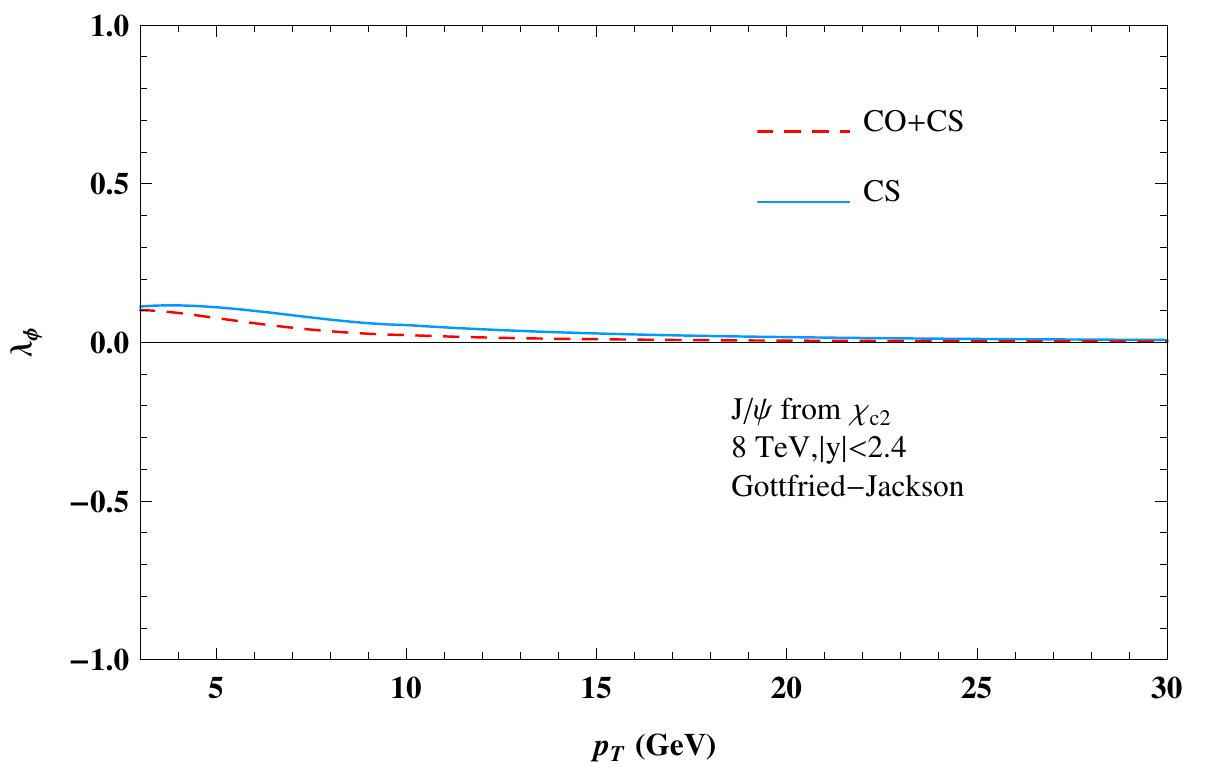}
(c) \caption{\label{fig:phi2jpsi}(color online). Transverse momentum dependence of
$\lambda_{\phi}$ of $\chict\to\jpsi\gamma$ angular distribution, in the (a) HX, (b) Collins-Soper, and (c) Gottfried-Jackson frames.}
\end{figure}
\begin{figure}
\includegraphics[width=8.5cm]{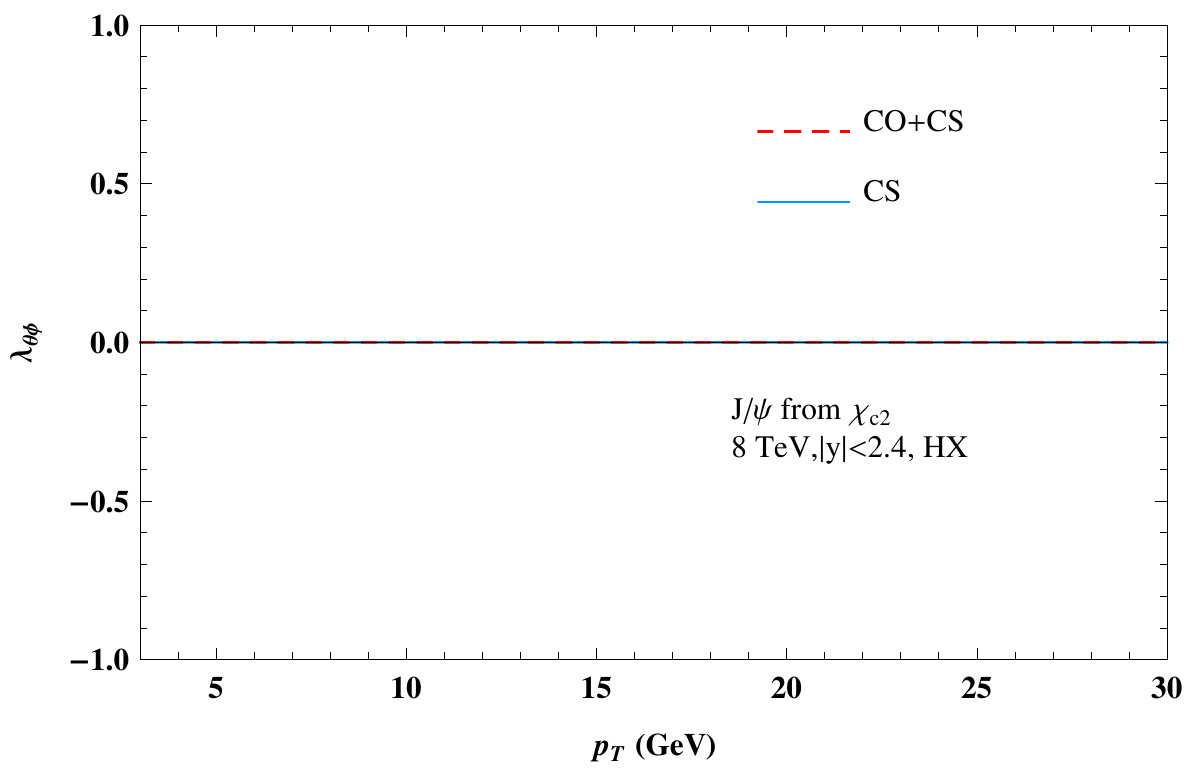}
(a)
\includegraphics[width=8.5cm]{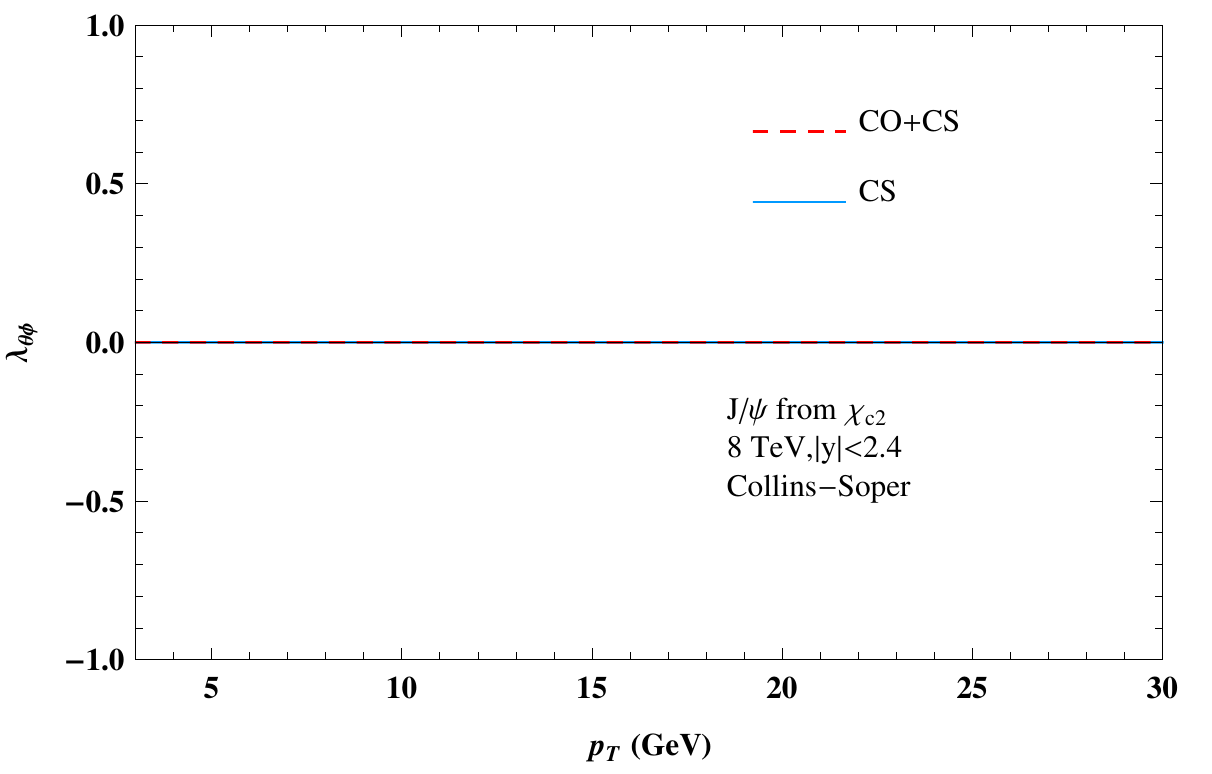}
(b)
\includegraphics[width=8.5cm]{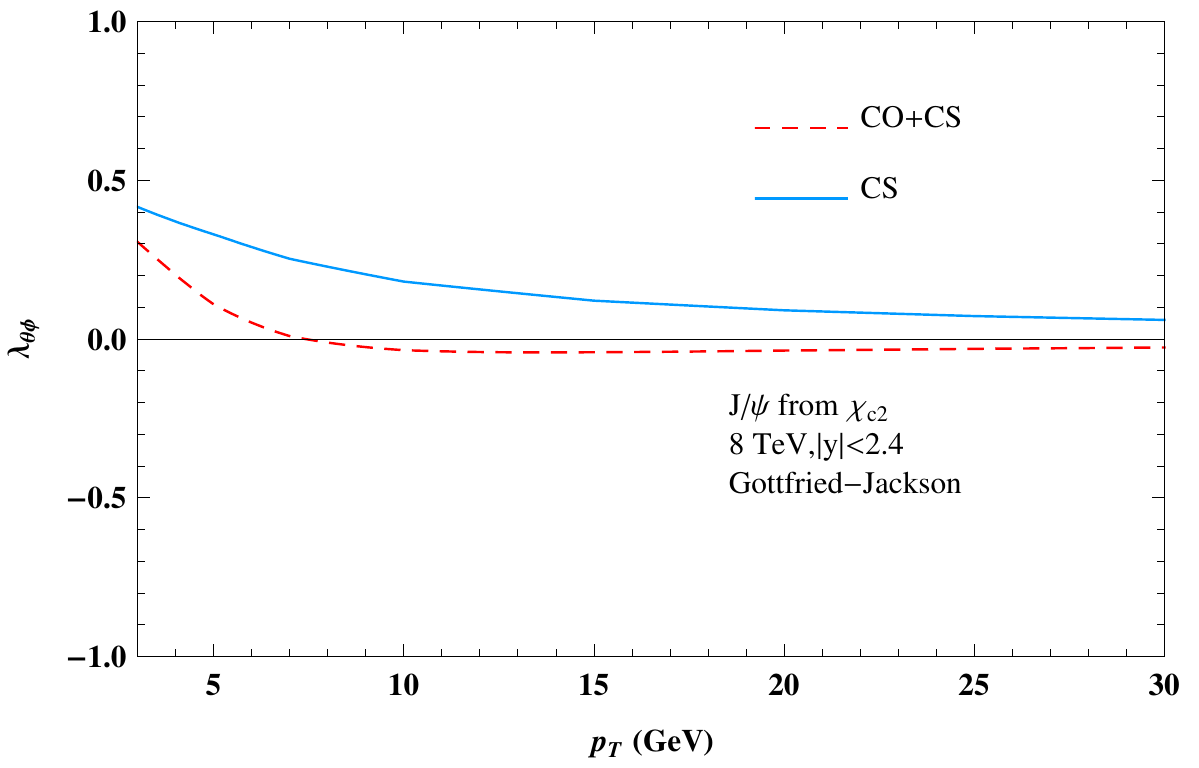}
(c) \caption{\label{fig:thphi2jpsi}(color online). Transverse momentum dependence
of $\lambda_{\th\phi}$ of $\chict\to\jpsi\gamma$ angular
distribution, in the (a) HX, (b) Collins-Soper, and (c) Gottfried-Jackson frames.}
\end{figure}

\begin{figure}
\includegraphics[width=8.5cm]{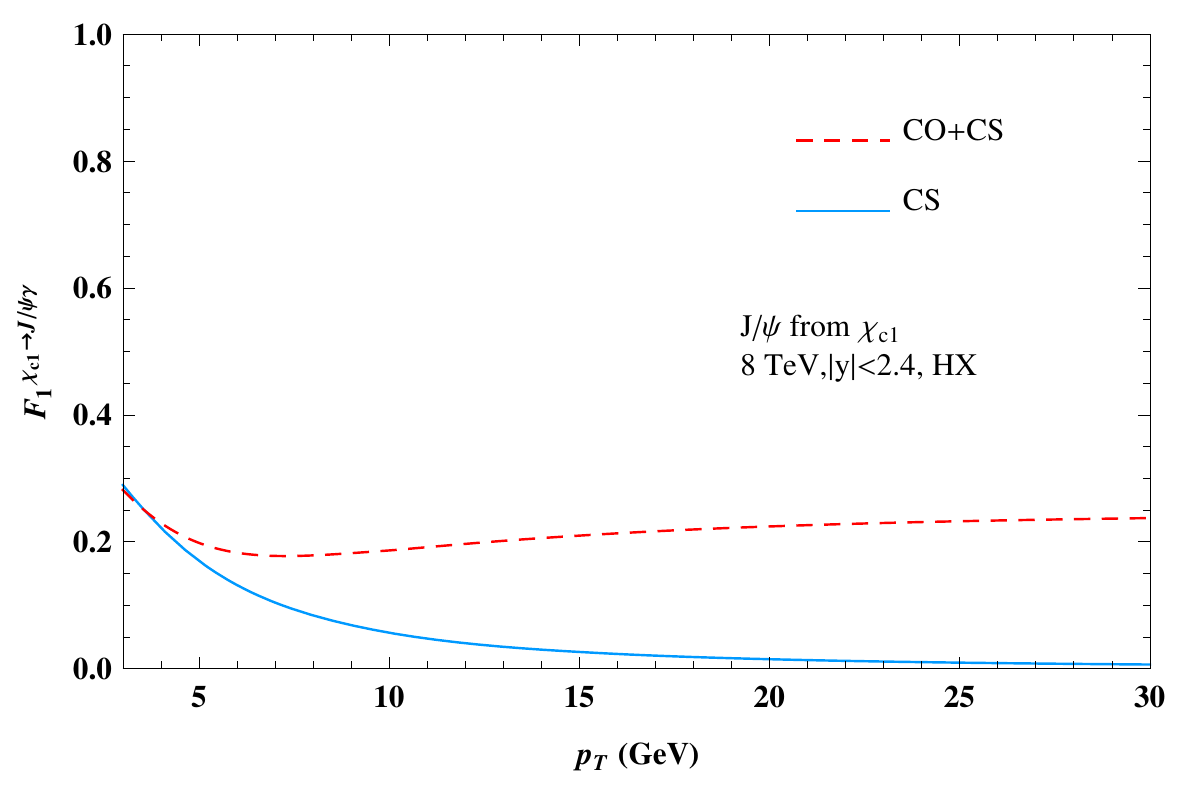}
(a)
\includegraphics[width=8.5cm]{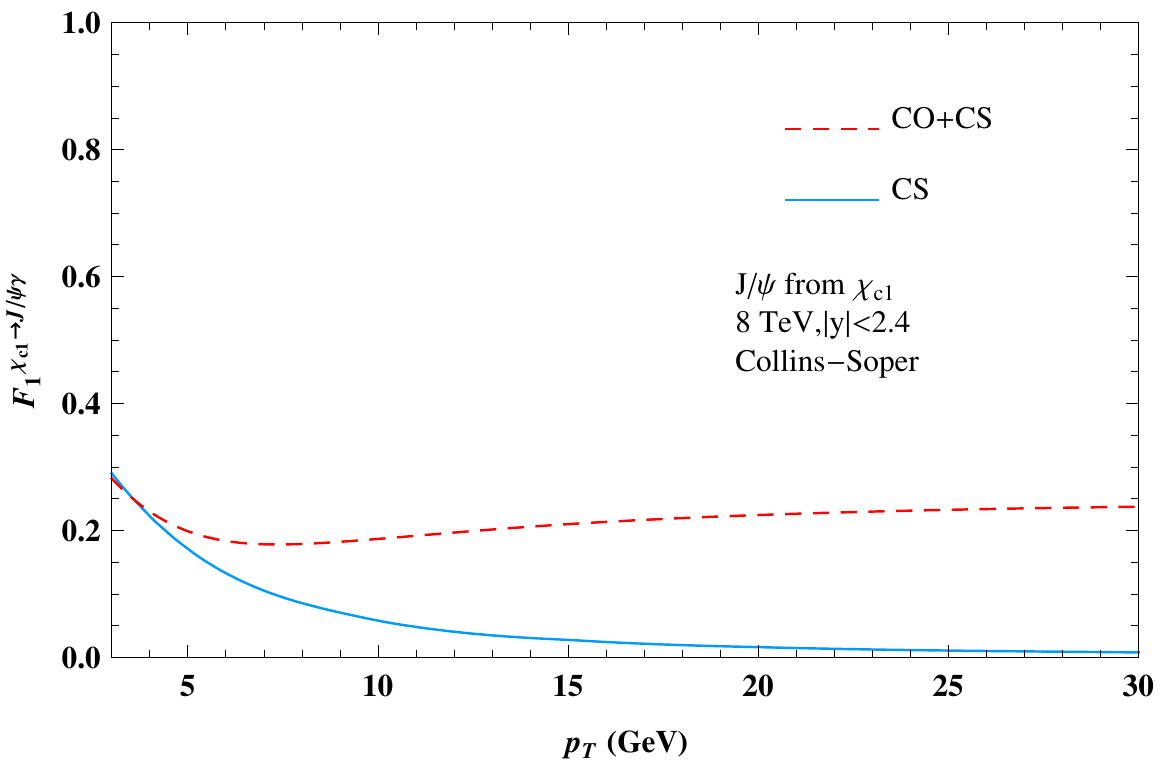}
(b)
\includegraphics[width=8.5cm]{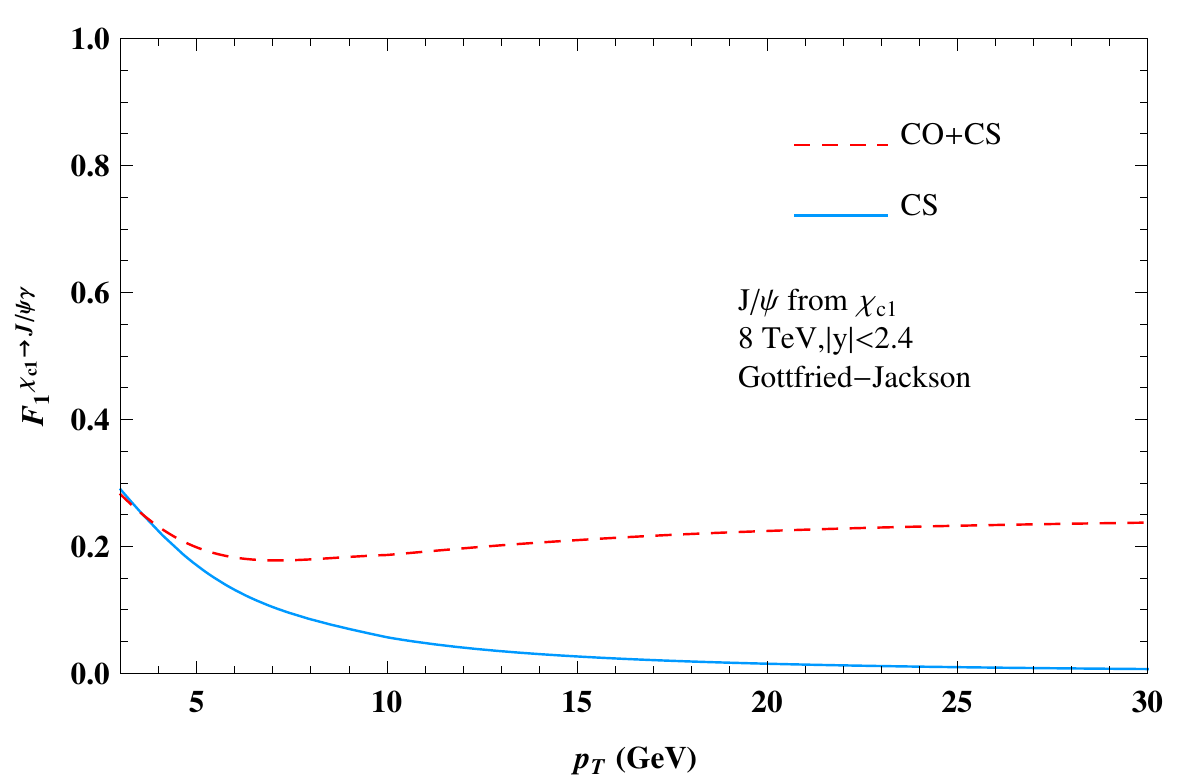}
(c) \caption{\label{fig:F1jpsi}(color online). Transverse momentum dependence of
the frame-independent parameter $F^{\chico\to\jpsi\gamma}_1$
[defined in Eq.(\ref{eq:F1})], in the (a) HX, (b) Collins-Soper, and (c) Gottfried-Jackson frames.}
\end{figure}

\begin{figure}
\includegraphics[width=8.5cm]{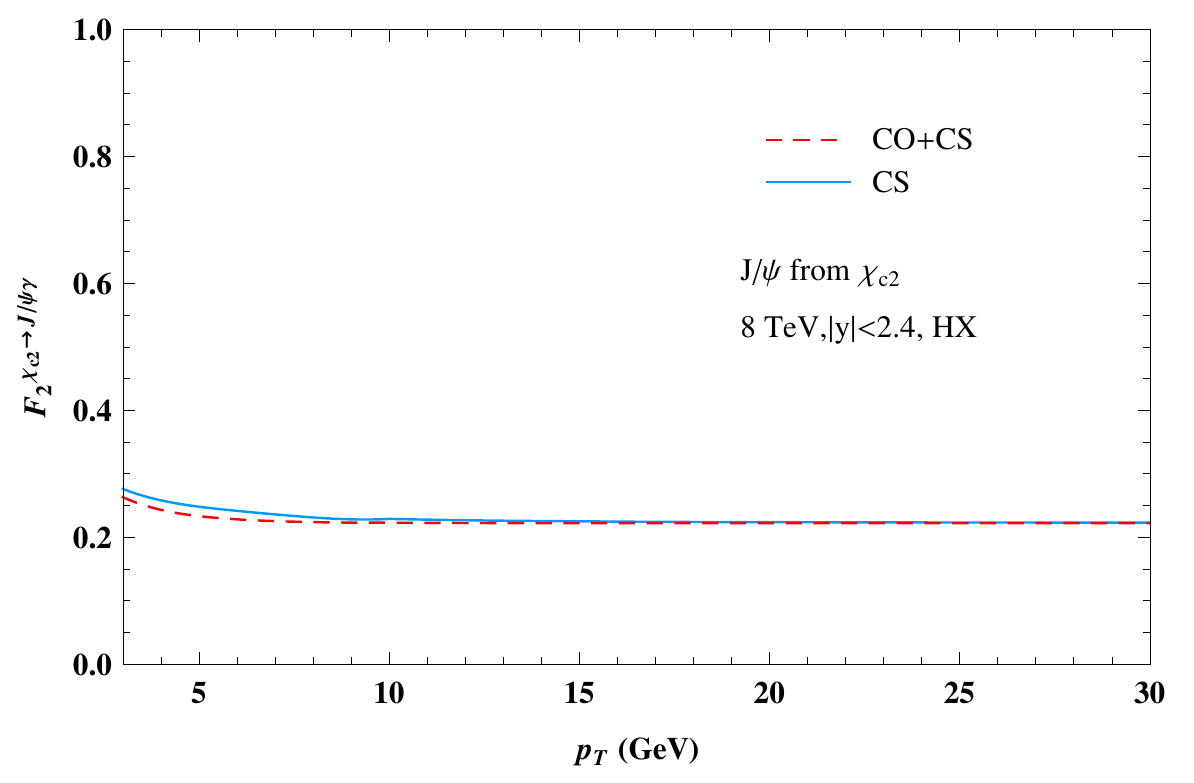}
(a)
\includegraphics[width=8.5cm]{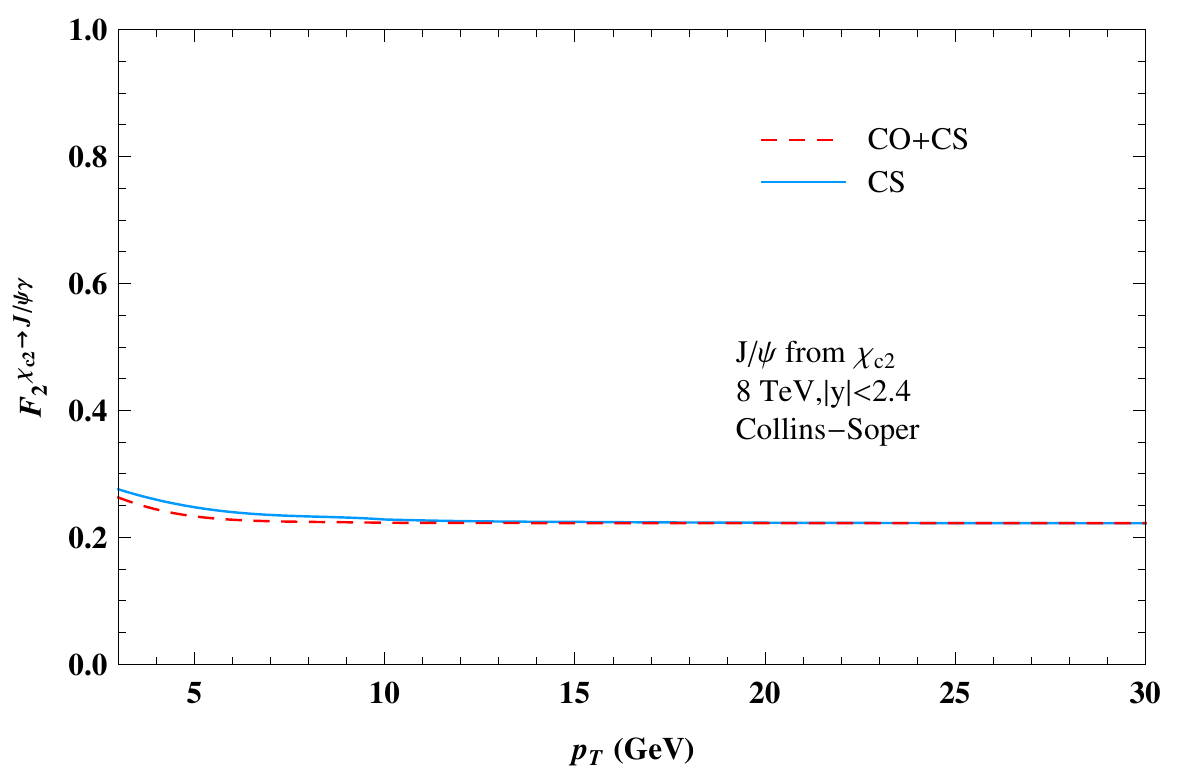}
(b)
\includegraphics[width=8.5cm]{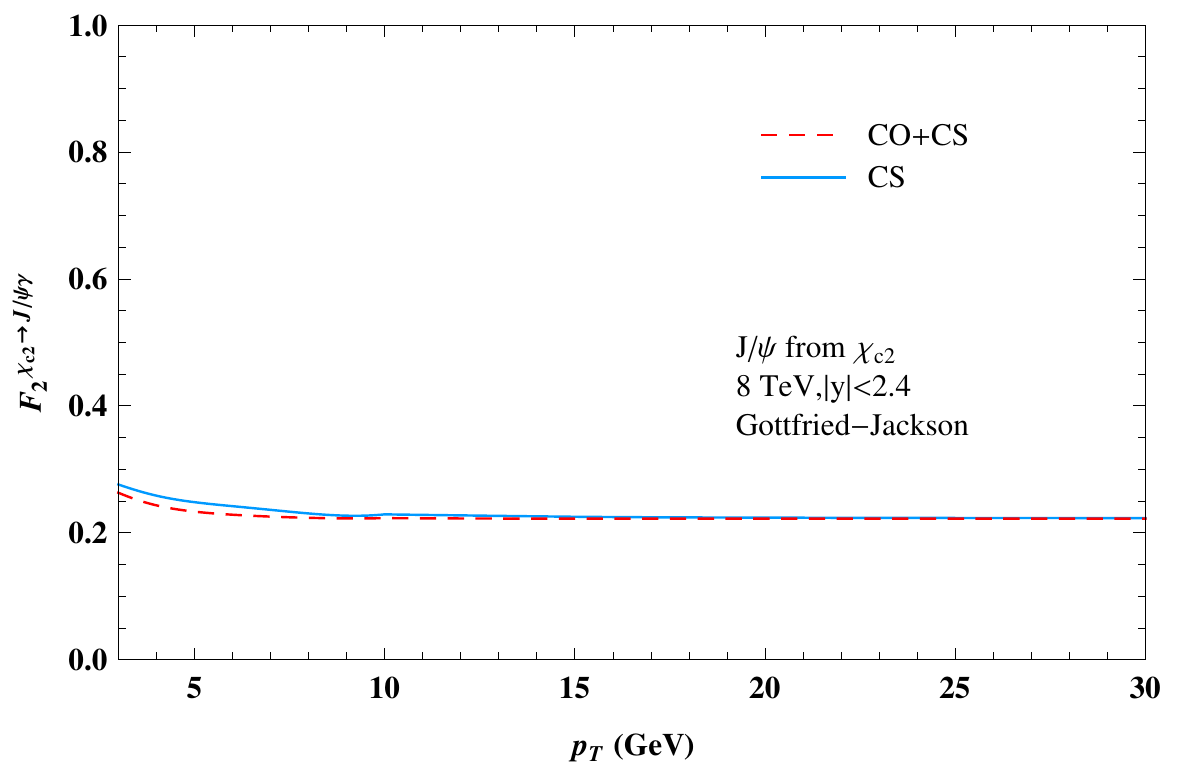}
(c) \caption{\label{fig:F2jpsi}(color online). Transverse momentum dependence of
the frame-independent parameter $F^{\chict\to\jpsi\gamma}_2$
[defined in Eq.(\ref{eq:F2})], in the (a) HX, (b) Collins-Soper, and (c) Gottfried-Jackson frames.}
\end{figure}

%
\providecommand{\href}[2]{#2}\begingroup\raggedright\endgroup

\end{document}